\documentclass[epj]{svjour}
\usepackage{graphics,epsfig}
\newcommand{\nn}{{\nonumber}\\}
\def\ket #1{|#1\rangle}
\def\tens{\otimes}

\def\btens{\bigotimes}
\def\Id{{\rm 1\kern-.3em I}}
\newcommand{\R}{{\rm I\kern-.2emR}}
\newcommand{\C}{{\ifmmode\mathchoice{{\rm C\kern-.4em\raisebox{.0ex}{\rule{.05ex}{.67em}}\kern.4em}}%
{{\rm C\kern-.4em\raisebox{.03ex}{\rule{.05ex}{.67em}}\kern.4em}}%
{{\rm C\kern-.3em\raisebox{.06ex}{\rule{.05ex}{.45em}}\kern.3em}}%
{{\rm C\kern-.3em\raisebox{.06ex}{\rule{.05ex}{.3em}}\kern.3em}}\else
{{\rm C\kern-.4em\raisebox{.03ex}{\rule{.05ex}{.67em}}\kern.4em}}\fi}}
\def\bra #1{\langle #1|}
\def\ket #1{|#1\rangle}
\def\SP #1 #2{\langle #1|#2\rangle}
\def\VacBra{\bra 0}
\def\VacKet{\ket 0}
\def\VacExpect #1{\VacBra #1\VacKet}

\def\SpinorComp(#1,#2,#3){\Psi^{#1}_{#3}(#2_{#1})}
\def\AdSpinorComp(#1,#2,#3){\overline\Psi^{#1}_{#3}(#2_{#1})}
\def\spinorComp(#1,#2,#3){\Psi_{#3}(#2_{#1})}
\def\AdspinorComp(#1,#2,#3){\overline\Psi_{#3}(#2_{#1})}
\def\Spinor(#1,#2){\Psi^{#1}(#2_{#1})}
\def\AdSpinor(#1,#2){\bar\Psi^{#1}(#2_{#1})}
\def\SpinorI(#1,#2){\Psi_{Ip}^{#1}(#2_{#1})}
\def\AdSpinorI(#1,#2){\overline\Psi_{Ip}^{#1}(#2_{#1})}
\def\pri {^\prime}

\def\FeyProp(#1,#2,#3){S^{#1}_F(#2_{#1},#3_{#1})}
\def\FeyPropComp(#1,#2,#3,#4,#5){S^{#1}_{F\;#4 #5}(#2_{#1},#3_{#1})}
\def\metric(#1,#2){\langle #1, #2\rangle}
\def\bfgrk #1{\mbox{\boldmath$#1$}}
\def\MSD (#1,#2,#3,#4){[\Delta\frac{#1}{2}^#2]_{#3}(#4)}
\def\MSN (#1,#2,#3,#4){[N\frac{#1}{2}^#2]_{#3}(#4)}
\def\MSL (#1,#2,#3,#4){[\Lambda\frac{#1}{2}^#2]_{#3}(#4)}
\def\MSS (#1,#2,#3,#4){[\Sigma\frac{#1}{2}^#2]_{#3}(#4)}
\def\MSX (#1,#2,#3,#4){[\Xi\frac{#1}{2}^#2]_{#3}(#4)}
\def\MSO (#1,#2,#3,#4){[\Omega\frac{#1}{2}^#2]_{#3}(#4)}

\begin{document}
\onecolumn
\title{Relativistic quark models of baryons with instantaneous forces}
\subtitle{Theoretical background}
\author{Ulrich L\"oring\thanks{e-mail: {\tt loering@itkp.uni-bonn.de}}, Klaus Kretzschmar, Bernard Ch.~Metsch \and  Herbert R.~Petry
}                     
%
%
\institute{Institut f\"ur Theoretische Kernphysik, Universit\"at Bonn, Nu{\ss}allee 14--16, D--53115 Bonn, Germany}
%
\date{}
%
\authorrunning{U.~L\"oring {\it et al.}}  

\abstract{This is the first of a series of three papers treating light baryon resonances (up to 3 GeV)
  within a relativistically covariant quark model based on the
  three-fermion Bethe-Salpeter equation with instantaneous two-
  and three-body forces.  In this paper we give a unified description of the
  theoretical background and demonstrate how to solve the Bethe-Salpeter
  equation by a reduction to the Salpeter equation. The specific new 
  features of our covariant Salpeter model with respect to the usual nonrelativistic
  quark model are discussed in detail. The purely theoretical results obtained in this
  paper will be applied numerically to explicit quark models for light baryons in two
  subsequent papers \cite{Loe01b,Loe01c}.
} 
\maketitle
%
\section{Introduction}
\label{intro}
The classification of baryon resonances as three-quark states within
nonrelativistic potential models has a long and very successful history. It is
however unclear how to relate such models to QCD. Some ingredients of nonrelativistic quark
models emerge from QCD, {\it e.g.} massive quarks as a consequence of chiral symmetry
breaking, linear confinement potentials (on the lattice) due to the nonabelian gauge coupling
and some candidates for spin-dependent residual interactions like
one-gluon-exchange or instanton-induced quark forces. For light quark flavors
it is however unclear, how to unite these features in a common picture.  The
main obstacle is the nonrelativistic approach which seems to be completely
inadequate for small constituent quark masses and strong quark binding.

Quantum field theory seems to offer a solution to this problem,
replacing the nonrelativistic wavefunctions by Bethe-Salpeter
amplitudes obeying a suitable Bethe-Salpeter equation. In the case of
QCD none of the basic ingredients of these equations is reliably
known, {\it i.e.} we have no reliable prescription to calculate the
full quark propagators and interaction vertices. Moreover we meet a
serious problem with gauge invariance because the Bethe-Salpeter
amplitudes are gauge-dependent.  Nonetheless the general framework of
quantum field theory can be used for a reasonable phenomenological
description.  If we want to remain as close as possible to the
features of nonrelativistic quark models the Bethe-Salpeter equation
should contain free quark propagators with constituent quark masses
and instantaneous, unretarded interactions only. Both requirements are
purely phenomenological assumptions but reasonably justified by the
apparent success of nonrelativistic quark models. In this way these
Bethe-Salpeter amplitudes form a more suitable basis for quark models,
but respecting, in particular, relativistic covariance. As such it was
already successfully used for the description of light mesons
\cite{ReMu94,MuRe94,MuRe95,Mu96,MePe96,KoRi00,RiKo00}.  The baryon
Bethe-Salpeter equation with genuine instantaneous three-quark forces
is solved as in the mesonic calculations by a reduction to a
three-dimensional integral equation (Salpeter equation) which is
very similar to the Schr\"odinger equation. The spectrum
contains however also antiparticle solutions corresponding to particles
with charge conjugated quantum numbers. This situation is new and
needs a special discussion. Another complication arises when genuine
two-particle interactions are taken into account. In quark models this
is natural, when the (three-body) confinement forces are supplemented
by a two-body residual interaction (one-gluon-exchange, instanton
induced forces). In this case an effective three-body interaction
kernel has to be derived.

None of these features is entirely new, but there is no reference in the
literature which presents this theoretical background in a unified way.  The
purpose of this paper is to fill this gap. In two consecutive papers \cite{Loe01b,Loe01c} we
will use these purely theoretical results for specific calculations of the
baryon spectrum up to 3 GeV.

This paper is organized as follows: In section \ref{sec:bsequation} we
briefly recall how in quantum field theory bound states of three fermions
occur as poles in the six-point Green's function defining the Bethe-Salpeter
amplitudes as the corresponding residua at these poles. This property of the
Green's function is used in section \ref{sec:BSE} do derive simultaneously the
Bethe-Salpeter equation for the Bethe-Salpeter amplitudes and their
normalization condition in a simple and appealing way by a Laurent expansion
of the integral equation for the six-point Green's function in the vicinity of
this pole.  Section \ref{sec:sequation} is concerned with the reduction of
the full eight-dimensional Bethe-Salpeter equation to a six-dimensional
Salpeter equation by integrating out the relative energy dependence of the full
Bethe-Salpeter amplitudes. To this end we use a covariant formulation of the
instantaneous approximation for three- and two-body interaction kernels
and assume that the full quark propagators can be suitably approximated by their free
forms introducing effective constituent quark masses. In a first step, taking
only the genuine (instantaneous) three-body kernels into account, we show how a
straightforward reduction can then be performed, thus yielding a reduced equation
which may be formulated as an ordinary eigenvalue problem in Hamiltonian form,
where the Hamiltonian is hermitean with respect to a scalar product induced
by the normalization condition of the Salpeter amplitudes. Complications arise for
the more general case when also genuine two-particle interactions are taken
into account. This case needs a special discussion and we demonstrate that a
reduction to a Salpeter equation in the same Hamiltonian form can nevertheless
be achieved by deriving an effective instantaneous three-body kernel which
parameterizes all retardation effects of the unconnected two-body interactions.
In section \ref{bound_states_born} we present the Salpeter equation in Born
approximation of the quasi potential which constitutes the basic covariant
equation of our model. We discuss the structure and main features of the
Salpeter equation and its solutions with respect to the ordinary
nonrelativistic quark model. Special features discussed in this section are
the one-to-one correspondence of the Salpeter amplitudes to the states of the
nonrelativistic quark model and the additional anti-particle solutions of the
Salpeter equation.  Finally we give a summary and conclusion in section
\ref{sec:concl}.

\section{Green's functions and Bethe-Salpeter amplitudes}
\label{sec:bsequation}
In non-relativistic quantum mechanics, a bound state of three
particles is described by a normalized wave function satisfying the
three-body Schr\"odinger equation. This is in general the
underlying equation for the description of baryons as bound states of
three quarks in the framework of the various phenomenological
non-relativistic potential models.
A more profound basis for describing
bound states in relativistic quantum field theory is the 
Bethe-Salpeter equation \cite{SaBe51} for the so-called 
Bethe-Salpeter amplitudes, which might be considered as the
covariant analogues of 'wave functions' in the non-relativistic case. 
The Bethe Salpeter equation has been first derived for the
two-particle system by Salpeter and Bethe \cite{SaBe51}. Taylor \cite{Tay66}
investigated the application of the Bethe-Salpeter
equation to the three-body system.

In this section, we outline a method to treat the three fermion bound
state problem in relativistic quantum field theory by using Green's
function techniques.  This allows to derive the Bethe-Salpeter
equation for three bound fermions simultaneously with the normalization
condition of the corresponding Bethe-Salpeter amplitudes.  The method
is based on the fact that in general a bound state of elementary
particles, whose fields appear in the underlying interaction
Lagrangian, corresponds to a pole in the total energy of the Feynman
propagator (Green's function) of the many particle system.  These
poles do not arise from single perturbative Feynman diagrams, but
rather from an infinite series of diagrams. In this context the
Bethe-Salpeter amplitude is then defined as the residuum of the
bound-state pole of the Green's function.  This connection between
bound states and the singularities of Green's functions was originally
the basis of the first rigorous proof of the two-particle
Bethe-Salpeter equation given by Gell-Mann and Low
\cite{GeLo51}. However, this non-perturbative approach is clearly
general and can be applied generically to the n-body Green's function
as shown for instance in the textbook of Weinberg \cite{Wein95}.
As mentioned above, we apply this method to the case of three
fermions (quarks) only. It consists of the following three steps
\cite{Mey74,BoMe79,FlSc82}:
\begin{enumerate}
\item
Starting point is the six-point Green's function describing the
propagation of three interacting fermions.  In section
\ref{sec:SixGreen} we analyze the structure of the usual perturbative
power series expansion of the three-quark Feynman propagator:
introducing the concept of irreducible interaction kernels for the
case of three particles in a manner similar to that of Salpeter
and Bethe in the two-particle case \cite{SaBe51}, we outline, how
the infinite power series can be rearranged into an inhomogeneous
integral equation.
\item
In section \ref{sec:ContrBoundStates} we examine the analytical
structure of the six-point Green's function: we isolate the
contribution of a three-fermion bound state to the six-point Green's
function and show, how the bound state gives rise to a pole in the
total energy variable (or in the invariant 
total four-momentum squared). This procedure defines the Bethe-Salpeter
amplitudes of a specific bound state by the residue of the
corresponding bound-state pole which factorizes into the Bethe-Salpeter amplitude and its adjoint.
\item
Finally, by a Laurent expansion of the Green's function in the
vicinity of this bound-state pole and using the results from
sects. \ref{sec:SixGreen} and \ref{sec:ContrBoundStates}, we will
derive a homogeneous integral equation for bound states, {\it i.e.} the
Bethe-Salpeter equation along with the normalization
condition of the corresponding amplitudes.  This will be done in
section \ref{sec:BSE}.
\end{enumerate}
\subsection{The six-point Green's function for three fermions}
\label{sec:SixGreen}
The fundamental quantity describing three interacting fermions in quantum
field theory is the six-point Greens's function (or three-fermion Feynman propagator), which is the vacuum expectation value of a time
ordered product of three fermion field operators $\Psi^i$ and their adjoints
$\overline\Psi^i := {\Psi^i}^\dagger\gamma^0$ in the Heisenberg picture:
\begin{eqnarray}
\label{greensfunction}
\lefteqn{G_{a_1 a_2 a_3;\; a'_1 a'_2 a'_3 }(x_1,x_2,x_3;x_1',x_2',x_3'):=}&&\\
& & 
-\VacExpect{\;
T\;
\SpinorComp(1,x,a_1)
\SpinorComp(2,x,a_2)
\SpinorComp(3,x,a_3)
\AdSpinorComp(1,x',a'_1)
\AdSpinorComp(2,x',a'_2)
\AdSpinorComp(3,x',a'_3)}\nonumber.
\end{eqnarray}
Here the $a_i = (\alpha_i, f_i, c_i)$ denote multi-indices combining
the indices of the quark fields $\alpha_i$ in Dirac, $f_i$ in flavor and
$c_i$ in color space.  $\VacKet$ denotes the true physical vacuum
state and $T$ is the time ordering operator acting on a general $n$-fold
product of Heisenberg fermion field operators $A^i = \Psi$ or
$\overline \Psi$, $(i=1, \ldots, n)$ defined as
\begin{eqnarray}
\lefteqn{T \left\{A^1(x_1)A^2(x_2)\cdots A^n(x_n)\right\}}&&\nn
&=&
\textrm{sign}(\sigma)\;
T \left\{
A^{\sigma(1)} (x_{\sigma(1)}) A^{\sigma(2)} (x_{\sigma(2)})\cdots A^{\sigma(n)} (x_{\sigma(n)})\right\}
\nn
&:=&
\sum_{\sigma \in S_n} \textrm{sign}(\sigma)\;
A^{\sigma(1)} (x_{\sigma(1)})  A^{\sigma(2)} (x_{\sigma(2)})\; 
\cdots A^{\sigma(n)} (x_{\sigma(n)})\;\theta(x^0_{\sigma(1)}, x^0_{\sigma(2)}, \ldots, x^0_{\sigma(n)}), 
\end{eqnarray}
where the sum runs over all permutations $\sigma \in S_n$ with signum
sign$(\sigma)$.  $\theta$ is a
generalization of the usual Heaviside function
\begin{equation}
\theta (x^0_1, x^0_2, \ldots, x^0_n) 
= 
\left\{
\begin{array}{ll}
1 & \textrm{for } x^0_1 \ge x^0_2 \ge \cdots \ge x^0_n\\[2mm]
0 & \textrm{otherwise}
\end{array}
\right. .
\end{equation} 

Among other possibilities (depending on the time ordering considered)
the six-point Green's function $G$ represents the probability
amplitude for three (generally off-shell) quarks to propagate from
space-time points $x'_1, x'_2, x'_3$ to $x_1, x_2,x_3$.
Using the technique of ordinary time-dependent perturbation theory,
the six-point Green's function $G$ may be expressed in the form of an
infinite power series (see any standard textbook of quantum field
theory, for instance \cite{Lur68}):
\begin{eqnarray}
  \label{pertub_series}
\lefteqn{G(x_1,x_2,x_3;x_1',x_2',x_3')=\frac{-1}{\VacExpect{T \exp\left(-\textrm{i} 
\int_{-\infty}^{+\infty} dt\; \hat H_{Ip}(t)\right)}}\sum_{k=1}^{\infty}\frac{(-\textrm{i})^k}{k!}\int
\textrm{d}^4 y_1\ldots \textrm{d}^4 y_k}\nn
&&\qquad\times
{\VacExpect{
T
\SpinorI(1,x)
\SpinorI(2,x)
\SpinorI(3,x)
\AdSpinorI(1,x\pri)
\AdSpinorI(2,x\pri)
\AdSpinorI(3,x\pri)\hat{\cal H}_{Ip}(y_1)\ldots\hat{\cal H}_{Ip}(y_k)}}.
\end{eqnarray}
Now the state $\VacKet$ represents the unperturbed vacuum and $\Psi_{Ip}$,
$\bar\Psi_{Ip}$, $\hat H_{Ip}$ and $\hat{\cal H}_{Ip}$ are the field
operators, the interaction Hamiltonian and the Hamiltonian density
operator in the interaction picture, respectively.

\subsection{The integral equation for the six-point Green's function}
\begin{figure*}[!h]
  \begin{center}
    \epsfig{file={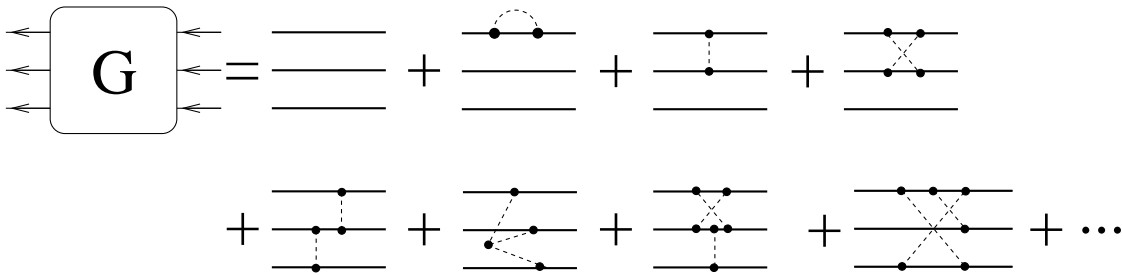},width=150mm}
  \end{center}
\caption{Finite order perturbative contributions to the six-point 
         Green's function $G$.}
\label{fig:greenperturb}
\end{figure*}

Using Wick's theorem for time ordered products of field operators, the
right hand side of eq. (\ref{pertub_series}) may be evaluated order by
order (in the coupling constant) to obtain a power series expansion which may
be represented in terms of ordinary Feynman graphs describing the
interaction of two or three fermions in finite order (see fig.
\ref{fig:greenperturb}).

In scattering processes (at high energy), where neither a three-body bound
state nor a two-body bound state in any of the two-particle subsystems occurs,
only a finite set of diagrams may be taken into consideration.  The
investigation of bound states, however, requires to go beyond such a
perturbative approach, {\it i.e.} an infinite sum of diagrams (or at least an
infinite subset of diagrams) has to be taken into account. The reason for this
is that {\it e.g.} a three-body bound state leads to a pole of the Green's
function in the total energy variable, as we will see in sect.
\ref{sec:ContrBoundStates}. But such a pole never arises from a finite set of
Feynman diagrams alone.  To go beyond perturbation theory, one recasts the
infinite power series expansion (\ref{pertub_series}) in the form of an
inhomogeneous integral equation, as it was done by Bethe and Salpeter
\cite{SaBe51} for the case of two particles.  Let us briefly sketch this
procedure for the case of three fermions:\\

\begin{figure*}[!h]
  \begin{center}
    \epsfig{file={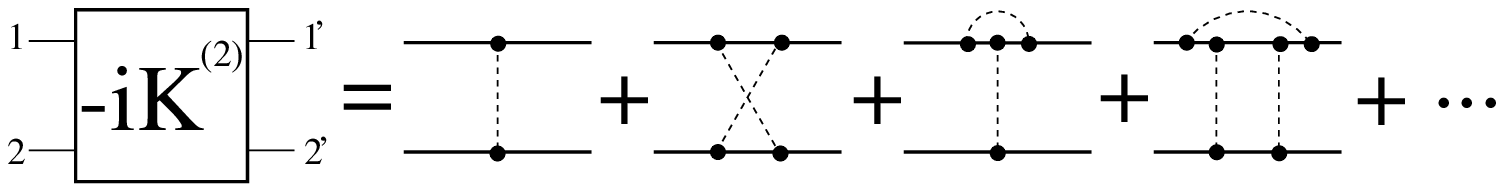},width=13cm}
  \end{center}
\caption{Graphical representation of the two-particle irreducible Bethe-Salpeter 
  kernel $K^{(2)}$ as sum of all possible connected irreducible two-particle
  interactions.}
\label{fig:K2}
\end{figure*}

\begin{figure*}[!h]
  \begin{center}
    \epsfig{file={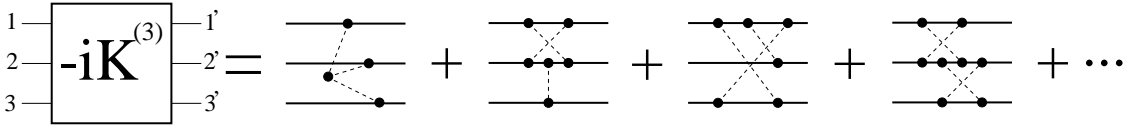},width=15cm}
  \end{center}
\caption{Diagrammatic picture of the three-particle irreducible Bethe-Salpeter 
  kernel $K^{(3)}$ as sum of all possible connected irreducible three-particle
  interactions.}
 \label{fig:K3}
\end{figure*}

{\bf 1.)} One introduces the concept of irreducibility, {\it i.e.} one classifies all
those diagrams appearing in the power expansion series
(\ref{pertub_series}) in reducible and irreducible graphs.  For the
definition of (ir)reducibility in the case of three interacting
particles we distinguish two- and three-particle interactions:
\begin{itemize}
\item
A connected two-fermion interaction graph is called irreducible, if it
cannot be split into two simpler graphs by cutting two fermion lines
only. Some examples of irreducible two-body diagrams are shown in
fig. \ref{fig:K2}.
\item
Correspondingly, a connected three-fermion interaction is called
irreducible, if it cannot be separated into two simpler graphs by just
cutting three fermion lines. Examples of such graphs are given in
fig. \ref{fig:K3}.
\item
All other interaction graphs are called reducible. Clearly, due to the
above definitions of irreducibility, reducible diagrams can
always be cut into irreducible parts.
\end{itemize}

{\bf 2.)} The (infinite) sum of all irreducible connected two-particle graphs is
collected into the so-called \textbf{irreducible two-particle interaction
kernel}
\begin{equation}
\label{K2}
K^{(2)}_{a_1 a_2;\; a'_1 a'_2}(x_1,x_2;x_1',x_2'). 
\end{equation}
See fig. \ref{fig:K2} for a diagrammatic representation of $K^{(2)}$.
Similarly, all irreducible connected three-particle graphs are added up to the
so-called \textbf{irreducible three-particle interaction kernel}
\begin{equation}
\label{K3}
K^{(3)}_{a_1 a_2 a_3;\; a'_1 a'_2 a'_3}(x_1,x_2,x_3;x_1',x_2',x_3').
\end{equation}
A graphical picture of $K^{(3)}$ is shown in fig. \ref{fig:K3}.  The
arguments $x'_i$, $x_i$ and multi-indices $a'_i$, $a_i$ in (\ref{K2}) and
(\ref{K3}) indicate the coordinate space and the Dirac, flavor and color
space dependences of the kernels, respectively.

\begin{figure*}[!h]
  \begin{center}
    \epsfig{file={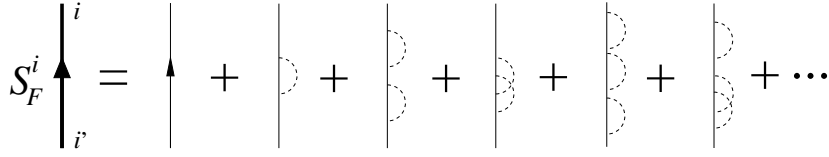},width=110mm}
  \end{center}
\caption{Perturbation series of the full dressed quark propagators defined in eq. (\ref{full_prop}).}
\label{fig:prop}
\end{figure*}

{\bf 3.)} Apart from the connected two- and three-particle interactions,
  applying Wick's theorem to the right hand side of eq.
  (\ref{pertub_series}) also generates unconnected terms, as {\it e.g.}
  the bare quark propagators, but moreover all kinds of self-energy
  contributions to the single fermion lines of each quark, summing up
  to the \textbf{full quark propagators}
\begin{equation}
\label{full_prop}
\FeyPropComp(i,x,x',a_i,a'_i)
  =\VacExpect{T\SpinorComp(i,x,a_i)\AdSpinorComp(i,x',a'_i)},
\end{equation}
as indicated in fig. \ref{fig:prop}.\\

\begin{figure*}[!h]
  \begin{center}
    \epsfig{file={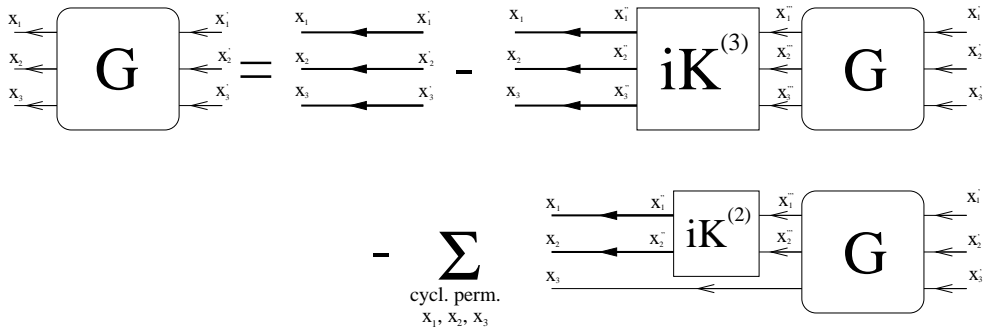},width=130mm}
  \end{center}
\caption{Graphical illustration of the inhomogeneous integral equation
  \ref{green_integral} for the six-point Green's function $G$. $K^{(3)}$ and
  $K^{(2)}$ denote the irreducible three- and two-body Bethe-Salpeter
  kernels, respectively, represented graphically in figs. \ref{fig:K2} and
  \ref{fig:K3}.  Thick arrows on quark lines indicate full dressed quark
  propagators as shown diagrammatically in fig. \ref{fig:prop}.}
\label{fig:green_2_3}
\end{figure*}

{\bf 4.)} All reducible interaction diagrams of any desired order
  appearing in the power series expansion can now be generated by
  iteration of the irreducible two-particle (in each quark pair) and
  three-particle interaction kernels $K^{(2)}$ and $K^{(3)}$ using the
  full quark propagators $S_F^i$ for the inner fermion lines.  This
  is accomplished to all orders by virtue of the following inhomogeneous
  integral equation, which uses the two- and three-particle
  interaction kernels as integral kernels \cite{Tay66,SaBe51}, {\it i.e.}
\begin{eqnarray}
  \label{green_integral}
  \lefteqn{G_{a_1 a_2 a_3;\; a'_1 a'_2 a'_3}(x_1,x_2,x_3;x_1',x_2',x_3')=}\nn[2mm]
  & &
  \FeyPropComp(1,x,x',a_1,a'_1)\;
  \FeyPropComp(2,x,x',a_2,a'_2)\;
  \FeyPropComp(3,x,x',a_3,a'_3)\nn[3mm]
  & & 
  -\textrm{i}
  \int \textrm{d}^4 y_1\; \textrm{d}^4 y_2\; \textrm{d}^4 y_3\;\;
  \FeyPropComp(1,x,y,a_1, b_1)\;
  \FeyPropComp(2,x,y,a_2, b_2)\;
  \FeyPropComp(3,x,y,a_3, b_3)\nn
  & &
  \phantom{-i}
  \int \textrm{d}^4 y_1'\; \textrm{d}^4 y_2'\; \textrm{d}^4 y_3'\;\;
  K^{(3)}_{b_1 b_2 b_3;\; b'_1 b'_2 b'_3}(y_1,y_2,y_3;y_1',y_2',y_3')\;
  G_{b'_1 b'_2 b'_3;\; a'_1 a'_2 a'_3}(y_1',y_2',y_3';x_1',x_2',x_3')\nn[3mm]
  & &
  -\textrm{i} \sum_{cycl. Perm.\atop (123)} 
  \int \textrm{d}^4 y_1\;\textrm{d}^4 y_2\; \;
  \FeyPropComp(1,x,y,a_1,b_1)\;
  \FeyPropComp(2,x,y,a_2,b_2)\nn[-3mm]
  & &
 \phantom{-i \sum_{cycl. Perm.\atop (123)}}
 \int \textrm{d}^4 y_1'\;\textrm{d}^4 y_2'\;\;
  K^{(2)}_{b_1 b_2;\; b'_1 b'_2}(y_1,y_2;y_1',y_2')\;
  G_{b'_1 b'_2 a_3;\; a'_1 a'_2 a'_3}(y_1',y_2',x_3;x_1',x_2',x_3'),
\end{eqnarray}
where our notation implies summation over indices $b_i$,
$b'_i$ occurring twice. For a diagrammatic illustration of this integral equation, see fig.
\ref{fig:green_2_3}. In fact, the Neumann iteration of this integral equation
reproduces all possible reducible interactions and thus precisely all the
terms of the power series expansion of eq. (\ref{pertub_series}).

Note that also the irreducible interaction kernels $K^{(2)}$ and $K^{(3)}$
consist already of an infinite number of graphs and in general cannot be
calculated exactly. They are basically unknown functions
and thus have to be parameterized phenomenologically.
However, the decisive advantage of the non-perturbative construction of
the Green's function $G$ from an inhomogeneous integral equation
(\ref{green_integral}) is that its solution automatically implies an
infinite number of interactions even if the kernels are approximated
by their lowest order Born terms, which constitutes the so-called ladder
approximation.  Such an approximation is sufficient in theories,
where the coupling constant is small and the interaction kernels may
be considered as an asymptotic series expanded in terms of the (small)
coupling constant. (In such a case one would expect most of the binding
of a bound state to come from the repeated action of the Born
diagrams alone.)\\

\begin{figure}[h]
  \begin{center}
    \epsfig{file={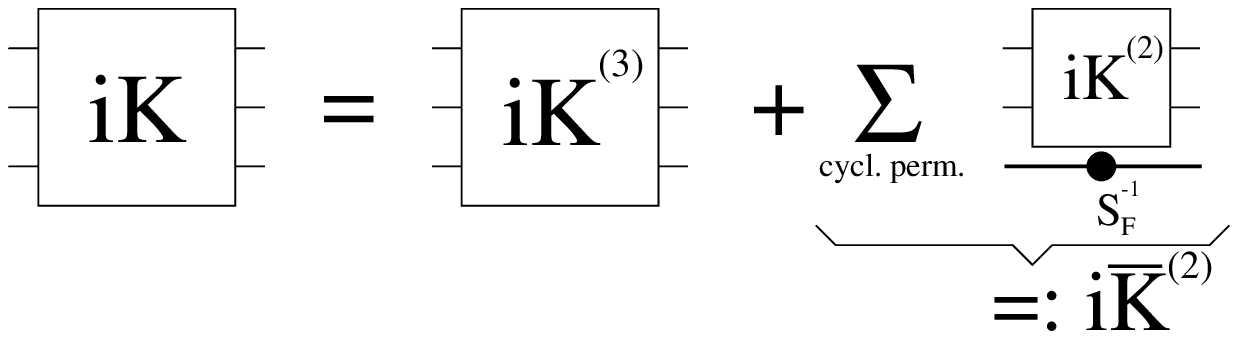},width=85mm}
  \end{center}
\caption{The integral kernel $K$ combining the three-body irreducible
kernel and the two-body irreducible kernels in each quark pair; the
filled circle denotes an inverse full quark propagator.}
\label{fig:K23}
\end{figure}

For further discussion of eq. (\ref{green_integral}) it is useful to
introduce an appropriate compact notation.  First let us combine the
irreducible two- and three-body kernels $K^{(2)}$ and $K^{(3)}$ to a single
integral kernel $K$. We introduce the inverse ${S_F^k}^{-1}$ of the
full quark propagator $S_F^k$ by
\begin{equation}
\label{S_S_inv}
\int \textrm{d}^4 y_k\;
\FeyPropComp(k,x,y,a_k,b)\;{S_F^k}^{-1}_{b a'_k}(y_k, x'_k)
=
\delta_{a_k a'_k}\;\delta^{(4)}(x_k-x'_k).
\end{equation}
This allows to rewrite the sum of the two-particle interactions $K^{(2)}$
in each quark pair in the form of a three-body kernel
\begin{equation}
\label{K2_lift}
\overline K^{(2)}_{a_1 a_2 a_3;\; a'_1 a'_2 a'_3}(x_1,x_2,x_3; x_1',x_2',x_3')
:=\hspace*{-5mm}
\sum_{\footnotesize 
\begin{array}{c}
(ijk) =\\
\textrm{cycl. perm. of}\\
(123)
\end{array}}
\hspace*{-5mm}
 K^{(2)}_{a_i a_j;\; a'_i a'_j}(x_i,x_j;x_i',x_j')\; {S_F^k}^{-1}_{a_k a'_k}(x_k, x_k').
\end{equation}
In this form we can combine the two-body interaction kernels with the
three-body kernel $K^{(3)}$ to a uniform integral kernel $K$ (see fig. \ref{fig:K23}):
\begin{equation}
K:=K^{(3)} + \overline K^{(2)}.
\end{equation} 
\\
Moreover, we introduce the symbol $G_0$ for the triple tensor product of
the single quark propagators $S_F^i$ which is the lowest order contribution
to $G$:
\begin{equation}
{G_{0\;}}_{a_1 a_2 a_3;\; a'_1 a'_2 a'_3}(x_1,x_2,x_3; x_1',x_2',x_3')
:=
 \FeyPropComp(1,x,x',a_1,a'_1)\;
 \FeyPropComp(2,x,x',a_2,a'_2)\;
 \FeyPropComp(3,x,x',a_3,a'_3).
\end{equation}
Finally, we define a shorthand operator product notation for the
summation over indices and the integral operation in coordinate space:
\begin{eqnarray}
\label{OP_ind}
\left[A\;B\right]_{a_1 a_2 a_3;\; a'_1 a'_2 a'_3}
&:=&
\sum_{b_1 b_2 b_3}
A_{a_1 a_2 a_3;\; b_1 b_2 b_3}\;
B_{b_1 b_2 b_3;\; a'_1 a'_2 a'_3},\\
\label{OP_int}
\left[A\; B\right](x_1,x_2,x_3;x_1',x_2',x_3')&:=&
\int \textrm{d}^4 y_1 \textrm{d}^4 y_2 \textrm{d}^4 y_3\;
A(x_1,x_2,x_3;y_1,y_2,y_3)\; B(y_1,y_2,y_3;x_1',x_2',x_3').
\end{eqnarray}
With these definitions the inhomogeneous integral equation for the six-point
Green's function can be represented in the more compact form of an operator
equation
\begin{equation}
\label{BSE_G}
G = G_0 - \textrm{i}\; G_0\;K\;G. 
\end{equation}
Note that this integral equation for the Green's function $G$ can also be
written in its equivalent adjoint form, where the operator product $G_0\;K\;G$
on the right hand side of eq. (\ref{BSE_G}) appears in reverse order:
\begin{equation}
\label{BSE_G_ad}
G = G_0 - \textrm{i}\; G\;K\;G_0.
\end{equation}
The equivalence of the integral equation (\ref{BSE_G}) and its adjoint
(\ref{BSE_G_ad}) is obvious, since both equations have the same Neumann
series.
\subsection{Space-time translational invariance}
The six-point Green's function $G$ as defined in eq.
(\ref{greensfunction}) is invariant under arbitrary space-time translations,
{\it i.e.}
\begin{equation}
\label{transl_sym_G}
G(x_1,x_2,x_3;x'_1,x'_2,x'_3) = G(x_1+a,x_2+a,x_3+a;x'_1+a,x'_2+a,x'_3+a)
\end{equation} 
for all $a\in \R^4$.  Due to this symmetry it is natural to introduce new
coordinates, namely an external 'center-of-mass' coordinate $X$ and
internal, {\it i.e.} translationally invariant, relative coordinates $\xi$ and
$\eta$, the so-called Jacobi coordinates. We choose:
\begin{equation}
\label{jacobi_spat}
\begin{array}{ccl}
X    &:=& \frac{1}{3}(x_1+x_2+x_3),\nn[3mm]
\xi  &:=& x_1-x_2,\nn[3mm]
\eta &:=& \frac{1}{2}(x_1+x_2-2x_3),
\end{array} 
\quad\Leftrightarrow\quad
\begin{array}{ccl}
x_1 &=& X+\frac{1}{2}\xi+\frac{1}{3}\eta,\nn[3mm]
x_2 &=& X-\frac{1}{2}\xi+\frac{1}{3}\eta,\nn[3mm]
x_3 &=& X-\frac{2}{3}\eta.
\end{array}
\end{equation}
The space-like components ${\bf X}$, ${\bfgrk \xi}$ and ${\bfgrk
\eta}$ of these variables can be interpreted in the non-relativistic
limit as usual center-of-mass and relative coordinates for a system of
three particles with equal mass. However, in a covariant framework
this choice is {\it a priori} arbitrary and the variables $X$, $\xi$
and $\eta$ have in general no direct physical meaning.  Choosing now
specifically $a := -\frac{1}{2}(X+X')$ in eq.
(\ref{transl_sym_G}) we find that in fact the six-point Green's
function $G$ depends only on translationally invariant coordinate
differences $X-X'$, $\xi$, $\eta$, $\xi'$ and $\eta'$, {\it i.e.}
\begin{equation}
G(x_1,x_2,x_3;\;x'_1,x'_2,x'_3)\equiv G(X-X';\xi,\eta;\xi',\eta'). 
\end{equation} 
Of course, the same holds also for the triple product $G_0$ of the free single
quark propagators, and the translation invariance of the Green's function
$G$ necessarily implies that in particular the interaction kernels $K^{(3)}$
and $\overline K^{(2)}$ must by themselves be translationally invariant
quantities. In momentum space space-time translation invariance is equivalent
to the conservation of the total four-momentum.  Consequently, as will be
shown in the following subsection, the twelve-dimensional integral equations
(\ref{BSE_G}) for the six-point Green's function in coordinate space and
its adjoint (\ref{BSE_G_ad}) after Fourier transformation become only
eight-dimensional integral equations in the momentum space representation. Due
to momentum conservation, these momentum space representations depend only
parametrically on the total four-momentum.  To perform the Fourier
transformation let us define the corresponding conjugate
momenta to $X$, $\xi$ and $\eta$ which are given by the total four-momentum
$P$ and the following relative four-momenta $p_\xi$ and $p_\eta$:
\begin{equation}
\label{jacobi_mom}
\begin{array}{ccl}
P      &:=& p_1+p_2+p_3,\nn[3mm]
p_\xi  &:=& \frac{1}{2}(p_1-p_2),\nn[3mm]
p_\eta &:=& \frac{1}{3}(p_1+p_2-2p_3),
\end{array} 
\quad\Leftrightarrow\quad
\begin{array}{ccl}
p_1 &=& \frac{1}{3}P+p_{\xi}+\frac{1}{2}p_{\eta},\nn[3mm]
p_2 &=& \frac{1}{3}P-p_{\xi}+\frac{1}{2}p_{\eta},\nn[3mm]
p_3 &=& \frac{1}{3}P-p_{\eta}.
\end{array}
\end{equation}
The new sets of coordinates (\ref{jacobi_spat}) and (\ref{jacobi_mom}) satisfy
the condition
\begin{equation}
\label{jacobi_property1}
\langle p_1, x_1\rangle + \langle p_2, x_2\rangle + \langle p_3, x_3\rangle 
=
\langle P, X\rangle + \langle p_\xi, \xi\rangle + \langle p_\eta, \eta\rangle,
\end{equation}
and a technical advantage of this special choice of variables is that the
Jacobians of the transformations  (\ref{jacobi_spat}) and (\ref{jacobi_mom})
are unity, {\it i.e.}
\begin{equation}
\label{jacobi_property2}
\left|\frac{\partial (X,\xi,\eta)}{\partial (x_1,x_2,x_3)}\right|=1 
\quad\textrm{and}\quad
\left|\frac{\partial (P,p_\xi,p_\eta)}{\partial (p_1,p_2,p_3)}\right|=1.
\end{equation}

\subsection{Momentum space representation of the integral equation}
\label{sub:mom_space_rep_G}
For any six-point
function $A= G$, $G_0$, $K$, ${\overline K}^{(2)}$ and $K^{(3)}$, {\it i.e.} the six-point Green's function, the triple
product of quark propagators or the interaction kernels, 
we define the Fourier transform by
\begin{eqnarray}
\label{fourier}
\lefteqn{\left[{\cal F} A\right] (p_1,p_2,p_3; p'_1,p'_2,p'_3):=}&&\nn
&&\quad\quad
\int \textrm{d}^4x_1 \;\textrm{d}^4x_2 \;\textrm{d}^4x_3 \; 
e^{+\textrm{\footnotesize i} \left(\langle p_1, x_1\rangle + \langle p_2, x_2\rangle + \langle p_3, x_3\rangle\right)}\\
&&\quad\quad
\times\;\int \textrm{d}^4x'_1 \;\textrm{d}^4x'_2 \;\textrm{d}^4x'_3\; 
e^{-\textrm{\footnotesize i} \left(\langle p'_1, x'_1\rangle + \langle p'_2, x'_2\rangle + \langle p'_3, x'_3\rangle\right)}\;
A(x_1,x_2,x_3; x_1',x_2',x_3').\nonumber
\end{eqnarray}
Using the properties (\ref{jacobi_property1}) and (\ref{jacobi_property2}) of the new coordinate sets, 
the Fourier transforms can be written in terms of relative
Jacobi momenta and the total four-momenta,
\begin{eqnarray}
\label{FT_A}
\left[{\cal F} A\right] (p_1,p_2,p_3; p'_1,p'_2,p'_3)&=& A_{P}(p_\xi,p_\eta; p_\xi',p_\eta')\; (2\pi)^4\delta^{(4)}(P-P').
\end{eqnarray}
Due to the translational invariance of the six-point functions $A$ the
$\delta$-function reflects the conservation $P'=P$ of total four-momentum. The
remaining part $A_P$, just depending parametrically on $P$, is defined by the
following Fourier transformation
\begin{eqnarray}
\label{A_fourier}       
\lefteqn{A(x_1,x_2,x_3;x_1',x_2',x_3')=
A(X-X';\xi,\eta;\xi',\eta')}\nn[3mm]
&=:&
\int \frac{\textrm{d}^4 P}{(2\pi)^4}\;
e^{-\textrm{\footnotesize i} \langle P, X-X'\rangle}\;
\int \frac{\textrm{d}^4 p_\xi}{(2\pi)^4}\;\frac{\textrm{d}^4 p_\eta}{(2\pi)^4}\;
e^{-\textrm{\footnotesize i} \langle p_\xi, \xi\rangle}\; 
e^{-\textrm{\footnotesize i} \langle p_\eta, \eta\rangle}
\int \frac{\textrm{d}^4 p_\xi'}{(2\pi)^4}\;\frac{\textrm{d}^4 p_\eta'}{(2\pi)^4}\;
e^{\textrm{\footnotesize i} \langle p_\xi', \xi'\rangle} 
e^{\textrm{\footnotesize i} \langle p_\eta',\eta'\rangle}\; A_{P}(p_\xi,p_\eta; p_\xi',p_\eta'),
\end{eqnarray}
which exhibits the exclusive dependence
on the relative coordinates $X-X'$, $\xi$, $\eta$, $\xi'$ and $\eta'$.
The momentum space representation ${G_0}_P$ of the quark propagators $G_0$
then reads explicitly
\begin{eqnarray}
\label{G0_mom}
{G_0}_{P}\;(p_\xi,p_\eta; p_\xi',p_\eta')&=&
S_F^1\left(\mbox{$\frac{1}{3}P\!+\!p_{\xi}\!+\!\frac{1}{2}p_{\eta}$}\right)
\tens
S_F^2\left(\mbox{$\frac{1}{3}P\!-\!p_{\xi}\!+\!\frac{1}{2}p_{\eta}$}\right)
\tens
S_F^3\left(\mbox{$\frac{1}{3}P\!-\!p_{\eta}$}\right)\nn
&& \hspace*{10mm}\times \;\;(2 \pi)^4\;\delta^{(4)}(p_\xi-p_\xi')\;\; (2 \pi)^4\;\delta^{(4)}(p_\eta-p_\eta'),
\end{eqnarray}
where (due to translational invariance) the Fourier transforms of the
single quark propagators are defined by
\begin{equation}
\label{S_fourier}
S_F^i(x_i,x_i')
=S_F^i(x_i-x_i')=: 
\int \frac{\textrm{d}^4 p_i}{(2\pi)^4}\;e^{-\textrm{\footnotesize i} \langle
  p_i, x_i-x_i'\rangle}\;S_F^i(p_i). 
\end{equation}
For the sake of completeness we should also specify the explicit form of the
Fourier transform $\overline K^{(2)}_{P}$ of the two-particle term
$\overline K^{(2)}$ defined in eq. (\ref{K2_lift}). To this end we first
have to define the Fourier transform of the two-particle
interaction kernel $K^{(2)}(x_i,x_j;\;x'_i,x'_j)$. According to
translational invariance, it is useful to introduce two-particle 'center of
mass' and relative coordinates $X_k$ and $\xi_k$ for each possible quark pair
$(ij)$, {\it i.e.}
\begin{equation}
\begin{array}{lcl}
X_k   &:=& \frac{1}{2} (x_i + x_j)\nn[2mm]
\xi_k &:=&  x_i - x_j
\end{array}
\quad \textrm{for}\; (ijk) = \textrm{cycl. perm. of}\; (123),
\end{equation}
as well as their corresponding conjugate variables, the total
two-particle momenta $P_k$ and the relative momenta $p_{\xi_k}$, {\it i.e.}
\begin{equation}
\begin{array}{lclclc}
P_k   &:=& p_i + p_j = \frac{2}{3}P + p_{\eta_k}\nn[1mm]
p_{\eta_k} &:=&\frac{1}{3}(p_i+p_j - 2 p_k)\nn[3mm]
p_{\xi_k} &:=&\frac{1}{2}(p_i-p_j)&\textrm{for}\; (ijk) = \textrm{cycl. perm. of}\; (123).
\end{array}
\end{equation}
Note that we have expressed $P_k$ by the total three-particle momentum $P$ and an
additional variable $p_{\eta_k}$ in order to relate the sets $(P_k,p_{\xi_k})$
of two-particle momenta to the set (\ref{jacobi_mom}) of relative
three-particle momenta $(p_{\xi},p_{\eta}) = (p_{\xi_3},p_{\eta_3})$ in the
case $(ijk)=(123)$ and to the equivalent cyclically permuted sets
$(p_{\xi_1},p_{\eta_1})$ and $(p_{\xi_2},p_{\eta_2})$ in the cases
$(ijk)=(231)$ and $(312)$, respectively.  The cyclically permuted sets of the
relative momenta are obtained by linear transformations of the existing set
(\ref{jacobi_mom}) according to
\begin{equation}
\label{cycl_perm_jacobi_mom}
\left(
\begin{array}{c}
p_\xi\\[3mm]
p_\eta
\end{array}
\right)
=
\left(
\begin{array}{c}
p_{\xi_3}\\[3mm]
p_{\eta_3}
\end{array}
\right)
=
\left(
\begin{array}{cc}
-\frac{1}{2} & -\frac{3}{4}\\[3mm]
\phantom{-}1 & -\frac{1}{2} 
\end{array}
\right)
\left(
\begin{array}{c}
p_{\xi_1}\\[3mm]
p_{\eta_1} 
\end{array}
\right)
=
\left(
\begin{array}{cc}
-\frac{1}{2} & \phantom{-}\frac{3}{4}\\[3mm]
-1 & -\frac{1}{2} 
\end{array}
\right)
\left(
\begin{array}{c}
p_{\xi_2}\\[3mm]
p_{\eta_2} 
\end{array}
\right).
\end{equation}
When $K^{(2)}$ depends on translationally invariant two-particle
variables only, the Fourier transform of $K^{(2)}$ is given as
\begin{eqnarray}
\label{K2_fourier}
\lefteqn{K^{(2)}(x_i,x_j; x'_i,x'_j)=
K^{(2)}(X_k-X'_k; \xi_k, \xi'_k)}&&\\
&&=:
\int \frac{\textrm{d}^4 P_k}{(2\pi)^4}\;
e^{-\textrm{\footnotesize i} \langle P_k, X_k-X'_k\rangle}
\int 
\frac{\textrm{d}^4 p_{\xi_k}}{(2\pi)^4}\;
e^{-\textrm{\footnotesize i} \langle p_{\xi_k}, \xi_k\rangle}
\int
\frac{\textrm{d}^4 p'_{\xi_k}}{(2\pi)^4}\;
e^{ \textrm{\footnotesize i} \langle p'_{\xi_k}, \xi'_k\rangle}\;
K_{P_k}^{(2)}(p_{\xi_k},p'_{\xi_k}).\nonumber
\end{eqnarray}
Using the definition (\ref{K2_lift}) of $\overline K^{(2)}$ and the
definitions (\ref{A_fourier}) and (\ref{K2_fourier}) of the Fourier
transforms of $\overline K^{(2)}$ and $K^{(2)}$, we find the following
explicit form for $\overline K^{(2)}_P$:
\begin{eqnarray}
\label{K2_lift_fourier}
 \lefteqn{\overline K^{(2)}_{P\;a_1 a_2 a_3;\;a'_1 a'_2
  a'_3}(p_\xi,p_\eta;\;p_\xi',p_\eta')=}\\[3mm]
  & &
  \sum_{(ijk)=\atop (123),(231),(312)}
  K^{(2)}_{(\frac{2}{3}P+p_{\eta_k})\;  a_i a_j;\; a'_i a'_j}(p_{\xi_k},p'_{\xi_k})\;
  {S_F^k}^{-1}_{a_k a'_k}\left(\mbox{$\frac{1}{3}P-p_{\eta_k}$}\right)\;
  (2\pi)^4 \; \delta^{(4)}(p_{\eta_k}-p'_{\eta_k}),\nonumber
\end{eqnarray}
where ${S_F^k}^{-1}(p_k)$ is the momentum space
representation of the inverse of the full quark propagator defined in eq. (\ref{S_S_inv}),
which obeys
\begin{equation}
\sum_{b_k} {S_F^k}_{a_k b_k}(p_k)\;{S_F^k}^{-1}_{b_k a'_k}(p_k) = \delta_{a_k a'_k}. 
\end{equation}
With definition (\ref{A_fourier}) of the Fourier transforms of $G$, $G_0$ and $K$, the
properties (\ref{jacobi_property1}) and (\ref{jacobi_property2}) of the Jacobi coordinates and the
explicit form (\ref{G0_mom}) of ${G_0}_P$, we are now in the position to write
the inhomogeneous integral equation (\ref{BSE_G}) for the six-point Green's function $G$ in its momentum
space representation,
\begin{eqnarray}
\label{green_momentum}
G_{P}(p_\xi,p_\eta; p_\xi',p_\eta') &=&
S_F^1\left(\mbox{$\frac{1}{3}P+p_{\xi}+\frac{1}{2}p_{\eta}$}\right)
\tens
S_F^2\left(\mbox{$\frac{1}{3}P-p_{\xi}+\frac{1}{2}p_{\eta}$}\right)
\tens
S_F^3\left(\mbox{$\frac{1}{3}P-p_{\eta}$}\right)\;\nn[1mm]
&&\quad\quad\times\; (2 \pi)^4\;\delta^{(4)}(p_\xi-p_\xi')\;\;(2 \pi)^4\;\delta^{(4)}(p_\eta-p_\eta')\nn[5mm]
&&+\;\;
S_F^1\left(\mbox{$\frac{1}{3}P+p_{\xi}+\frac{1}{2}p_{\eta}$}\right)
\tens
S_F^2\left(\mbox{$\frac{1}{3}P-p_{\xi}+\frac{1}{2}p_{\eta}$}\right)
\tens
S_F^3\left(\mbox{$\frac{1}{3}P-p_{\eta}$}\right)\nn
&&\quad\quad\times\; (-\textrm{i})\;
\int\frac{\textrm{d}^4 p_\xi''}{(2\pi)^4}\;\frac{\textrm{d}^4 p_\eta''}{(2\pi)^4}\;\;
K_P(p_\xi,p_\eta;p_\xi'',p_\eta'')\;
G_P(p_\xi'',p_\eta'';p_\xi',p_\eta'),
\end{eqnarray}
where we suppressed the dependences on the indices using the shorthand
tensor notation and the definition (\ref{OP_ind}) of the operator
product.  Note that, due to the conservation of the total four
momentum, the inhomogeneous integral equation depends only parametrically
on the total four momentum $P$, while the integral operation involves only the relative 
momenta $p_\xi$ and $p_\eta$. Let us therefore introduce the momentum space representation
of the operator product corresponding to (\ref{OP_int}) as
\begin{eqnarray}
\label{operator_prod_mom}
\left[A_P\;B_P\right](p_\xi,p_\eta;p_\xi',p_\eta') := 
\int\frac{\textrm{d}^4 p_\xi''}{(2\pi)^4}\;\frac{\textrm{d}^4 p_\eta''}{(2\pi)^4}\;\;
A_P(p_\xi,p_\eta;p_\xi'',p_\eta'')\;
B_P(p_\xi'',p_\eta'';p_\xi',p_\eta')
\end{eqnarray}
which again allows to write the momentum space representations of
the integral equation (\ref{BSE_G}) and its adjoint
(\ref{BSE_G_ad}) in a concise operator notation:
\begin{eqnarray}
\label{BSEmomentum_G}
G_P = {G_0}_P - \textrm{i}\; {G_0}_P\;K_P\;G_P,\\[3mm]
\label{BSEmomentum_G_ad}
G_P = {G_0}_P - \textrm{i}\; G_P\;K_P\;{G_0}_P,
\end{eqnarray}
with the subscript $P$ indicating the parametrical dependence on the
total four momentum, which becomes important in the next two sections
for the investigation of bound-state contributions to $G$.
Accordingly, we will evaluate eqs. (\ref{BSEmomentum_G}) and
(\ref{BSEmomentum_G_ad}) at the positions $P=\bar P$, where bound states with
mass $M^2 = \bar P^2$ occur, allowing the derivation of the bound state
Bethe-Salpeter equation and the normalization condition of the
corresponding amplitudes. But to this end we first have to know how the
six-point Green's function behaves at these bound-state pole positions.
This is the topic of the next subsection.
\subsection{Bound-state contributions -- Bethe-Salpeter amplitudes}
\label{sec:ContrBoundStates}
In quantum field theory bound states are related to the
occurrence of poles of Green's functions in the total energy variable
$P^0$ or, equivalently, in the invariant four-momentum squared $P^2$. 
Here we verify this statement for the case of the
six-point Green's function $G_P$.\\

In the following we consider bound
states of three quarks with (positive) mass $M$ and positive energy
$\omega_{\bf P} := \sqrt{{\bf P}^2+M^2}$.  The corresponding Fock states with
total four momentum $\bar P = (\omega_{\bf P}, {\bf P})$ and mass $\bar P^2 =
M^2$ are denoted by $\ket{\bar P}$. They are eigenstates of the total
four-momentum operator $\hat P = \hat p_1 + \hat p_2 + \hat p_3$, {\it i.e.}
\begin{equation}
\label{Fock_P}
\hat P\; \ket{\bar P} = \bar P\; \ket{\bar P},
\end{equation}
and are normalized covariantly according to
\begin{equation}
\label{lor_inv_norm}
\SP {\bar P} {\bar P'} = {(2\pi)}^3\; 2\omega_{\bf P}\;
\delta^{(3)}({\bf P} - {\bf P}').
\end{equation}

The six-point Green's function (\ref{greensfunction}) in general
describes all possible kinds of processes with three incoming and three
outgoing fermions. The 'initial' and 'final' fermion
or anti-fermion lines, however, are not yet fixed, until a particular time-ordering has
been chosen. Here we are interested in the extraction of 'baryon'
contributions to $G$, {\it i.e.} real bound states of three quarks with
positive energy that propagate forward in time.  Therefore we shall consider
those specific contributions to the six-point Green's function $G$ which
have the particular time orderings $x^0_1,x^0_2,x^0_3 > x'^0_1,x'^0_2,x'^0_3$,
{\it i.e.} which contain
\begin{equation}
\label{spez_TO}
\theta\left(\min(x^0_1,x^0_2,x^0_3)-\max({x'}^0_1,{x'}^0_2,{x'}^0_3)\right)
=
\cases{
1\;\; \textrm{for}\; x^0_1,x^0_2,x^0_3 > {x'}^0_1,{x'}^0_2,{x'}^0_3\cr
0\;\; \textrm{otherwise}.
}
\end{equation}
Isolating this part of the Green's function defined in eq.
(\ref{greensfunction}), we have
\begin{eqnarray}
\label{green_spez_TO}
\lefteqn{G_{a_1 a_2 a_3;\; a'_1 a'_2 a'_3 }(x_1,x_2,x_3;x_1',x_2',x_3')=}\nn[5mm]
& &
-\VacExpect{\;
T\;
\left\{
\SpinorComp(1,x,a_1)
\SpinorComp(2,x,a_2)
\SpinorComp(3,x,a_3)
\right\}\;\;
T\;
\left\{
\AdSpinorComp(1,x',a'_1)
\AdSpinorComp(2,x',a'_2)
\AdSpinorComp(3,x',a'_3)\right\}}
\nn[3mm]
& &\quad\times\;\theta\left(\min(x^0_1,x^0_2,x^0_3)-\max({x'}^0_1,{x'}^0_2,{x'}^0_3)\right)\nn[5mm]
& & 
+ \quad \textrm{other terms arising from different time-orderings}.
\end{eqnarray}
Now we can evaluate that contribution to the Green's function which
arises from three-quark bound states (\ref{Fock_P}) with mass M, by inserting the
complete set of the intermediate states $\ket{\bar P}$ in between the
two time-ordered products in the matrix element (\ref{green_spez_TO}):
\begin{eqnarray}
  \label{green_bound}
  \lefteqn{G_{a_1 a_2 a_3;\; a'_1 a'_2 a'_3 }(x_1,x_2,x_3;x_1',x_2',x_3')=}\nn[5mm]
  & &
  -\int \frac{\textrm{d}^3 P}{(2\pi)^3\; 2 \omega_{\bf P}}\;
  \VacBra\;
  T\;
  \SpinorComp(1,x,a_1)
  \SpinorComp(2,x,a_2)
  \SpinorComp(3,x,a_3)
  \ket{\bar P}\;
  \bra{\bar P}\; 
  T\;
  \AdSpinorComp(1,x',a'_1)
  \AdSpinorComp(2,x',a'_2)
  \AdSpinorComp(3,x',a'_3)\VacKet
  \nn[3mm]
  & &\quad\quad\times\;\theta\left(\min(x^0_1,x^0_2,x^0_3)-\max({x'}^0_1,{x'}^0_2,{x'}^0_3)\right)\nn[5mm]
  & &
  + 
  \quad \textrm{other terms}.
\end{eqnarray}
Here 'other terms' now denotes the terms not only arising from other
time-orderings, but also from other intermediate states.\\

We define the \textbf{Bethe-Salpeter amplitude} $\chi_{\bar P}$ for the bound state $\ket{\bar P}$ and its 
adjoint $\overline\chi_{\bar P}$ by the following transition amplitudes between the state $\ket{\bar P}$ and the vacuum
$\VacKet$,
\begin{eqnarray}
  \label{BS_amplitude}
  \chi_{\bar P\; a_1 a_2 a_3}(x_1,x_2,x_3)
  &:=&
  \VacBra\;
  T\;
  \SpinorComp(1,x,a_1)
  \SpinorComp(2,x,a_2)
  \SpinorComp(3,x,a_3)\;
  \ket{\bar P},\\[2mm]
  \label{BS_amplitude_ad}
  \overline\chi_{\bar P\; a'_1 a'_2 a'_3}(x'_1,x'_2,x'_3)
  &:=&
  \bra{\bar P}\;
  T\;
  \AdSpinorComp(1,x',a'_1)
  \AdSpinorComp(2,x',a'_2)
  \AdSpinorComp(3,x',a'_3)\;\VacKet,
\end{eqnarray}
which appear in the bound-state contribution (\ref{green_bound}) to
the Green's function $G$.  Due to translational invariance we can
factorize the total momentum dependence of the Bethe-Salpeter
amplitude $\chi_{\bar P}$ and its adjoint $\overline\chi_{\bar P}$
which contributes just by a trivial phase factor:
\begin{eqnarray}
\label{BSA_sep_four}
\chi_{\bar P}(x_1,x_2,x_3)
&=&
e^{-{\rm i} \langle \bar P, X\rangle }\;\;\chi_{\bar P}(\xi,\eta)\nn[2mm]
&=:&
e^{-{\rm i} \langle \bar P X, \rangle}\;
\int \frac{\textrm{d}^4 p_\xi}{(2\pi)^4}\;\frac{\textrm{d}^4 p_\eta}{(2\pi)^4}\;
e^{-{\rm i} \langle p_\xi, \xi\rangle}\; e^{-{\rm i} \langle p_\eta, \eta\rangle}\;
\chi_{\bar P}(p_\xi,p_\eta),\\[3mm]
\label{BSA_sep_four_ad}
\overline\chi_{\bar P}(x_1',x_2',x_3')
&=&
e^{{\rm i} \langle \bar P, X'\rangle}\;\;\overline\chi_{\bar P}(\xi',\eta')\nn[2mm]
&=:&
e^{{\rm i} \langle \bar P, X'\rangle}\;
\int\frac{\textrm{d}^4 p_\xi'}{(2\pi)^4}\;\frac{\textrm{d}^4p_\eta'}{(2\pi)^4}\;
e^{{\rm i} \langle p_\xi', \xi'\rangle}\; e^{{\rm i} \langle p_\eta',\eta'\rangle}\;
\overline\chi_{\bar P}(p'_\xi,p'_\eta).
\end{eqnarray}
Thus, we obtain translationally invariant Bethe-Salpeter amplitudes
and their Fourier transforms which depend only on the internal relative
coordinates $\xi$, $\eta$ and $p_\xi$, $p_\eta$, respectively.\\

The $\theta$-function in eq. (\ref{green_bound}), which dictates the specific time ordering
$x^0_1,x^0_2,x^0_3 > x'^0_1,x'^0_2,x'^0_3$, gives rise to a pole of $G$ in the
total energy variable $P^0$ and we finally arrive at the following Laurent expansion of $G_P$ in momentum space near the pole at
$P^0=\omega_{\bf P}$:
\begin{equation}
\label{green_pol_p0}
G_P(p_\xi,p_\eta;p_\xi',p_\eta')
=
\frac{-\textrm{i}}{2\omega_{\bf P}}\;
\frac{\chi_{\bar P}(p_\xi,p_\eta)\;\overline\chi_{\bar P}(p'_\xi,p'_\eta)}
{P^0-\omega_{\bf P} + \textrm{i}\epsilon}
+\;\textrm{regular terms for } P^0\rightarrow \omega_{\bf P},
\end{equation}
or written covariantly
\begin{equation} 
  \label{green_pol_p2}
  G_{P}(p_\xi,p_\eta; p_\xi',p_\eta')
  =
  - {\rm i}\;\frac{\chi_{\bar P}(p_\xi,p_\eta)
    \overline\chi_{\bar P}(p'_\xi,p'_\eta)}{P^2 - M^2 + {\rm i}\epsilon} 
    +\;\textrm{regular terms for } P^2\rightarrow M^2,
\end{equation}
where we have introduced a six-point function $\left[\chi_{\bar
    P}\;\overline\chi_{\bar P}\right]$ by the separable product of the
    Bethe-Salpeter 
amplitudes allowing us to suppress the dependence on indices in
(\ref{green_pol_p0}) and (\ref{green_pol_p2}):
\begin{equation}
\label{abbre_sep_prod_BSA}
\bigg[
\chi_{\bar P}(p_\xi,p_\eta)
\;
\overline\chi_{\bar P}(p'_\xi,p'_\eta)
\bigg]_{a_1 a_2 a_3;\; a'_1 a'_2 a'_3}
:= 
\chi_{\bar P\;a_1 a_2 a_3}(p_\xi,p_\eta)
\;
\overline\chi_{\bar P\;a'_1 a'_2 a'_3}(p'_\xi,p'_\eta).
\end{equation}

\begin{figure}[!h]
  \begin{center}
    \epsfig{file={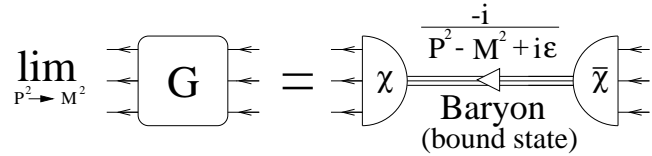},width=85mm}
  \end{center}
\caption{The behavior of the six-point Green's function $G$ in the
  vicinity of a three-quark bound-state pole of a baryon with mass $M$: Via
  the adjoint Bethe-Salpeter amplitude $\overline\chi_{\bar P}$ as 'vertex'
  (right halved bubble) the three (off shell) quarks form a bound state
  (baryon), which then propagates by means of the propagator $\sim (P^2 - M^2
  + {\rm i} \epsilon)^{-1}$ (denoted by the threefold line) and finally 'decays'
  again via the 'vertex' given by the Bethe-Salpeter amplitude $\chi_{\bar P}$ (left halved bubble)
  into three off shell quarks.}
\label{fig:residue}
\end{figure}

This typical analytical structure of the six-point Green's function
$G_P$ in the vicinity of the bound-state $P\approx\bar P$ enables us to
isolate the three-fermion bound-state contributions and to extract the
relevant quantity describing the bound states, namely the Bethe-Salpeter amplitude $\chi_{\bar P}$.
In summary:
\begin{itemize}
\item
We see that a three-fermion bound state with mass $M$ indeed gives
rise to a first order pole in the total three-body
energy $P^0$ at the bound-state energy $P^0 \rightarrow \omega_{\bf
P}= \sqrt{{\bf P}^2+M^2}$ or, equivalently, $P^2\rightarrow M^2$, $P^0 > 0$.  This
analytical dependence of $G_P$ on $P$ is a useful criterion to
identify bound states.  Note that it is just the Fourier transform
of the $\theta$-function, due to the particular
time-ordering (\ref{spez_TO}), which causes this singularity.
\item
A further striking feature is that the Green's function becomes
separable on the mass shell of the bound state, {\it i.e.}  the dependence
on the relative momenta and also in the indices for the three incoming
and outgoing quarks separates; the product of both parts, which
corresponds to the residuum of the six-point-Green's function at the
baryon pole $P^0 \rightarrow \omega_{\bf P}$, just defines the
Bethe-Salpeter amplitude and its adjoint, see also fig.
\ref{fig:residue}:
\begin{eqnarray}
\label{green_res_p0}
\mathrm{Res}_{|_{P^0=\omega_{\bf P}}} 
G_{P\;a_1 a_2 a_3;\;a'_1 a'_2 a'_3}(p_\xi,p_\eta;p_\xi',p_\eta') 
&=&
\frac{-\textrm{i}}{2\omega_{\bf P}}\;
\chi_{\bar P\;a_1 a_2 a_3}(p_\xi,p_\eta)\;\overline\chi_{\bar P\;a'_1 a'_2 a'_3}(p'_\xi,p'_\eta).
\end{eqnarray}
\end{itemize}

Evaluating the inhomogeneous integral equation (\ref{BSEmomentum_G})
for the six-point Green's function $G_P$ at the bound-state pole and
using this special behavior (\ref{green_res_p0}) of $G_P$ at this
pole position will allow to derive the Bethe-Salpeter equation for
the Bethe-Salpeter amplitudes and the corresponding normalization
condition. This will be shown in the next section.

\section{Bethe-Salpeter equation and normalization condition}
\label{sec:BSE}
With the results of the foregoing sections, we are now in the
position to derive
\begin{itemize}
\item
the Bethe-Salpeter equation for the Bethe-Salpeter amplitudes, which is an homogeneous
integral equation describing the bound states relativistically, 
\item
the normalization of the Bethe-Salpeter amplitudes.
\end{itemize}
This can be done simultaneously in a simple and appealing way by a
Laurent expansion of the inhomogeneous integral equations
(\ref{BSEmomentum_G}) and (\ref{BSEmomentum_G_ad}) for the six-point
Green's function $G_P$ in the total energy variable $P^0$ around the
bound state pole at $P=\bar P$. To this end it is convenient to bring
the integral equation (\ref{BSEmomentum_G}) and its adjoint
(\ref{BSEmomentum_G_ad}) into the equivalent forms
\begin{eqnarray}
\label{green_mom}
\left[{G_0}_P^{-1} + \textrm{i}\;K_P\right]\;G_P &=& \Id,\\[2mm]
\label{green_mom_ad}
G_P\;\left[{G_0}_P^{-1} + \textrm{i}\;K_P\right] &=& \Id,
\end{eqnarray}
where the dependence on the four momentum $P$ appears only on the left
hand side.  Here $\Id$ is the identity for the operator product
(\ref{operator_prod_mom}), which reads explicitly
\begin{eqnarray}
\Id_{a_1 a_2 a_3;\; a'_1 a'_2 a'_3}(p_\xi,p_\eta;p_\xi',p_\eta') 
:= 
\delta_{a_1 a'_1} \delta_{a_2 a'_2} \delta_{a_3 a'_3}
\;\;(2 \pi)^4\;\delta^{(4)}(p_\xi-p_\xi')\;\; (2 \pi)^4\;\delta^{(4)}(p_\eta-p_\eta'),
\end{eqnarray}
and the operator ${G_0}_P^{-1}$ is the inverse of ${G_0}_P$ with respect to
this operator product, which thus obeys ${G_0}_P^{-1}\;G_0 =
G_0\;{G_0}_P^{-1} = \Id$. It is given by the triple product of the
inverse quark propagators
\begin{eqnarray}
{G_0}_P^{-1}(p_\xi,p_\eta; p_\xi',p_\eta')\!&=&\! 
{S_F^1}^{-1}\!\left(\mbox{$\frac{1}{3}P+p_{\xi}+\frac{1}{2}p_{\eta}$}\right)
\tens
{S_F^2}^{-1}\!\left(\mbox{$\frac{1}{3}P-p_{\xi}+\frac{1}{2}p_{\eta}$}\right)
\tens
{S_F^3}^{-1}\!\left(\mbox{$\frac{1}{3}P-p_{\eta}$}\right)\nn[3mm]
&& \quad\times \;\;(2 \pi)^4\;\delta^{(4)}(p_\xi-p_\xi')\;\; (2 \pi)^4\;\delta^{(4)}(p_\eta-p_\eta').
\end{eqnarray}
Equations (\ref{green_mom}) and (\ref{green_mom_ad}) imply that $G_P$ is the resolvent of a pseudo-Hamiltonian
\begin{equation}
\label{pseudo_H}
H_P := {G_0}_P^{-1} + \textrm{i}\;K_P,
\end{equation}
{\it i.e.}
\begin{equation}
\label{green_very_compact}
H_P\;G_P = 
G_P\;H_P = \Id.
\end{equation}
In order to obtain an equation for the Bethe-Salpeter amplitudes and their
normalization condition from (\ref{green_very_compact}), we use the analytical
dependence of the six-point Green's function $G_P$ on $P$ in the vicinity of
the bound-state pole at $\bar P$ derived in the preceding subsection.
Therefore we perform an expansion of the Green's function $G_P$ and the
pseudo-Hamiltonian $H_P$ in the variable $P^0$ around the bound-state energy
$\omega_{\bf P}$.  Due to eq. (\ref{green_pol_p0}) we find a Laurent expansion
of the Green's function $G_P$ beginning with the first order
singularity\footnote{Note that with our notation $\bar P=(\omega_{\bf P}, {\bf
    P})$ and $P=(P^0,{\bf P})$ the Bethe-Salpeter amplitudes $\chi_{\bar P}$
  and $\overline\chi_{\bar P}$ do not depend on $P^0$, as they are on shell
  amplitudes by definition.},
\begin{equation}
  \label{laurent_G}
  G_P= \frac{- {\rm i}}{2\omega_{\bf P}}\;\frac{\chi_{\bar P}\;\overline\chi_{\bar P}}
  {P^0-\omega_{\bf P}+ {\rm i}\epsilon}\;
  +\;\frac{\partial}{\partial P^0}\; \left(P^0-\omega_{\bf P}\right)\;G_P \bigg|_{P^0=\omega_{\bf P}}\;
  +\;{\cal O}\left(P^0-\omega_{\bf P}\right)\;.
\end{equation}
and analogously for the operator $H_P$ we have the Taylor series expansion
\begin{equation}
  \label{taylor_H}
  H_P= H_{\bar P}\; 
  +\; \frac{\partial}{\partial P^0}\; H_P \bigg|_{P^0=\omega_{\bf P}}
  \hspace*{-5mm}\left(P^0-\omega_{\bf P}\right)\;
  +\;{\cal O}\left(\left(P^0-\omega_{\bf P}\right)^2\right).
\end{equation}
Inserting both expansions (\ref{laurent_G}) and (\ref{taylor_H}) into
eq. (\ref{green_very_compact}) then yields the following Laurent
expansion of the operator equation $H_P\;G_P=\Id$ up to the first order:
\begin{equation}
\label{expansion_HG}
\begin{array}{rl|ll}
&\displaystyle{
-\;\frac{\rm i}{2\omega_{\bf P}}\;H_{\bar P}\; \left[\chi_{\bar P}\;\overline\chi_{\bar P}\right] \;\left(P^0-\omega_{\bf P}+i\epsilon\right)^{-1}}
&
\;\textrm{order} &-1\\[5mm]
&\displaystyle{
+\; H_{\bar P}\;\left[\frac{\partial}{\partial P^0}\left[\left(P^0-\omega_{\bf P}\right) G_P\right]\right]_{P^0=\omega_{\bf P}}\;
\!\!-\;\frac{\rm i}{2\omega_{\bf P}}\;\left[\frac{\partial}{\partial P^0}\;H_P\right]_{P^0=\omega_{\bf P}}\chi_{\bar P}\;\overline\chi_{\bar P}}
&
\;\textrm{order} &0\\[5mm]
&
+\;\displaystyle{ \mathcal{O}\left(P^0-\omega_{\bf P}\right)}
& 
\;\textrm{orders} &\geq 1  \\[5mm]
= &\displaystyle{\Id} 
&
\;\textrm{order} &0
\end{array}
\end{equation}
Comparing the expansion coefficients of each order in (\ref{expansion_HG}) we
obtain simultaneously the equation for the amplitudes $\chi_{\bar P}$, {\it
  i.e.} the Bethe-Salpeter equation, and the normalization
condition.

\subsection{The Bethe-Salpeter equation for three bound fermions}
\begin{figure}[h]
  \begin{center}
    \epsfig{file={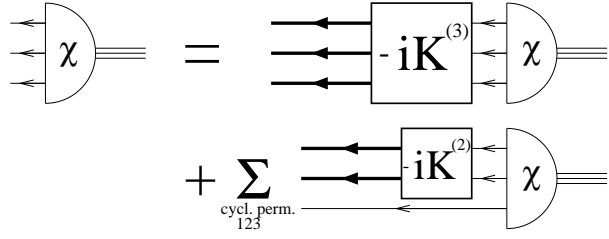},width=80mm}
  \end{center}
\caption{Graphical illustration of the three-fermion Bethe-Salpeter
equation for the Bethe-Salpeter amplitude $\chi_{\bar P}$. $K^{(3)}$
and $K^{(2)}$ denote the irreducible three- and two-body
interaction kernels, respectively. Thick arrows on quark lines indicate
full quark propagators.}
\label{fig:BSE}
\end{figure}

The expansion coefficients in the Laurent series (\ref{expansion_HG})
of the order $(P^0-\omega_{\bf P})^{-1}$ yield
\begin{equation}
H_{\bar P}\; \left[\chi_{\bar P}\;\overline\chi_{\bar P}\right] = 0.
\end{equation}
Now the factorization property of the pole residue becomes crucial: due to the separability (\ref{green_res_p0}) of the
product $\chi_{\bar P}\;\overline\chi_{\bar P}$ the operation
(\ref{operator_prod_mom}) of $H_P$ acts only on the indices and
relative momenta of $\chi_{\bar P}$, while $\overline\chi_{\bar P}$
remains unaffected, thus producing the \textbf{Bethe-Salpeter equation}
for the Bethe-Salpeter amplitude $\chi_{\bar P}$: 
\begin{equation}
H_{\bar P}\; \chi_{\bar P} = 0.
\end{equation}
Here the operator product of a six-point function $H_P$ with a
three-point function $\chi_{\bar P}$ is defined, analogous to
(\ref{operator_prod_mom}), as
\begin{equation}
\label{operator_prod_6_3_mom}
\left[H_{\bar P}\;\chi_{\bar P}\right]_{a_1 a_2 a_3}(p_\xi,p_\eta)
:=
\int\frac{\textrm{d}^4 p_\xi'}{(2\pi)^4}\;\frac{\textrm{d}^4 p_\eta'}{(2\pi)^4}\;
H_{\bar P\;a_1 a_2 a_3;\;a'_1 a'_2 a'_3}(p_\xi,p_\eta;\;p_\xi',p_\eta')\;
\chi_{\bar P\;a'_1 a'_2 a'_3}(p_\xi',p_\eta').
\end{equation}
In the same fashion the corresponding Laurent expansion of the adjoint
equation $G_P H_P=\Id$ gives the \textbf{adjoint Bethe-Salpeter
equation} for the adjoint amplitude\footnote{Here the operator product
is defined similar to (\ref{operator_prod_6_3_mom}) but with summation
and integration over indices and momenta that appear on the left in
$H_{\bar P}$} $\overline\chi_{\bar P}$:
\begin{equation}
\label{HbarChi}
\overline\chi_{\bar P}\;H_{\bar P} = 0.
\end{equation}
Inserting definition (\ref{pseudo_H}) for $H_{\bar P}$ and
multiplying by ${G_0}_{\bar P}$, we bring the Bethe-Salpeter equation
and its adjoint into their more conventional form
\begin{equation}
\begin{array}{rcl}
\chi_{\bar P} &=&  -\textrm{i}\;{G_0}_{\bar P}\;K_{\bar P}\;\chi_{\bar P},\\[3mm]
\overline\chi_{\bar P} &=&  -\textrm{i}\;\overline\chi_{\bar P}\;K_{\bar P}\;{G_0}_{\bar P},
\end{array}
\quad \textrm{with}\;\; K_{\bar P} = K^{(3)}_{\bar P} + \overline K^{(2)}_{\bar P}.
\end{equation}
The three-body Bethe-Salpeter equation is a covariant eight-dimensional
homogeneous integral equation in the variables $p_\xi = (p_\xi^0, {\bf
p_\xi})$ and $p_\eta = (p_\eta^0, {\bf p_\eta})$ describing the
properties of bound states. It reads explicitly:
\begin{eqnarray}
\label{BSEquation}
\lefteqn{\chi_{\bar P\;a_1 a_2 a_3}(p_\xi,p_\eta)=}\nn\nn
& &
S^1_{F\;a_1 a'_1}\left(\mbox{$\frac{1}{3}\bar P+p_{\xi}+\frac{1}{2}p_{\eta}$}\right)\;
S^2_{F\;a_2 a'_2}\left(\mbox{$\frac{1}{3}\bar P-p_{\xi}+\frac{1}{2}p_{\eta}$}\right)\;
S^3_{F\;a_3 a'_3}\left(\mbox{$\frac{1}{3}\bar P-p_{\eta}$}\right)\nn
& &\quad\times\; (-i)\;
\int\frac{\textrm{d}^4 p_\xi'}{(2\pi)^4}\;\frac{\textrm{d}^4 p_\eta'}{(2\pi)^4}\;
K^{(3)}_{\bar P\; a'_1 a'_2 a'_3;\; a''_1 a''_2 a''_3}
(p_\xi,p_\eta;\; p_\xi',p_\eta')\;
\chi_{\bar P\; a''_1 a''_2 a''_3}(p_\xi',p_\eta')\nn\nn\nn
& &+\;
S^1_{F\;a_1 a'_1}\left(\mbox{$\frac{1}{3}\bar P+p_{\xi}+\frac{1}{2}p_{\eta}$}\right)\;
S^2_{F\;a_2 a'_2}\left(\mbox{$\frac{1}{3}\bar P-p_{\xi}+\frac{1}{2}p_{\eta}$}\right)\nn
& &\quad\times\; (-i)\;
\int\frac{\textrm{d}^4 p'_{\xi }}{(2\pi)^4}\;
K^{(2)}_{(\frac{2}{3}\bar P+p_\eta)\; a'_1 a'_2;\; a''_1 a''_2}
(p_{\xi}, p'_{\xi})
\;\chi_{\bar P\; a''_1 a''_2 a_3}(p_\xi',p_\eta)\nn\nn
& &+\;
S^1_{F\;a_1 a'_1}\left(\mbox{$\frac{1}{3}\bar P+p_{\xi}+\frac{1}{2}p_{\eta}$}\right)\;
S^3_{F\;a_3 a'_3}\left(\mbox{$\frac{1}{3}\bar P-p_{\eta}$}\right)\nn
& &\quad\times\; (-i)\;
\int\frac{\textrm{d}^4 p'_{\xi_2}}{(2\pi)^4}\;
K^{(2)}_{(\frac{2}{3}\bar P+p_{\eta_2})\;a'_1 a'_3; \;a''_1  a''_3}
(p_{\xi_2}, p'_{\xi_2})
\;\chi_{\bar P\; a''_1 a_2 a''_3}\left(\mbox{$\!-\frac{1}{2} p'_{\xi_2}  +
    \frac{3}{4} p_{\eta_2}$}, \mbox{$- p'_{\xi_2}  - \frac{1}{2} p_{\eta_2}$}\right)\nn\nn
& &+\;
S^2_{F\;a_2 a'_2}\left(\mbox{$\frac{1}{3}\bar P-p_{\xi}+\frac{1}{2}p_{\eta}$}\right)\;
S^3_{F\;a_3 a'_3}\left(\mbox{$\frac{1}{3}\bar P-p_{\eta}$}\right)\nn
& &\quad\times\; (-i)\;
\int\frac{\textrm{d}^4 p'_{\xi_1}}{(2\pi)^4}\;
K^{(2)}_{(\frac{2}{3}\bar P+p_{\eta_1})\; a'_2 a'_3;\; a''_2 a''_3}
(p_{\xi_1},p'_{\xi_1})
\;\chi_{\bar P\; a_1 a''_2 a''_3}\left(\mbox{$\!-\frac{1}{2} p'_{\xi_1} -
    \frac{3}{4} p_{\eta_1}$}, \mbox{$p'_{\xi_1} - \frac{1}{2}p_{\eta_1}$} \right).
\end{eqnarray}
Recall that the two sets $(p_{\xi_1},p_{\eta_1})$ and
$(p_{\xi_2},p_{\eta_2})$ of relative momenta are related to
the standard set $(p_\xi,p_\eta)=(p_{\xi_3},p_{\eta_3})$ by cyclic permutations of the
quark momenta represented by the linear transformations (\ref{cycl_perm_jacobi_mom}).
The Bethe-Salpeter equation is represented diagrammatically in fig.
\ref{fig:BSE}.

\subsection{The normalization condition}
Comparing the expansion coefficients of order $(P^0-\omega_{\bf
P})^0$ in the Laurent series (\ref{expansion_HG}) gives
\begin{equation}
\label{HG_order_0}
\; H_{\bar P}\;\left[\frac{\partial}{\partial P^0}\left[\left(P^0-\omega_{\bf P}\right)\;G_P\right] \right]_{P^0=\omega_{\bf P}}\;
-\;\frac{\rm i}{2\omega_{\bf P}}\;\left[\frac{\partial}{\partial P^0}\;H_P \right]_{P^0=\omega_{\bf P}}\;\chi_{\bar P}\;\overline\chi_{\bar P}
= \Id 
\end{equation}
which expresses the requirement that the product of $\chi_{\bar P}$ and
$\overline \chi_{\bar P}$ is the residue of the bound-state pole in $G_P$.  If
we multiply this equation from the left hand side with the adjoint amplitude
$\overline\chi_{\bar P}$, the first term in (\ref{HG_order_0}) vanishes
according to the adjoint Bethe-Salpeter equation (\ref{HbarChi}) and we find the \textbf{normalization
  condition} for the Bethe-Salpeter amplitudes \cite{Tom83}
\begin{equation}
\label{norm_condition_compact}
-{\rm i}\;\overline\chi_{\bar P}\;
\bigg[\frac{\partial}{\partial P^0}\;H_P \bigg]_{P^0=\omega_{\bf P}}\!
\chi_{\bar P}
= 2 \omega_{\bf P}.
\end{equation}
The full explicit expression then reads
\begin{eqnarray}
\lefteqn{
-\textrm{i}\int\frac{\textrm{d}^4 p_\xi'}{(2\pi)^4}\;\frac{\textrm{d}^4 p_\eta'}{(2\pi)^4}\;
\int\frac{\textrm{d}^4 p_\xi}{(2\pi)^4}\;\frac{\textrm{d}^4 p_\eta}{(2\pi)^4}
}\\
& &
\overline\chi_{\bar P}(p_\xi',p_\eta')
\;
\bigg[
\frac{\partial}{\partial P^0}
\left({G_0}_P^{-1}(p_\xi',p_\eta',p_\xi,p_\eta)+{\rm i} K_P(p_\xi',p_\eta',p_\xi,p_\eta)\right)\bigg]_{P^0=\omega_{\bf P}}\;
\chi_{\bar P}(p_\xi,p_\eta)
\;\;=\;\;2 \omega_{\bf P}.\nonumber
\end{eqnarray}
{\it A priori} the normalization condition provides the correct relation
between the amplitudes $\chi_{\bar P}$ and the six-point Green's
function $G$. But furthermore, this additional boundary condition is essential in selecting
the proper solutions $\chi_{\bar P}$ of the three-fermion Bethe-Salpeter
equation  (\ref{BSEquation})
thus providing a discrete spectrum $\bar P^2 = M^2$ of bound states.\\

\begin{figure*}[h]
  \begin{center}
    \epsfig{file={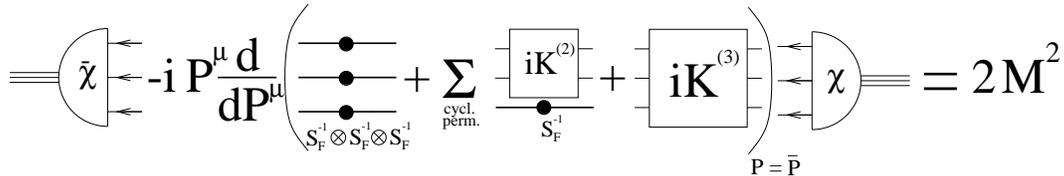},width=140mm}
  \end{center}
\caption{The normalization condition for the Bethe-Salpeter
amplitudes.The filled circles denote the inverse quark propagators,
$K^{(2)}$ and $K^{(3)}$ are the irreducible two- and three-quark
interaction kernels.}
\label{fig:normBS}
\end{figure*}

Note that the normalization condition for the amplitudes as written in
the form of eq. (\ref{norm_condition_compact}) is not manifestly
covariant in contrast to the Bethe-Salpeter equation. But it holds in
any frame since both sides of eq. (\ref{norm_condition_compact})
transform like the time component of a four-vector, if the amplitudes
transform properly under Lorentz transformations. However, we would
like to remark here that the normalization
(\ref{norm_condition_compact}) may also be rewritten in explicit
covariant form as
\begin{equation}
\label{norm_condition_compact_cov}
-{\rm i}\;\overline\chi_{\bar P}\;
\bigg[P^\mu\frac{\partial}{\partial P^\mu}\;H_P \bigg]_{P =\bar P}
\chi_{\bar P}
= 2 M^2.
\end{equation}
For a diagrammatic illustration of eq.
(\ref{norm_condition_compact_cov}) see fig. \ref{fig:normBS}.

\section{Reduction to the Salpeter equation}
\label{sec:sequation}
\subsection{Motivation and general remarks}
In principle, the Bethe-Salpeter equation (\ref{BSEquation}) for three
fermions, derived in the foregoing section, provides a suitable starting point
for the covariant description of baryons as bound states of three quarks in
the framework of QCD. Solving this equation for given single quark propagators
$S_F$ and interaction kernels $K^{(2)}$ and $K^{(3)}$, the discrete spectrum
of states is then determined by the normalization condition
(\ref{norm_condition_compact_cov}).  However, an exact solution of the
Bethe-Salpeter equation within the framework of QCD is impossible, since the
quark propagator $S_F$ and the irreducible interaction kernels $K^{(2)}$ and
$K^{(3)}$ are only formally defined in perturbation theory as an infinite sum
of Feynman diagrams. Moreover it is unclear, which particular approximation
will provide quark confinement in hadrons.

But even if the exact kernels and propagators were known in QCD, the
dependence on the relative energy (or the corresponding relative time)
variables leads to a complicated analytic pole structure, which so far
could be treated rigorously only in the case of two scalar particles
interacting through a (massless) scalar exchange (the so-called Wick-Cutkosky
model, see \cite{Wi54,Cu54}).  Thus, the use of general
two-quark and three-quark interaction kernels, that depend on the
relative energy variables, leads to serious conceptional and practical
problems. To our knowledge the only attempt to solve (approximately) a
full four-dimensional three-quark Bethe-Salpeter equation
in Euclidean space has been performed by Meyer and B\"ohm
\cite{Mey74,Mey75,BoMe79} and subsequently by Kielanowski
\cite{Kie80} and Falkensteiner \cite{Falk82} in an approach where
baryons were considered as extremely strongly bound systems of three
quite heavy constituent quarks $m \gg 1 GeV$, that interact via
harmonic oscillator interactions, so that a solution can be
obtained by an expansion in powers of $\frac{1}{m}$. However, from
a modern point of view, the crude approximations and especially the large
constituent quark masses are questionable and not suited for
phenomenologically successful applications.

Thus, the use of the full eight-dimensional Bethe-Salpeter
equation is of rather limited practical value and the lack of a confinement
kernel that could be rigorously derived from QCD anyhow requires an
appropriate phenomenological parameterization: so far, the only ansatz that
can give a realistic description of the quark confinement and thus can account
for the gross features of the whole baryon spectrum up to highest orbital
excitations, is the nonrelativistic quark
model, which uses {\it static} two- and three-quark potentials.

For these reasons we will not treat the full three-quark Bethe-Salpeter
equation. Instead we try to eliminate the difficult relative
energy dependence in order to get a six-dimensional reduction of the full
eight-dimensional Bethe-Salpeter equation, the so-called Salpeter equation
\cite{Sa52}, with the aim to obtain a 
framework that is still covariant. At the same time we want to keep as close as possible to the quite
successful nonrelativistic quark potential model in order to obtain at least
this model as a non-relativistic limit.  In this spirit, a covariant quark
model for mesons based on the instantaneous $q\bar q$-Bethe-Salpeter 
equation has been developed already and has been successfully applied
to the calculation of mass spectra and various transition
matrix elements up to high momentum transfers, see \cite{ReMu94,MuRe94}. To
extend this model for calculations of baryons, we make the
same simplifying assumptions and approximations in the three-quark Bethe-Salpeter equation ({sect. \ref{sec:approx}}): The full quark
propagators $S_F$ are assumed to be given by their free forms with effective
constituent quark masses. Moreover, the kernels $K^{(2)}$ and $K^{(3)}$ are
approximated by effective interactions that are instantaneous in the rest
frame of the bound state, which thus corresponds to the neglect of retardation
effects. We should mention here that the
instantaneous approximation can be formulated in a frame independent way
\cite{WaMa89}, so that formal covariance is preserved, which becomes important
for the calculation of transitions between baryon states,
where at least one of the baryon has to be boosted.

In the meson case these approximations allow for a direct and
straightforward reduction to the $q\bar q$-Salpeter equation
\cite{Sa52,ReMu94,MuRe94} by an analytical integration over the
relative energy variable, since the connected instantaneous $q\bar
q$-kernel cuts the whole relative energy dependence of the
Bethe-Salpeter equation. The same applies also to the three-quark
Bethe-Salpeter equation, if only an instantaneous, connected
three-quark kernel $K^{(3)}$ is taken into account and two-particle
kernels are neglected ($K^{(2)}=0$). In this case the Salpeter
equation can be formulated in a concise Hamiltonian form with some
characteristic projector properties that reduce the number of
independent functions necessary to describe a baryon state. For the
sake of conceptual simplicity such an approach has been used in our
former investigations \cite{Met97a,Met99a,Met99b},
where all kinds of interactions have been parameterized in a kind of
collective instantaneous three-body kernel.  In section
\ref{sec_reduc_V3} we will first give a summary of the reduction
procedure in this simple and instructive case and discuss the 
specific structure of the resulting Salpeter equation.

However, as soon as genuine two-quark kernels $K^{(2)}$ are considered, new
difficulties arise since the two-body terms are unconnected within the
three-quark system: despite an instantaneous approximation of $K^{(2)}$ there
remains a relative energy dependence due to retardation effects of the third
non-interacting spectator quark, which is off-shell in general.  In this
respect the elimination of the relative energies is technically and
conceptually much more involved and an enhanced reduction procedure is needed.
In {section \ref{sec:reduc_V3V2}} we give a procedure that nevertheless allows
for the reduction to a Salpeter equation. The crucial point is the
existence of a genuine instantaneous connected part of the interaction
$K^{(3)}$, right from start. In our model this part will be given by a
convenient form of a static three-body confinement potential that must be
present for all baryon states in all sectors due to the confinement
hypothesis.  Recasting the Bethe-Salpeter equation into a more convenient form
with all two-particle effects collected into a six-point Green's function thus
provides a similar reduction procedure as in the case of vanishing two-body
interactions. Extending a kind of quasi-potential approach as it was first
proposed by Logunov and Tavkhelidze \cite{LoTa63} for the equal-time Green's
function of two scalar particles, all effects of the unconnected two-body
interactions can then be transformed into an effective instantaneous
potential that adds to the genuine three-body kernel $K^{(3)}$ and we finally
end up with a reduced equation that exhibits the same expedient projector
structure as in the case where the dynamics of the quarks is given by an
instantaneous three-body kernel alone. The effective potential, however,
consists of an infinite perturbation series of time-ordered Feynman diagrams,
which needs to be truncated for explicit calculations. In the subsequent sect.
\ref{bound_states_born} we will analyze the structure of the resulting baryon
Salpeter equation and its corresponding Salpeter amplitudes in detail: a
remarkable substantial property of our covariant Salpeter approach will turn
out to be that it exhibits a one-to-one correspondence with the states of the
nonrelativistic quark model.

\subsection{Approximations}
\label{sec:approx}
In order to transform the Bethe-Salpeter equation into
a solvable integral equation several simplifying approximations
have to be made. To start, we follow the prescription of \cite{ReMu94}
and assume free quark propagators and instantaneous interaction kernels.
\subsubsection{Free quark propagators}
First, we make the assumption that the full quark propagators can be
approximated by the usual free fermion propagators with effective
constituent quark masses $m_i$ for each quark\footnote{For a
simplified notation we suppress the explicit flavor- and color
dependencies for the moment.}
\begin{equation}
\label{free_prop_approx}
S^i_{F}\left(p_i\right) 
\approx \frac{\rm i}{\not p_i - m_i + {\rm i}\epsilon }. 
\end{equation}
This approximation is consistent with the picture of a hadron mainly built out
of {\it constituent quarks} analogous to the non-relativistic quark model.
The effective constituent quark masses $m_i$ enter as free parameters in our
model.

\subsubsection{Instantaneous approximation}
Moreover, we choose the irreducible two- and three-body interaction
kernels to be instantaneous in the rest frame of the baryon, meaning
that in the center-of-mass system there is no dependence on the
relative energy variables $p_\xi^0$ and $p_\eta^0$:
\begin{eqnarray}
\label{inst_approx_CMS_V3}
K^{(3)}_{P}(p_\xi,p_\eta;\;p_\xi',p_\eta')\bigg|_{P=(M,{\bf 0})}
&\stackrel{!}{=}&
V^{(3)}({\bf p_\xi},{\bf p_\eta};\;{\bf p_\xi'},{\bf p_\eta'}),\\[3mm]
\label{inst_approx_CMS_V2}
K^{(2)}_{(\frac{2}{3}P+p_{\eta_k})}(p_{\xi_k},p'_{\xi_k})
\bigg|_{ P=(M,{\bf 0})}
&\stackrel{!}{=}&
V^{(2)}({{\bf p}_{\xi_k}},{{\bf p}_{\xi_k}'}).
\end{eqnarray}
This approximation corresponds to the neglect of retardation effects in the
rest-frame. To preserve the formal covariance of the Bethe-Salpeter equation, however, we need a covariant description of the
instantaneous approximation which holds in any arbitrary reference frame of
the bound state. We follow an idea of Wallace and Mandelzweig \cite{WaMa89} and
introduce for arbitrary time-like total four-momenta $P$, $P^2>0$, the
following covariant decomposition of any four-dimensional four-vector $p$,
\begin{equation}
\label{cov_decomp}
p = p_\|\; \frac{P}{\sqrt{P^2}} + p_\bot 
\end{equation}
into components parallel and perpendicular to the total four-momentum $P$:
\begin{equation}
\label{p_par_per_P}
p_\| := \frac{\metric(p, P)}{\sqrt{P^2}},\quad
p_\bot := p - \frac{\metric(p, P)}{ P^2}\; P.
\end{equation}
This is a decomposition into a time-like vector $p_\|\;P/\sqrt{P^2}$ and a
space-like vector $p_\bot$ which effectively is three-dimensional in content.
Now the instantaneous approximation, which has been formulated in eqs.
(\ref{inst_approx_CMS_V3}) and (\ref{inst_approx_CMS_V2}) within the
center-of-mass frame of the three-body system, can be formulated
in any reference frame (which is specified by the four-momentum $P$): we
assume that the kernels do not depend on the time-like parallel components of
the relative momenta, {\it i.e.}
${p_\xi}_\|,{p_\eta}_\|,{p_\xi'}_\|,{p_\eta'}_\|$, but only on the space-like
perpendicular components:
\begin{eqnarray}
\label{inst_approx_COV_V3}
K^{(3)}_{P}(p_\xi,p_\eta;\;p_\xi',p_\eta') 
&\stackrel{!}{=}&
V^{(3)}({p_\xi}_\bot,{p_\eta}_\bot;\;{p_\xi'}_\bot,{p_\eta'}_\bot),\\[3mm]
\label{inst_approx_COV_V2}
K^{(2)}_{(\frac{2}{3}P+p_{\eta_k})}(p_{\xi_k},p'_{\xi_k})
&\stackrel{!}{=}&
V^{(2)}({p_{\xi_k}}_\bot,{p_{\xi_k}'}_\bot).
\end{eqnarray}
For interaction kernels of this type we have 
\begin{eqnarray}
\label{no_contrib_V3_to_norm}
P^\mu\frac{\partial}{\partial P^\mu} K^{(3)}_{P}(p_\xi,p_\eta;\;p_\xi',p_\eta')\bigg|_{P=\bar P}&=&
P^\mu\frac{\rm d}{{\rm d} P^\mu}\;V^{(3)}({p_\xi}_\bot,{p_\eta}_\bot;\;{p_\xi'}_\bot,{p_\eta'}_\bot)\bigg |_{P=\bar P } = 0,\\[3mm]
P^\mu\frac{\partial}{\partial P^\mu}K^{(2)}_{(\frac{2}{3}P+p_{\eta_k})}(p_{\xi_k},p'_{\xi_k})\bigg|_{P=\bar P}&=&
P^\mu\frac{\rm d}{{\rm d}
  P^\mu}\;V^{(2)}({p_{\xi_k}}_\bot,{p_{\xi_k}'}_\bot)\bigg |_{P= \bar P} = 0
\end{eqnarray}
and, consequently, these give no direct contributions to the
normalization condition (\ref{norm_condition_compact_cov}) for the
Bethe-Salpeter amplitudes. In the rest frame of the
baryon where $P = \bar P = (M,{\bf 0})$ we find
\begin{equation}
p_\| = p^0\qquad {\rm and}\qquad
p_\bot = (0,{\bf p}),
\end{equation}
so that the covariant formulation of the instantaneous approximation given in
(\ref{inst_approx_COV_V3}) and (\ref{inst_approx_COV_V2}) indeed recovers the
conditions (\ref{inst_approx_CMS_V3}) and (\ref{inst_approx_CMS_V2}) in the
center-of-mass frame.\\

In the two-fermion case \cite{ReMu94} it was shown that the assumptions of free
quark propagators and instantaneous interaction kernels are sufficient to
completely eliminate the dependence on the relative energy dependence in order
to arrive at a reduced equation which can be solved with standard techniques. In
the three-fermion problem, however, this is in general not possible, unless we
consider systems without two-body interactions. In the more general case new
difficulties arise from the property of the two-body terms that these
are unconnected within the three-body system. Despite the instantaneous
approximation of the two-particle interactions, the kernel $\overline
K^{(2)}_{P = (M,{\bf 0})}$ defined by eq. (\ref{K2_lift_fourier}) remains
retarded, since (in the CMS) it maintains the dependence on the relative
energies $p_\xi^0$ and $p_\eta^0$ due to retardation effects of the third
noninteracting spectator quark which is off-shell in general. Accordingly, in
$\overline K^{(2)}_{P = (M,{\bf 0})}$ this remaining relative energy
dependence is given explicitly by the inverse single quark propagator of the
spectator together with its four-momentum conserving $\delta$-function:
\begin{eqnarray}
\label{bar_K2_inst_approx}
 \lefteqn{\overline K^{(2)}_{P}(p_\xi,p_\eta;\;p_\xi',p_\eta')\bigg|_{ P=(M,{\bf 0})}=}\nn[3mm]
  & &
\sum_{{\rm cycl. perm}\atop{{\rm of } (123)}}       
  V^{(2)}({{\bf p}_{\xi_3}}, {{\bf p}_{\xi_3}'}) \tens
  {S_F^3}^{-1}\left(\mbox{$\frac{1}{3}P-p_{\eta_3}$}\right)\bigg|_{P = (M,{\bf 0})}\;
  (2\pi)^4 \; \delta^{(4)}(p_{\eta_3}-p'_{\eta_3}).
\end{eqnarray}
Thus, the consideration of unconnected two-particle terms in the three-body
Bethe-Salpeter equation makes a reduction technically much more
involved, despite the instantaneous approximation of the two-body kernels.
With regard to the goal of finding a convenient reduction procedure it is therefore
instructive to consider first the conceptually much easier case of vanishing
two-particle kernels, where the dynamics of the
quarks is determined by a connected instantaneous three-body kernel alone.  
In this case the reduction of the eight-dimensional Bethe-Salpeter 
equation to an equivalent six-dimensional equation -- the
so-called Salpeter equation -- is straightforward (as in the two-fermion
case with a connected instantaneous two-body kernel \cite{ReMu94}).

\subsection{The reduction without two-particle kernels}
\label{sec_reduc_V3}
Neglecting the irreducible two-particle interaction kernels, {\it i.e.}
$K^{(2)} =0$, and taking only an instantaneous three-body kernel
(\ref{inst_approx_COV_V3}) into account, the Bethe-Salpeter
equation and its adjoint
in the center-of-mass frame\footnote{Due to the formally covariant formulation (\ref{inst_approx_COV_V3}) of the instantaneous
approximation of the irreducible three-body kernel (which preserves the formal
covariance of the Bethe-Salpeter equation), it is sufficient (and convenient) to
go into the center-of-mass (CMS) frame}
of the baryon with $\bar P = (M,{\bf 0})\equiv M$
 are given by
\begin{eqnarray}
\label{BSE_V3}
\chi_{M} &=&  
-\textrm{i}\;{G_0}_{M}\; V^{(3)}  \;\chi_{M},\\[3mm]
\label{ad_BSE_V3}
\overline\chi_{M} &=&  
-\textrm{i}\;\overline\chi_{M}\; V^{(3)}\;{G_0}_{M}.
\end{eqnarray}
The crucial point is now that $V^{(3)}$ being instantaneous truncates 
the $p_\xi^0$, $p_\eta^0$ dependences of the Bethe-Salpeter equations
(\ref{BSE_V3}) and (\ref{ad_BSE_V3}). This has the following
consequences:\\

{\bf 1.)} The $p_\xi^0$, $p_\eta^0$ integration within the operator product
$V^{(3)}\;\chi_M$ on the right hand side of eq. (\ref{BSE_V3})
acts on $\chi_{M}$ directly and thus can be used to reduce this
eight-dimensional Bethe-Salpeter amplitude to a
six-dimensional amplitude $\Phi_M$, {\it i.e.} in detail
\begin{eqnarray}
\label{Vchi_eq_Vphi}
\left[V^{(3)}\;\chi_M\right]({p_\xi},{p_\eta})
&=&
\int 
\frac{\textrm{d}^4 p_\xi'}{(2\pi)^4}\;
\frac{\textrm{d}^4 p_\eta'}{(2\pi)^4}\;
V^{(3)}({\bf p_\xi},{\bf p_\eta};\;{\bf p_\xi'},{\bf
  p_\eta'})\;\chi_M(p_\xi',p_\eta')\nn[3mm]
&=&
\int 
\frac{\textrm{d}^3 p_\xi'}{(2\pi)^3}\;
\frac{\textrm{d}^3 p_\eta'}{(2\pi)^3}\;
V^{(3)}({\bf p_\xi},{\bf p_\eta};\;{\bf p_\xi'},{\bf
  p_\eta'})\;
\int\frac{\textrm{d} {p_\xi'}^0}{2\pi}\;
\frac{\textrm{d} {p_\eta'}^0}{2\pi}\;
\chi_M(p_\xi', p_\eta')\nn[3mm]
&=&
\int 
\frac{\textrm{d}^3 p_\xi'}{(2\pi)^3}\;
\frac{\textrm{d}^3 p_\eta'}{(2\pi)^3}\;
V^{(3)}({\bf p_\xi},{\bf p_\eta};\;{\bf p_\xi'},{\bf
  p_\eta'})\;
\Phi_M({\bf p_\xi'}, {\bf p_\eta'})\nn[4mm]
&=&\left[V^{(3)}\;\Phi_M\right]({\bf p_\xi},{\bf p_\eta}).
\end{eqnarray}
Consequently, there remains a six-dimensional integral
operation\footnote{Notice that we do not introduce a new 
product notation for this six-dimensional integral operation to
distinguish it from the eight-dimensional one. The difference between
the two products should be obvious from the context in which they are
used.} of $V^{(3)}$ on the reduced six-dimensional amplitude $\Phi_M$,
which is the so-called \textbf{Salpeter amplitude}:
\begin{equation}
\label{def:salpeter_ampl_CMS}
\Phi_M ({\bf p_\xi},{\bf p_\eta}) := 
\int\frac{\textrm{d} p_\xi^0}{2\pi}\;
\frac{\textrm{d} p_\eta^0}{2\pi}\;
\chi_{M}\left((p_\xi^0, {\bf p_\xi}), (p_\eta^0, {\bf p_\eta})\right).
\end{equation}
In the same way one proceeds with the operator product
$\overline\chi_M\;V^{(3)}$ in the adjoint Bethe-Salpeter equation (\ref{ad_BSE_V3}), {\it i.e.}
\begin{eqnarray}
\label{Vchiad_eq_Vphiad}
\left[\overline\chi_M\;V^{(3)}\right]({p_\xi},{p_\eta})
&=&\left[\overline\Phi_M\;V^{(3)}\right]({\bf p_\xi},{\bf p_\eta}),
\end{eqnarray}
which accordingly defines the
\textbf{adjoint Salpeter amplitude}
\begin{equation}
\label{def:salpeter_ampl_ad_CMS}
\overline\Phi_M ({\bf p_\xi},{\bf p_\eta}) := 
\int\frac{\textrm{d} p_\xi^0}{2\pi}\;
\frac{\textrm{d} p_\eta^0}{2\pi}\;
\overline\chi_{M}\left((p_\xi^0, {\bf p_\xi}), (p_\eta^0, {\bf p_\eta})\right).
\end{equation}

{\bf 2.)} Inserting (\ref{Vchi_eq_Vphi}) and (\ref{Vchiad_eq_Vphiad}) into the
Bethe-Salpeter equations (\ref{BSE_V3}) and (\ref{ad_BSE_V3}), respectively, we have
\begin{eqnarray}
\label{reconstruction_V3}
\chi_M 
&=&  
-\textrm{i}\;{G_0}_M\;V^{(3)}\;\Phi_M,\\[2mm]
\label{reconstruction_V3_ad}
\overline\chi_M 
&=&
-\textrm{i}\;\overline\Phi_M\;V^{(3)}\;{G_0}_M,
\end{eqnarray}
which gives a prescription how to reconstruct the full Bethe-Salpeter
amplitudes from the Salpeter amplitudes for any on-shell total momentum.
Consequently, in the instantaneous approximation it is sufficient to know the
reduced six-dimensional Salpeter amplitudes $\Phi_M$ and $\overline\Phi_M$ to
get the full eight-dimensional Bethe-Salpeter amplitudes $\chi_M$ and
$\overline\chi_M$, {\it i.e.}  the solutions of the Bethe-Salpeter equation
(\ref{BSE_V3}) and (\ref{ad_BSE_V3}), respectively.  The next step is to get
an equation which determines
$\Phi_M$ and $\overline\Phi_M$.\\

{\bf 3.)} As shown in eqs.
(\ref{Vchi_eq_Vphi}) and (\ref{Vchiad_eq_Vphiad}),
the quantities
\begin{equation}
\label{vertex_funcV3}
\begin{array}{c}
\Gamma_M({p_\xi},{p_\eta})
:=
\left[{G_0}_M^{-1}\;\chi_M\right]({p_\xi},{p_\eta})
=
\left[V^{(3)}\;\chi_M\right]({p_\xi},{p_\eta})
=
\left[V^{(3)}\;\Phi_M\right]({\bf p_\xi},{\bf p_\eta})
\equiv\Gamma_M({\bf p_\xi},{\bf p_\eta}),\\[2mm]
\overline \Gamma_M({p_\xi},{p_\eta})
:=
\left[\overline\chi_M\;{G_0}_M^{-1}\right]({p_\xi},{p_\eta})
=
\left[\overline\chi_M\;V^{(3)}\right]({p_\xi},{p_\eta})
=\left[\Phi_M\;V^{(3)}\right]({\bf p_\xi},{\bf p_\eta})
\equiv \overline \Gamma_M({\bf p_\xi},{\bf p_\eta}),
\end{array}
\end{equation} 
which are usually called amputated Bethe-Salpeter amplitudes or
three-quark \textbf{vertex functions}, do not depend on the relative energies
${p_\xi}^0$ and ${p_\eta}^0$ in the center-of-mass frame of the baryon.
Consequently, the analytical dependence of the Bethe-Salpeter
amplitudes $\chi_M = {G_0}_M\; \Gamma_M$ and $\overline\chi_M =
\overline\Gamma_M\; {G_0}_M $ on the variables ${p_\xi}^0$ and
${p_\eta}^0$ stems exclusively from the triple tensor product ${G_0}_M$ of the
free quark propagators.  This enables us to reduce the eight-dimensional
Bethe-Salpeter equations for the Bethe-Salpeter
amplitudes to six-dimensional integral equations for the Salpeter
amplitudes by integrating out the ${p_\xi}^0$, ${p_\eta}^0$ dependence on both
sides of eqs. (\ref{reconstruction_V3}) and (\ref{reconstruction_V3_ad}).
The Bethe-Salpeter amplitudes on the left hand side reduce to the
corresponding Salpeter amplitudes and on the right hand side the
relative energy integration affects merely the free propagator ${G_0}_M$, {\it i.e.} in detail
\begin{eqnarray}
\lefteqn{\Phi_M({\bf p_\xi},{\bf p_\eta})
=
\int\frac{\textrm{d} p_\xi^0}{2\pi}\;
\frac{\textrm{d} p_\eta^0}{2\pi}\;
\chi_{M}\left((p_\xi^0, {\bf p_\xi}), (p_\eta^0, {\bf p_\eta})\right)}\nn[3mm] 
&=&
-\textrm{i}\;
\int\frac{\textrm{d} p_\xi^0}{2\pi}\;
\frac{\textrm{d} p_\eta^0}{2\pi}\;
\int 
\frac{\textrm{d}^4 p_\xi'}{(2\pi)^4}\;
\frac{\textrm{d}^4 p_\eta'}{(2\pi)^4}\;
{G_0}_M(p_\xi,p_\eta;\;p_\xi',p_\eta')\;
\left[V^{(3)}\;\Phi_M\right]({\bf p_\xi'},{\bf p_\eta}')\nn[2mm]
&=&
-\textrm{i}\;
\int 
\frac{\textrm{d}^3 p_\xi'}{(2\pi)^3}\;
\frac{\textrm{d}^3 p_\eta'}{(2\pi)^3}\;
\int
\frac{\textrm{d} p_\xi^0}{2\pi}\;
\frac{\textrm{d} p_\eta^0}{2\pi}\;
\int
\frac{\textrm{d} p_\xi'^0}{2\pi}\;
\frac{\textrm{d} p_\eta'^0}{2\pi}\;
{G_0}_M(p_\xi,p_\eta;\;p_\xi',p_\eta')\;
\left[V^{(3)}\;\Phi_M\right]({\bf p_\xi'},{\bf p_\eta}')\nn[2mm]
&=&
-\textrm{i}\;
\int 
\frac{\textrm{d}^3 p_\xi'}{(2\pi)^3}\;
\frac{\textrm{d}^3 p_\eta'}{(2\pi)^3}\;
\langle{G_0}_M\rangle({\bf p_\xi},{\bf p_\eta};\;{\bf p_\xi'},{\bf p_\eta'})\;
\left[V^{(3)}\;\Phi_M\right]({\bf p_\xi'},{\bf p_\eta}')\nn[3mm]
&=&
-\textrm{i}\;\left[\left\langle{G_0}_M\right\rangle\;V^{(3)}\;\Phi_M\right]({\bf p_\xi},{\bf p_\eta}). 
\end{eqnarray}
Thus, we end up with the so-called \textbf{Salpeter equation} and its adjoint
for the Salpeter amplitudes $\Phi_M$ and $\overline \Phi_M$ 
\begin{eqnarray}
\label{salpeter_V3}
\Phi_M &=&  
-\textrm{i}\;\left\langle{G_0}_M\right\rangle\;V^{(3)}\;\Phi_M,\\[2mm]
\overline\Phi_M &=&
\label{ad_salpeter_V3}
-\textrm{i}\;\overline\Phi_M\;V^{(3)}\;\left\langle{G_0}_M\right\rangle.
\end{eqnarray}
Here we introduced the notation
\begin{equation}
\label{reduc_sixpoint}
\langle A\rangle ({\bf p_\xi},{\bf p_\eta};\;{\bf p_\xi'},{\bf p_\eta'})
:= 
\int\frac{\textrm{d} p_\xi^0}{2\pi}\;
\frac{\textrm{d} p_\eta^0}{2\pi}\;
\int\frac{\textrm{d} p_\xi'^0}{2\pi}\;
\frac{\textrm{d} p_\eta'^0}{2\pi}\;
A (p_\xi,p_\eta;\;p_\xi',p_\eta')
\end{equation}
for the six-dimensional reduction of any eight-dimensional six-point function $A$.
Accordingly, $\langle{G_0}_M\rangle=\langle S_F^1\tens S_F^2\tens S_F^3\rangle $ denotes the reduction of the
free three-quark propagator ${G_0}_M$ defined in eq. (\ref{G0_mom}).
Due to the approximative choice of bare quark propagators with effective
constituent quark masses, the analytical
structure of ${G_0}_M$ in the relative energy variables $p_\xi^0$ and
$p_\eta^0$ is rather simple and consequently, the $p_\xi^0$, $p_\eta^0$ integration 
in $\langle{G_0}_M\rangle$ can be performed
analytically by applying Cauchy's theorem. To this end it is
convenient to use the following partial fraction decomposition of the free
one-particle propagators into positive and negative energy contributions \cite{IZ80},
\begin{equation}
\label{S_PFD_CMS}
S_F^i(p_i)
=
\frac{\rm i}{\not p_i - m_i + {\rm i}\epsilon}
= 
{\rm i}
\left(
\frac{\Lambda_i^+({\bf p_i})}{p_i^0-\omega_i({\bf p_i}) + {\rm i}\epsilon}\;
+
\frac{\Lambda_i^-({\bf p_i})}{p_i^0+\omega_i({\bf p_i}) - {\rm i}\epsilon}\;
\right)
\gamma^0
\end{equation}
which isolates the pole positions in the energy variable $p_i^0$
located at the relativistic on-shell kinetic energies 
\begin{equation}
\label{omega_kin_CMS}
\omega_i({\bf p_i}) := \sqrt{|{\bf p_i}|^2 + m_i^2}
\end{equation}
of the quarks.  The operators $\Lambda_i^{\pm}({\bf p_i})$ are the projectors
onto positive and negative energy solutions of the free Dirac
equation, written explicitly as
\begin{equation}
\label{Lambda_CMS}
\Lambda_i^{\pm}({\bf p_i}) := \frac{\omega_i({\bf p_i})\;\Id \pm H_i({\bf p_i})}{2\omega_i({\bf p_i})}\;,
\end{equation}
where $H_i$ is the usual free single particle Dirac-Hamiltonian given by
\begin{equation}
\label{H_Dirac_CMS}
H_i({\bf p_i}) := \gamma^0\left({\bfgrk\gamma}{\bf\cdot p_i} + m_i\right)
= 
{\bfgrk\alpha}\cdot {\bf p_i} + m_i\;\beta.
\end{equation}
Performing the $p_\xi^0$, $p_\eta^0$ integration we obtain the
three-fermion \textbf{Salpeter propagator}:
\begin{eqnarray}
\label{reduction_G0_CMS}
\lefteqn{\langle {G_0}_M\rangle ({\bf p_\xi},{\bf p_\eta};\;{\bf p_\xi'},{\bf
    p_\eta'})=}&&\nn[4mm]
&&
{\rm i}
\left[ 
\frac
{
\Lambda_1^{+}({\bf p_1})
\tens
\Lambda_2^{+}({\bf p_2})
\tens
\Lambda_3^{+}({\bf p_3})
}
{
M 
- \omega_1({\bf p_1})
- \omega_2({\bf p_2})
- \omega_3({\bf p_3}) 
+ {\rm i}\epsilon
} 
+
\frac
{
\Lambda_1^{-}({\bf p_1})
\tens
\Lambda_2^{-}({\bf p_2})
\tens
\Lambda_3^{-}({\bf p_3})
}
{
M 
+ \omega_1({\bf p_1})
+ \omega_2({\bf p_2})
+ \omega_3({\bf p_3}) 
- {\rm i}\epsilon
} 
\right]\nn[2mm]
&& \hspace*{10mm}\times\;\;
\gamma^0 \tens\gamma^0 \tens\gamma^0\;\;
(2 \pi)^3\;\delta^{(3)}({\bf p_\xi-p_\xi'})\;\; 
(2 \pi)^3\;\delta^{(3)}({\bf p_\eta-p_\eta'})
\end{eqnarray}
with ${\bf p_i} = {\bf p_i}({\bf p_\xi}, {\bf p_\eta})$ defined as in eq.
(\ref{jacobi_mom}) with ${\bf P} = {\bf p_1} + {\bf p_2} + {\bf p_3} = {\bf 0}$.
Notice the remarkable property that due to the pole structure of ${G_0}_M$ in the
relative energy variables $p_\xi^0$ and $p_\eta^0$, the residue theorem merely
provides the projectors onto purely positive-energy and purely negative-energy
three-quark states. All mixed components vanish!

Finally, the Salpeter equation (\ref{salpeter_V3}) in the case of
vanishing two-quark kernels reads explicitly
\begin{eqnarray}
\label{salpeter_V3_explicit}
\lefteqn{\hspace*{-20mm}\Phi_M ({\bf p_\xi},{\bf p_\eta})=
\left[ 
\frac
{
\Lambda_1^{+}({\bf p_1})
\tens
\Lambda_2^{+}({\bf p_2})
\tens
\Lambda_3^{+}({\bf p_3})
}
{
M 
- \omega_1({\bf p_1})
\!-\! \omega_2({\bf p_2})
\!-\! \omega_3({\bf p_3}) 
+ {\rm i}\epsilon
} 
+
\frac
{
\Lambda_1^{-}({\bf p_1})
\tens
\Lambda_2^{-}({\bf p_2})
\tens
\Lambda_3^{-}({\bf p_3})
}
{
M 
+ \omega_1({\bf p_1})
\!+\! \omega_2({\bf p_2})
\!+\! \omega_3({\bf p_3}) 
- {\rm i}\epsilon
}\right]}\nn[3mm]
&&\times\;\;
\gamma^0 \tens\gamma^0 \tens\gamma^0\;
\int 
\frac{\textrm{d}^3 p_\xi'}{(2\pi)^3}\;
\frac{\textrm{d}^3 p_\eta'}{(2\pi)^3}\;
V^{(3)}({\bf p_\xi},{\bf p_\eta};\;{\bf p_\xi'},{\bf
  p_\eta'})\;
\Phi_M({\bf p_\xi'}, {\bf p_\eta'}).
\end{eqnarray}

Thus, we have seen that in the case, where the dynamics of the three
quarks (fermions) is described by an instantaneous, connected
three-body kernel alone, the reduction of the full eight-dimensional
three-fermion Bethe-Salpeter equation to the six-dimensional Salpeter
equation (in the CMS) is straightforward.  The Salpeter equation is
equivalent to the full Bethe-Salpeter equation since eq.
(\ref{reconstruction_V3}) allows an exact reconstruction of the
Bethe-Salpeter amplitude $\chi_M$ from the solution $\Phi_M$ of the
Salpeter equation in the rest frame. Finally, the formally covariant
framework provides the possibility to obtain the amplitude $\chi_{\bar
P}$ in any frame with $\bar P^2 = M^2$ by a kinematical Lorentz boost
of the rest-frame amplitude $\chi_M$.\\

According to eq. (\ref{salpeter_V3_explicit}) we find the remarkable fact
that the reduction in the case of pure instantaneous three-body kernel leads
to certain projection properties for the Salpeter amplitudes which effectively
reduce the number of independent functions necessary to describe a baryon
state.  Let us continue our discussion with some investigations of
this particular structure of the Salpeter equation (\ref{salpeter_V3_explicit}).
\subsubsection{The projector  structure of the Salpeter equation}
\label{subsec:proj_struct_SE_V3}
Due to the energy projectors appearing in the Salpeter propagator
$\langle {G_0}_M \rangle$, the Salpeter amplitudes are eigenstates of the
\textbf{Salpeter projectors}
\begin{eqnarray}
\label{salpeter_proj_CMS}
\Lambda({\bf p_\xi},{\bf p_\eta}) &:=&
\Lambda_1^{+}({\bf p_1})
\tens
\Lambda_2^{+}({\bf p_2})
\tens
\Lambda_3^{+}({\bf p_3})
+
\Lambda_1^{-}({\bf p_1})
\tens
\Lambda_2^{-}({\bf p_2})
\tens
\Lambda_3^{-}({\bf p_3}),\\[2mm]
\label{bar_salpeter_proj_CMS}
\overline \Lambda({\bf p_\xi},{\bf p_\eta}) 
&:=& \gamma^0\tens\gamma^0\tens\gamma^0\;\Lambda({\bf p_\xi},{\bf p_\eta})\;\gamma^0\tens\gamma^0\tens\gamma^0.
\end{eqnarray}
which project onto the subspace of purely positive and negative energy
components, {\it i.e.}
\begin{eqnarray}
\Phi_M &=& 
\Lambda\;\Phi_M
= \Phi_M^{+++} + \Phi_M^{---},\\[2mm]
\overline\Phi_M &=& 
\overline\Phi_M\overline\Lambda
= \overline\Phi_M^{+++} + \overline\Phi_M^{---}.
\end{eqnarray}
and accordingly the Salpeter equation only involves the amplitudes
\begin{equation}
\Phi_M^{+++} := 
\Lambda_1^{+}
\tens
\Lambda_2^{+}
\tens
\Lambda_3^{+}
\;\Phi_M\quad \textrm{and}\quad
 \Phi_M^{---} := 
\Lambda_1^{-}
\tens
\Lambda_2^{-}
\tens
\Lambda_3^{-}
\;\Phi_M,
\end{equation}
whereas all mixed components such as $\Phi_M^{++-}$ vanish. This property
reduces the Salpeter amplitudes effectively to only 16-component functions of
the six variables ${\bf p_\xi},{\bf p_\eta}$, in contrast to the full (in
Dirac space) 64-component Bethe-Salpeter amplitudes, which are functions of
eight variables. This projector structure implies that for the dynamics of the
three quarks in the bound state (baryon) not the full structure of the
instantaneous three-body kernel $V^{(3)}$ is relevant, but only its projected
part
\begin{equation}
V^{(3)}_{\Lambda}({\bf p_\xi},{\bf p_\eta};\;{\bf p_\xi'},{\bf p_\eta'}) := 
\overline \Lambda({\bf p_\xi},{\bf p_\eta})\; V^{(3)} ({\bf p_\xi},{\bf p_\eta};\;{\bf p_\xi'},{\bf p_\eta'})\;
\Lambda({\bf p_\xi'},{\bf p_\eta'}).
\end{equation}
Consequently the residual part $V^{(3)}_{\rm R} := V^{(3)} -
V^{(3)}_{\Lambda}$, which describes the coupling to the mixed energy states,
plays no role for spectroscopy ({\it i.e.} the determination of bound state
masses), although they become relevant for the reconstruction of the full
Bethe-Salpeter amplitude $\chi_M$ according to eq.
(\ref{reconstruction_V3}) and thus can contribute when calculating various
transition matrix elements \cite{Kr01}.
\begin{figure}[!h]
  \begin{center}
    \epsfig{file={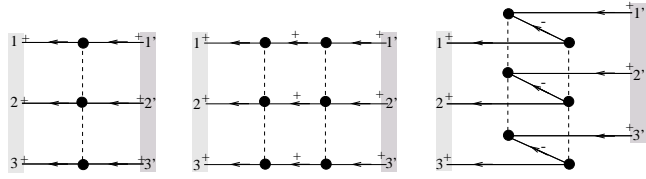},width=85mm}
  \end{center}
\caption{Time ordered graphs of an instantaneous three-body
interaction which contribute to the three-quark propagation in the
Salpeter equation.  The instantaneous three-body kernel is represented
by the dashed line.}
\label{fig:TOPT_V3inst}
\end{figure}

In the language of time-ordered perturbation theory this means that the
instantaneity of the kernel prevents the inclusion of single and double
Z-graphs which correspond to the mixed components $\sim
\Lambda^{++-},\;\Lambda^{--+}, \ldots$, etc. of the interaction kernel.
However, compared to a nonrelativistic ansatz, where all three quarks
propagate forward in time (corresponding here to the components $\sim
\Lambda^{+++}$), the Salpeter equation takes into account also those diagrams,
where all three quarks propagate backwards in time (triple Z-graphs
corresponding to the components $\sim \Lambda^{---}$ and their coupling to
components $\sim \Lambda^{+++}$ via $\overline
\Lambda^{---}\;V^{(3)}\;\Lambda^{+++}$), as shown in fig.
\ref{fig:TOPT_V3inst}.  We want to remark here that the appearance of these
negative energy components in the Salpeter equation is connected with the
particle-antiparticle symmetry due to the ${\cal CPT}$ invariance of the
underlying relativistic field theoretical framework.  We will come back to
this characteristic feature of the Salpeter equation and discuss
it in some more detail in sect. \ref{sec:sym_SE} after we have taken
also the two-particle interactions into account. The importance of the
negative energy contributions depends on the energy denominators $M \mp
(\omega_1 + \omega_2 + \omega_3 - {\rm i}\epsilon)$ of the positive and
negative energy components in (\ref{salpeter_V3_explicit}) as can be
illustrated by
the following two extreme cases:
\begin{itemize}
\item
For small binding energies, {\it i.e.} $M\approx m_1+m_2+m_3$ and $|{\bf p_i}|/m_i\ll1$ one has
\begin{equation}
\frac{1}{M + \omega_1 + \omega_2 + \omega_3} \ll \frac{1}{M - \omega_1 - \omega_2 - \omega_3}
\end{equation}
such that the negative energy component in (\ref{salpeter_V3_explicit})
becomes negligible compared to the positive component and one is led to the
so-called reduced Salpeter equation.
\item
For deeply bound states, {\it i.e.} $M \ll m_1+m_2+m_3$, both components are
of equal order of magnitude:
\begin{equation}
\frac{1}{M + \omega_1 + \omega_2 + \omega_3} \approx \frac{1}{M - \omega_1 - \omega_2 - \omega_3}
\end{equation}
\end{itemize}
In our case of baryons as a bound three-quark system we should definitely be
rather far away from the limit of deeply bounds states. Nevertheless the
negative energy term of the Salpeter amplitude might contribute to a certain
amount.

\subsubsection{Hamiltonian formulation of the Salpeter equation}
\label{subsec:hamilton_form_V3}
The special projector structure in connection with the particular energy
denominators $M \mp (\omega_1 + \omega_2 + \omega_3 - {\rm i}\epsilon)$
allows for the formulation of the Salpeter equation in Hamiltonian form,
{\it i.e.} as an eigenvalue problem 
\begin{equation}
\label{SE_V3_Hamilt}
{\cal H}\;\Phi_M\;\; =\;\; M\; \Phi_M\quad\quad\textrm{with}\quad
\Lambda\Phi_M = \Phi_M.
\end{equation}
Here we define the \textbf{Salpeter Hamiltonian} ${\cal H}$ by 
\begin{eqnarray}
\label{Salp_Hamilt_V3}
\left[{\cal H}\Phi_M\right]({\bf p_\xi}, {\bf p_\eta})
&=& 
{\cal H}_0({\bf p_\xi}, {\bf p_\eta})\;\Phi_M({\bf p_\xi}, {\bf p_\eta})\\[3mm]
&+&
\left[ 
\Lambda_1^{+}({\bf p_1})
\tens
\Lambda_2^{+}({\bf p_2})
\tens
\Lambda_3^{+}({\bf p_3})
+
\Lambda_1^{-}({\bf p_1})
\tens
\Lambda_2^{-}({\bf p_2})
\tens
\Lambda_3^{-}({\bf p_3})
\right]\nn[2mm]
&& \times\;\;
\gamma^0 \tens\gamma^0 \tens\gamma^0\;
\int 
\frac{\textrm{d}^3 p_\xi'}{(2\pi)^3}\;
\frac{\textrm{d}^3 p_\eta'}{(2\pi)^3}\;
V^{(3)}({\bf p_\xi},{\bf p_\eta};\;{\bf p_\xi'},{\bf
  p_\eta'})\;
\Phi_M({\bf p_\xi'}, {\bf p_\eta'})\nonumber
\end{eqnarray}
where the free three-fermion Hamiltonian 
\begin{equation}
\label{free_breit_hamilt_CMS}
{\cal H}_0({\bf p_\xi}, {\bf p_\eta}) 
\;:=\;
H_1({\bf p_1})\tens\Id\tens\Id
\;\;+\;\;
\Id\tens H_2({\bf p_2})\tens\Id
\;\;+\;\;
\Id\tens\Id\tens H_3({\bf p_3})
\end{equation}
represents the relativistic kinetic energy operator.\\

Of course, a similar representation of the adjoint Salpeter equation,
which determines the adjoint amplitude $\overline \Phi_M$, can also be
found.  Note however, that both equations are not independent, but
even are equivalent, since there is a general\footnote{ {\it i.e.} the
interconnection (\ref{interconnectPhi_Phiad_mom}) between $\Phi_M$ and
$\overline \Phi_M$ is independent of the so far considered assumption
of vanishing two-body kernels and other approximations of the
Bethe-Salpeter equation.} interconnection between the Salpeter
amplitude $\Phi_M$ and its adjoint $\overline \Phi_M$, which in
momentum space reads:
\begin{equation}
\label{interconnectPhi_Phiad_mom}
\overline \Phi_M({\bf p_\xi}, {\bf p_\eta})
=
- \Phi_M^\dagger ({\bf p_\xi}, {\bf p_\eta})\;
\gamma^0
\tens
\gamma^0
\tens
\gamma^0.
\end{equation}

To be consistent, one has to require: If $\Phi_M$ is a solution of the Salpeter 
equation (\ref{SE_V3_Hamilt}), then $\overline \Phi_M$, as defined
by relation (\ref{interconnectPhi_Phiad_mom}), has to be a solution of the
corresponding adjoint Salpeter equation (and vice versa). Using
${\cal H}_0^\dagger = {\cal H}_0$ and $\Lambda_i^\dagger = \Lambda_i$,
one easily shows that this equivalence of the Salpeter equation
(\ref{SE_V3_Hamilt}) and its adjoint implies the following condition for the
interaction kernel $V^{(3)}$ in the CMS:
\begin{equation}
\label{hermit_cond_V3}
V^{(3)}
({\bf p_\xi},{\bf p_\eta};\;
{\bf p_\xi'},{\bf p_\eta'})
\stackrel{!}{=}
\gamma^0
\tens
\gamma^0
\tens
\gamma^0\;
\left[{V^{(3)}}
({\bf p_\xi'},{\bf p_\eta'};\;
{\bf p_\xi},{\bf p_\eta})\right]^\dagger\;
\gamma^0
\tens
\gamma^0
\tens
\gamma^0
\end{equation}

\subsubsection{Normalization of Salpeter amplitudes -- Scalar product}
\label{subsec:norm_SA_V3}
The normalization condition (\ref{norm_condition_compact}) of
the Bethe-Salpeter amplitudes, which reads in the center-of-mass
frame with $\bar P = (M, {\bf 0})\equiv M$
\begin{equation}
\label{BSAnorm_V3_CMS}
-{\rm i}\;\overline\chi_{M}\;
\left[\frac{\partial}{\partial P^0}\left({G_0}_P^{-1} +{\rm i}V^{(3)}\right)\right]_{P^0= M}
\chi_{M}
= 2 M,
\end{equation}
induces a normalization condition of the corresponding Salpeter
amplitudes $\Phi_M$.  The instantaneous three-body kernel $V^{(3)}$ has
no explicit energy dependence and thus gives no contribution to the norm.  Using the
representation $\chi_M = {G_0}_M\;\Gamma_M$ and $\overline\chi_M
=\overline\Gamma_M\;{G_0}_M$ of the Bethe-Salpeter amplitudes,
where the vertex functions $\Gamma_M$ and $\overline\Gamma_M$ defined in
(\ref{vertex_funcV3}) do not depend on the relative energies $p_\xi^0$,
$p_\eta^0$, eq. (\ref{BSAnorm_V3_CMS}) becomes
\begin{equation}
\label{norm_V3_redA}
2 M
=
-{\rm i}\;\overline\Gamma_{M}\;{G_0}_M\bigg[
\frac{\partial}{\partial M}\;{G_0}_M^{-1}\bigg]
{G_0}_M\;\Gamma_{M}
=
\overline\Gamma_{M}
\left\langle
-{\rm i}\;
{G_0}_M\bigg[
\frac{\partial}{\partial M}\;{G_0}_M^{-1}\bigg]
{G_0}_M
\right\rangle
\Gamma_{M}.
\end{equation}
Here the angled brackets $\langle\ldots\rangle$ indicate the internal
integration over $p_\xi^0$ and $p_\eta^0$ which is used to reduce the
enclosed operator according to eq. (\ref{reduc_sixpoint}). With
${G_0}_M\;(\partial/\partial
M\;{G_0}_M^{-1})\;{G_0}_M\;=-\partial/\partial M\;{G_0}_M$ this
reduced operator may be rewritten as the derivative of the Salpeter
propagator (\ref{reduction_G0_CMS}) and we obtain
\begin{equation}
\left\langle
-{\rm i}\;
{G_0}_M\bigg[
\frac{\partial}{\partial M}\;{G_0}_M^{-1}\bigg]\;
{G_0}_M\;
\right\rangle 
=
{\rm i}\frac{\partial}{\partial M}\;\langle{G_0}_M\rangle  
=
- \langle{G_0}_M\rangle\;\gamma^0\tens\gamma^0\tens\gamma^0\;\langle{G_0}_M\rangle.
\end{equation}
Substitution into eq. (\ref{norm_V3_redA}) and replacing the vertex
functions according to the relations $\langle{G_0}_M\rangle\;\Gamma_M = \Phi_M$
and $\overline\Gamma_M\;\langle{G_0}_M\rangle = \overline\Phi_M =
-\Phi_M^\dagger \;\gamma^0\tens\gamma^0\tens\gamma^0$ then yields the following
\textbf{normalization condition of the Salpeter amplitudes} $\Phi_M$:
\begin{equation}
\label{SAnorm_V3_CMS}
\Phi_M^\dagger\;\Phi_M 
=
\int 
\frac{\textrm{d}^3 p_\xi}{(2\pi)^3}\;
\frac{\textrm{d}^3 p_\eta}{(2\pi)^3}\;
\sum_{a_1, a_2, a_3}
\Phi_{M\; a_1 a_2 a_3}^* ({\bf p_\xi}, {\bf p_\eta})\;
\Phi_{M\; a_1 a_2 a_3} ({\bf p_\xi}, {\bf p_\eta})
= 2 M.
\end{equation}
Thus, the solutions $\Phi_M$ of the Salpeter equation
have to be normalized according to the usual ${\cal L}^2$-norm just like the solutions of
the ordinary nonrelativistic Schr\"odinger equation.  This norm induces
a \textbf{positive definite scalar product} for arbitrary amplitudes $\Phi_1$ and
$\Phi_2$ that are restricted to positive and negative energy
components, {\it i.e.} $\Phi_i = \Lambda \Phi_i$:
\begin{eqnarray}
\label{SP_V3}
\SP {\Phi_1} {\Phi_2} &:=&
\int 
\frac{\textrm{d}^3 p_\xi}{(2\pi)^3}\;
\frac{\textrm{d}^3 p_\eta}{(2\pi)^3}\;
\sum_{a_1, a_2, a_3}
\Phi_{1\; a_1 a_2 a_3}^* ({\bf p_\xi}, {\bf p_\eta})\;
\Phi_{2\; a_1 a_2 a_3} ({\bf p_\xi}, {\bf p_\eta}).
\end{eqnarray}
Hence, the normalization condition (\ref{SAnorm_V3_CMS}) for solutions
$\Phi_M$ of the Salpeter equation is then given as
\begin{equation}
\SP {\Phi_M} {\Phi_M} = 2 M.
\end{equation} 
We want to remark here that a similar treatment of the 
fermion-antifermion (or the two-fermion) system does not lead to a positive definite scalar
product, owing to a relative sign between the positive and negative energy
contributions, see \cite{ReMu94}.\\

Note that the Salpeter Hamiltonian ${\cal H}$ is hermitean
with respect to the scalar product (\ref{SP_V3}), {\it i.e.}
\begin{equation}
\SP {\Phi_1} {{\cal H}\;\Phi_2} =  \SP {{\cal H}\;\Phi_1} {\Phi_2}
\end{equation}
which is a direct consequence of the condition (\ref{hermit_cond_V3})
on $V^{(3)}$ resulting from the interconnection
(\ref{interconnectPhi_Phiad_mom}) between $\Phi_M$ and $\overline \Phi_M$.
This guarantees, as in the case of the ordinary Schr\"odinger equation,
two important consequences, namely:
\begin{itemize}
\item   
  The eigenvalues $M$ of ${\cal H}$, {\it i.e.} the three-fermion bound-state masses are
  real, {\it i.e.} $M^* = M$.
\item Salpeter amplitudes $\Phi_{M_1}$ and $\Phi_{M_2}$ belonging to
  different eigenvalues $M_1 \not = M_2$ are mutually orthogonal, {\it i.e.}
  $\SP {\Phi_{M_1}} {\Phi_{M_2}} = 0$.
\end{itemize}

\subsection{The reduction with genuine two-particle kernels}
\label{sec:reduc_V3V2}
Now let us come back to the general case we are in fact interested in, where
in addition to the instantaneous three-body kernel $V^{(3)}$, the dynamics of
the quarks is also determined by the unconnected instantaneously approximated
two body-terms given by (\ref{inst_approx_CMS_V2}) and
(\ref{bar_K2_inst_approx}). Referring again to the formal covariance of the
instantaneous approximation as before, we choose for these considerations the
three-body rest-frame with $\bar P = (M,{\bf 0})\equiv M$. The
Bethe-Salpeter equation and its adjoint then read
\begin{eqnarray}
\label{BSE_V3_V2}
\chi_{M} 
&=& 
-\textrm{i}\;{G_0}_{M}\; V^{(3)} \;\chi_{M} 
-\textrm{i}\;{G_0}_{M}\; \overline K^{(2)}_{M}\;\chi_{M},\\[3mm]
\label{ad_BSE_V3_V2}
\overline\chi_{M} &=&  
-\textrm{i}\;\overline\chi_{M}\; V^{(3)}\;{G_0}_{M}
-\textrm{i}\;\overline\chi_{M}\; \overline K^{(2)}_{M}\;{G_0}_{M},
\end{eqnarray}
and now the difficulty stems from the circumstance that due to the second
term on the right hand side of (\ref{BSE_V3_V2}), which contains
$K^{(2)}_{M}$, the relative energy dependence in the Bethe-Salpeter 
equation can no longer be separated and thus, from the outset, the reduction
cannot be performed. Nevertheless, we still can take advantage of the fact
that the $p_\xi^0$, $p_\eta^0$ dependence at least is cut by the first
term, due to the instantaneity of $V^{(3)}$. Recasting the Bethe-Salpeter 
equation into a more convenient form, this feature will in fact provide a
possibility to perform a reduction, as we will see in the following
discussion. But let us emphasize that the way of how to perform the
reduction and consequently the final form of the Salpeter equation is
not unique, although the various resulting reduced equations are formally
equivalent.  However, in practice, even the reduced equations are not solvable
in general so that further approximations are indispensable and thus the
different reduced equations become practically non-equivalent. Therefore, the
reduced equation in its full exact form should, right from start, have an
expedient canonical structure allowing further approximations to be made in a
systematical way. Referring to this we will
orientate our considerations according to the instructive canonical form of the Salpeter
equation as given in the previously discussed case of vanishing
two-body kernels by eqs. (\ref{SE_V3_Hamilt}) and (\ref{Salp_Hamilt_V3}).
Before we present this specific method for the reduction of the Bethe-Salpeter
equation (\ref{BSE_V3_V2}) and its adjoint (\ref{ad_BSE_V3_V2}) in practice,
let us generally discuss in a first attempt,
\begin{itemize}
\item
how  in principle it becomes possible to reduce the eight-dimensional
three-fermion Bethe-Salpeter equation to an equivalent
six-dimensional Salpeter equation, provided that the full
interaction kernel $K_{P}$ contains at least one connected
instantaneous part, as given in our case by the instantaneous
three-body kernel,
\item
what changes at all in the structure and the properties of the Salpeter
equation and thus of the Salpeter amplitudes $\Phi_M$ in
comparison to the case discussed previously, where the dynamics was given
by an instantaneous three-body part alone.
\end{itemize}

\subsubsection{A first attempt -- concepts, ideas and problems}
\label{subsec:reduction_V2_first}
In a first attempt we now want to sketch a procedure showing that a reduction
of the Bethe-Salpeter equation (\ref{BSE_V3_V2}) can indeed be achieved,
utilizing that $V^{(3)}$ cuts the relative energy dependence
in one term of the Bethe-Salpeter equation. The crucial idea and concept of
this procedure is to get rid of the problematical second term
$-\textrm{i}\;{G_0}_{M}\; \overline K^{(2)}_{M}\;\chi_{M}$ appearing on the
right hand side of eq. (\ref{BSE_V3_V2}), where the unconnected two-body
term $\overline K^{(2)}_{M}$ acts on the Bethe-Salpeter amplitude
$\chi_{M}$ directly. This can be reached by recasting the Bethe-Salpeter
equation (\ref{BSE_V3_V2}) in the following manner. First we separate the
terms of the Bethe-Salpeter equation into $p_\xi^0$, $p_\eta^0$ dependent and
independent parts, as follows:
\begin{equation}
\label{BSE_V3_V2_a}
\left[{G^{-1}_0}_M\; + \textrm{i}\; \overline K^{(2)}_M\right]\;\chi_M =
-\textrm{i}\;V^{(3)}\;\chi_M
\end{equation}
Remember that $-\textrm{i}\;V^{(3)}\;\chi_M$ on the right hand side indeed has
no relative energy dependence due to eq. (\ref{vertex_funcV3}). Now
let us introduce the resolvent $\overline {\cal G}_M^{(2)}$ of the operator
$[{G^{-1}_0}_M\; + \textrm{i}\; \overline K^{(2)}_M]$ appearing on the left
hand side of the eqs. (\ref{BSE_V3_V2_a}), {\it i.e.}
\begin{equation}
\overline{\cal G}_M^{(2)}\;
\left[{G^{-1}_0}_M\; + \textrm{i}\; \overline K^{(2)}_M\right] 
= 
\left[{G^{-1}_0}_M\; + \textrm{i}\; \overline K^{(2)}_M\right]
\;\overline{\cal G}_M^{(2)}
= \Id.
\end{equation}
This Green's function $\overline{\cal G}_M^{(2)}$ is the solution of the
inhomogeneous eight-dimensional integral equation
\begin{eqnarray}
\label{BSE_mathcalG}
\overline{\cal G}_M^{(2)} 
&=&
{G_0}_M -\textrm{i}\;{G_0}_M\; \overline K^{(2)}_M\;\overline{\cal G}_M^{(2)}
\end{eqnarray}
and thus describes, apart from the free propagation ${G_0}_M$, also the propagation of
the three quarks via the unconnected two-particle interactions alone.
Multiplying eq. (\ref{BSE_V3_V2_a}) by this resolvent
$\overline{\cal G}_M^{(2)}$ we obtain the Bethe-Salpeter equation
in a form similar to (\ref{BSE_V3}), {\it i.e.} the case where we neglected the
two-particle forces, but with ${G_0}_M$ now replaced by $\overline{\cal G}_M^{(2)}$
which additionally collects all remaining retardation effects concerning the unconnected
two-quark interactions within the baryon:
\begin{equation}
\label{BSE_V3_V2_b}
\chi_M = -{\rm i}\;\overline{\cal G}_M^{(2)}\;V^{(3)}\;\chi_M.
\end{equation}
This form enables us again to exploit the crucial property of the instantaneous kernel
$V^{(3)}$ to separate the dependence on the relative energy variables ${p_\xi}^0$ and
${p_\eta}^0$. Consequently we can proceed in the same way to reduce the
Bethe-Salpeter equations (\ref{BSE_V3_V2_b}) as we did when reducing
eq. (\ref{BSE_V3}) in the case of vanishing two-quark kernels.
According to eq. (\ref{Vchi_eq_Vphi}) the eight-dimensional integral 
operation $V^{(3)}$ on $\chi_M$  on the right
hand side of eq. (\ref{BSE_V3_V2_b}) can be reduced to a
six-dimensional operation on the Salpeter amplitude
$\Phi_M$, 
\begin{equation}
\left[V^{(3)}\;\chi_M\right]({p_\xi},{p_\eta})
=
\left[V^{(3)}\;\Phi_M\right]({\bf p_\xi},{\bf p_\eta}).
\end{equation}
This (in principle) provides us again the possibility to reconstruct the full
eight-dimensional Bethe-Salpeter amplitude $\chi_M$ from the Salpeter
amplitude $\Phi_M$ according to
\begin{equation}
\label{BSE_V3_V2_c}
\chi_M = -{\rm i}\;\overline{\cal G}_M^{(2)}\;V^{(3)}\;\Phi_M
\end{equation}
and shows that the analytical $p_\xi^0$, $p_\eta^0$ dependence of the
Bethe-Salpeter amplitude is completely determined by
the analytical structure of $\overline{\cal G}_M^{(2)}$ in these variables.
Thus, performing the $p_\xi^0, p_\eta^0$ integration on both sides of
eq. (\ref{BSE_V3_V2_c}), the Bethe-Salpeter amplitude
$\chi_M$ on the left reduces to the Salpeter amplitude $\Phi_M$
and on the right hand side only $\overline{\cal G}_M^{(2)}$ is
affected by this integration and reduces to
$\langle\overline{\cal G}_M^{(2)}\rangle$.  Consequently, we finally
end up with the reduced equation which determines the Salpeter
amplitude $\Phi_M$, \textit{i.e.}
\begin{equation}
\label{SE_V3_V2}
\Phi_M \;=\; -{\rm i}\;\left\langle\overline{\cal G}_M^{(2)}\right\rangle\;V^{(3)}\;\Phi_M.
\end{equation}
All the difficulties, arising from retardation effects due to
the unconnected two-body terms, are now transferred to the
reduction $\langle \overline{\cal G}_M^{(2)}\rangle$ of
$\overline{\cal G}_M^{(2)}$.  Corresponding to the inhomogeneous
integral equation (\ref{BSE_mathcalG}) this reduction of the Green's function is determined by
\begin{eqnarray}
\label{red_G_M_2}
\left\langle\overline{\cal G}_M^{(2)}\right\rangle 
&=& 
\left\langle {G_0}_M\right\rangle -\textrm{i}\;\left\langle {G_0}_M\; \overline
K^{(2)}_M\;\overline{\cal G}_M^{(2)}\right\rangle\nn[4mm]
&=& 
\left\langle {G_0}_M\right\rangle
\;+\;
\left\langle {G_0}_M
\left[-\textrm{i}\;\overline K^{(2)}_M\right]{G_0}_M\right\rangle
\;+\;
\left\langle {G_0}_M 
\left[-\textrm{i}\;\overline K^{(2)}_M\right]{G_0}_M
\left[-\textrm{i}\;\overline K^{(2)}_M\right]{G_0}_M\right\rangle\nn
& &
\phantom{\langle {G_0}_M\rangle}
\;+\;\ldots\;.
\end{eqnarray}

\begin{figure}[!h]
  \begin{center}
    \epsfig{file={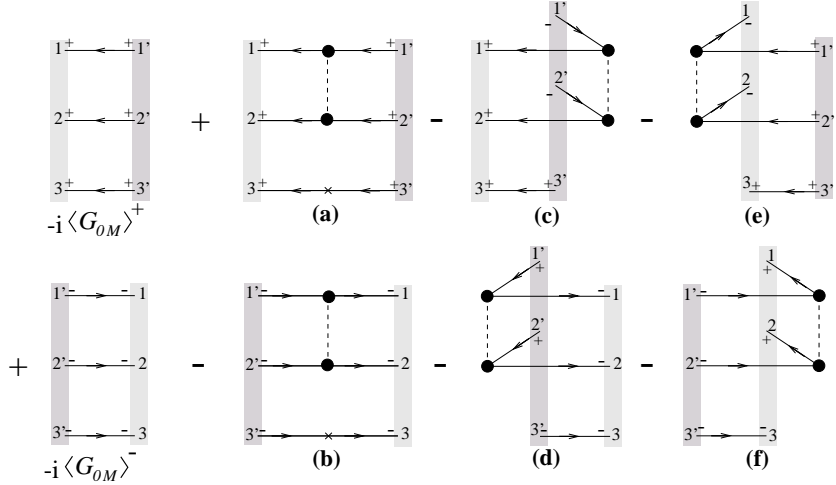},width=110mm}
    \end{center}
\caption{Time ordered graphs of the reduced Green's function
$\langle\overline{\cal G}_M^{(2)}\rangle$ up to the Born term
(only the interactions in the quark pair (12) are  shown.)
The instantaneous two-body interaction between two quarks is
represented by the vertical dashed lines and dots on the affected
quark lines.  The Born graphs correspond to the expressions of
eq. (\ref{reduction_Born}).}
\label{fig:TOPTbarG2bornInst}
\end{figure}

Thus, we have shown that, even with unconnected two-particle kernels, it is in
principle possible to reduce the eight-dimensional three-fermion
Bethe-Salpeter equation to a six-dimensional Salpeter equation,
provided we choose at least one part of the full interaction kernel to be
instantaneous.  Due to the interconnection (\ref{BSE_V3_V2_c}) of the full
eight-dimensional Bethe-Salpeter amplitude $\chi_M$ and its
six-dimensional reduction, {\it i.e.} the Salpeter amplitude $\Phi_M$,
the Salpeter equation is equivalent to the full Bethe-Salpeter 
equation since eq. (\ref{BSE_V3_V2_c}) provides in principle an
equally exact reconstruction of the Bethe-Salpeter amplitude.
Unfortunately, the analytical structure of the Green's function
$\overline{\cal G}_M^{(2)}$ in the complex planes of the relative energy
variables $p_\xi^0$ and $p_\eta^0$ is rather complicated, so that in practice
neither its reduction $\langle\overline{\cal G}_M^{(2)}\rangle$ required
for solving the Salpeter equation (\ref{SE_V3_V2}), nor
$\overline{\cal G}_M^{(2)}$ itself, required for the reconstruction
(\ref{BSE_V3_V2_c}) of the Bethe-Salpeter amplitude, is manageable
in its full, exact form. The determination of
$\langle\overline{\cal G}_M^{(2)}\rangle$ for example requires in principle
the calculation of an infinite number of reduced diagrams due to the Neumann series of $\overline{\cal G}_M^{(2)}$, see eq.
(\ref{red_G_M_2}). We do not want to bother about that complexity at the
moment and first consider $\langle\overline{\cal G}_M^{(2)}\rangle$ in
eq. (\ref{SE_V3_V2}) only up to the Born term,
\begin{equation}
\Phi_M\;=\;  
\left[-\textrm{i}\;\left\langle {G_0}_M\right\rangle
\;-\;
\left\langle {G_0}_M\;\overline K^{(2)}_M\;{G_0}_M\right\rangle
\;+\;\ldots\right]\;
V^{(3)}\;\Phi_M
\end{equation}
in order to inspect what changes basically in the structure of the Salpeter
equation and the corresponding Salpeter amplitudes. A tedious but
straightforward calculation, using the residue theorem for performing the
relative energy integration, yields the following contributions to the reduced
Born term:
\begin{eqnarray}
\lefteqn{
-\left\langle {G_0}_M\;\overline K^{(2)}_M\;{G_0}_M\right\rangle
({\bf p_\xi}, {\bf p_\eta};\;{\bf p_\xi'}, {\bf p_\eta'})\;\; = \;\;}&&\nn[1mm]
\Bigg\{
&+&
\frac
{\Lambda^+_1\tens\Lambda^+_2\tens\Lambda^+_3\;
\gamma^0\tens\gamma^0\tens\gamma^0}
{M-\omega_1-\omega_2-\omega_3+{\rm i}\;\epsilon}\;
\left[V^{(2)}({\bf p_\xi}, {\bf p_\xi'})\tens\gamma^0\right]\;
\frac
{{\Lambda^+_1}'\tens{\Lambda^+_2}'\tens{\Lambda^+_3}'\;
\gamma^0\tens\gamma^0\tens\gamma^0}
{M-\omega_1'-\omega_2'-\omega_3'+{\rm i}\;\epsilon}\;{\bf(a)}\nn[1mm]
&-&
\frac
{\Lambda^-_1\tens\Lambda^-_2\tens\Lambda^-_3\;
\gamma^0\tens\gamma^0\tens\gamma^0}
{M+\omega_1+\omega_2+\omega_3-{\rm i}\;\epsilon}\;
\left[V^{(2)}({\bf p_\xi}, {\bf p_\xi'})\tens\gamma^0\right]\;
\frac
{{\Lambda^-_1}'\tens{\Lambda^-_2}'\tens{\Lambda^-_3}'\;
\gamma^0\tens\gamma^0\tens\gamma^0}
{M+\omega_1'+\omega_2'+\omega_3'-{\rm i}\;\epsilon}\;{\bf(b)}\nn[1mm]
&-&
\frac
{\Lambda^+_1\tens\Lambda^+_2\tens\Lambda^+_3\;
\gamma^0\tens\gamma^0\tens\gamma^0}
{M-\omega_1-\omega_2-\omega_3+{\rm i}\;\epsilon}\;
\left[V^{(2)}({\bf p_\xi}, {\bf p_\xi'})\tens\gamma^0\right]\;
\frac
{{\Lambda^-_1}'\tens{\Lambda^-_2}'\tens{\Lambda^+_3}'\;
\gamma^0\tens\gamma^0\tens\gamma^0}
{\omega_1+\omega_2+\omega_1'+\omega_2'}\;{\bf(c)}\nn[1mm]
&-&
\frac
{\Lambda^-_1\tens\Lambda^-_2\tens\Lambda^-_3\;
\gamma^0\tens\gamma^0\tens\gamma^0}
{M+\omega_1+\omega_2+\omega_3-{\rm i}\;\epsilon}\;
\left[V^{(2)}({\bf p_\xi}, {\bf p_\xi'})\tens\gamma^0\right]\;
\frac
{{\Lambda^+_1}'\tens{\Lambda^+_2}'\tens{\Lambda^-_3}'\;
\gamma^0\tens\gamma^0\tens\gamma^0}
{\omega_1+\omega_2+\omega_1'+\omega_2'}\;{\bf(d)}\nn[1mm]
&-&
\frac
{\Lambda^-_1\tens\Lambda^-_2\tens\Lambda^+_3\;
\gamma^0\tens\gamma^0\tens\gamma^0}
{\omega_1+\omega_2+\omega_1'+\omega_2'}\;
\left[V^{(2)}({\bf p_\xi}, {\bf p_\xi'})\tens\gamma^0\right]\;
\frac
{{\Lambda^+_1}'\tens{\Lambda^+_2}'\tens{\Lambda^+_3}'\;
\gamma^0\tens\gamma^0\tens\gamma^0}
{M-\omega_1'-\omega_2'-\omega_3'+{\rm i}\;\epsilon}\;{\bf(e)}\nn[1mm]
&-&
\frac
{\Lambda^+_1\tens\Lambda^+_2\tens\Lambda^-_3\;
\gamma^0\tens\gamma^0\tens\gamma^0}
{\omega_1+\omega_2+\omega_1'+\omega_2'}\;
\left[V^{(2)}({\bf p_\xi}, {\bf p_\xi'})\tens\gamma^0\right]\;
\frac
{{\Lambda^-_1}'\tens{\Lambda^-_2}'\tens{\Lambda^-_3}'\;
\gamma^0\tens\gamma^0\tens\gamma^0}
{M+\omega_1'+\omega_2'+\omega_3'-{\rm i}\;\epsilon}\;{\bf(f)}\nn[1mm]
\label{reduction_Born}
&&\Bigg\}\;(2 \pi)^3\;\delta^{(3)}({\bf p_\eta} - {\bf p_\eta'})
\;+\;
\begin{array}{l}
\textrm{cyclic. perm. of $(12)\;3$ corresponding to the}\\
\textrm{interacting quark pairs (23) and (31)} 
\end{array}
\end{eqnarray}
For the sake of clarity we suppressed partially the explicit coordinate
dependences by using the more compact notation $\Lambda^\pm_i \equiv
\Lambda^\pm_i ({\bf p_i})$, $\omega_i \equiv \omega_i ({\bf p_i})$ and
${\Lambda^\pm_i}' \equiv \Lambda^\pm_i ({\bf p_i'})$, $\omega_i'
\equiv \omega_i ({\bf p_i'})$. The time-ordered Feynman graphs
corresponding to the six different terms in eq. (\ref{reduction_Born})
are shown in fig. \ref{fig:TOPTbarG2bornInst}. 

The first two terms (a) and (b) have the same projector structure and
corresponding energy denominators as the Salpeter propagator $\langle
{G_0}_M\rangle$ and thus are of a similar form as the reduction of a genuine
instantaneous three-body interaction diagram.  The decisive difference to the
previously discussed case is due to the occurrence of the mixed energy
components \mbox{$(++-)$}, \mbox{$(--+)$}, etc. in the remaining four terms
(c) - (f), which result from retardation effects of the unconnected
two-particle interactions.  In other words: The propagator
$\langle\overline{\cal G}_M^{(2)}\rangle$ which has been substituted for
$\langle {G_0}_M\rangle$ (compared to the case of neglected two-body kernels)
does not exhibit the particular projection properties of $\langle
{G_0}_M\rangle$, {\it i.e.} the restriction to purely positive and purely
negative energy components only.  This implies (compare the discussion in
subsect. \ref{subsec:proj_struct_SE_V3}):
\begin{itemize}
\item
The Salpeter amplitudes $\Phi_M$ are no longer eigenstates of
the Salpeter projector $\Lambda$, but also possess mixed energy components
according to the terms (e) and (f) in (\ref{reduction_Born}).
\item
In connection with the unconnected, retarded two-particle kernels,
also the 'residual' part $V_{\rm R}^{(3)} = V^{(3)}- V_\Lambda^{(3)}$
of the instantaneous three-body kernel $V^{(3)}$, that couples to the
mixed components, now contributes to the three-fermion bound state and
therefore to the spectroscopic results.
\end{itemize}
Note however, that (assuming weakly bound states) the important dominant terms of this Born
contribution are given by the purely positive and negative contributions (a)
and (b) due to their particular structure of the energy denominators:
\begin{itemize}
\item In the case ${\bf M>0}$ the terms (c) and (e) are suppressed with respect
  to the dominant term (a), since $(M-\omega_1-\omega_2-\omega_3)^{-1} \gg
  (\omega_1+\omega_2+\omega_1'+\omega_2')^{-1}$.  All other contributions have
  denominators $(M+\omega_1+\omega_2+\omega_3)$ and thus are anyhow
  suppressed.
\item In the case\footnote{We should note here already that the
Salpeter equation generally possesses both positive and negative mass
solution due to the ${\cal CPT}$-symmetry (see subsect. \ref{subsubsec:CPT}). 
In this respect both cases $M>0$ and $M<0$ have to be considered.} ${\bf M<0}$ the dominant term
is (b), whereas (d) and (f) are suppressed against (b), since
$(M+\omega_1+\omega_2+\omega_3)^{-1} \gg
(\omega_1+\omega_2+\omega_1'+\omega_2')^{-1}$ All other contributions
have denominators $(M-\omega_1-\omega_2-\omega_3)$ and hence are
anyway negligible.
\end{itemize}
The more complex structure of the Salpeter equation (\ref{SE_V3_V2})
restricts its applicability to explicit three-body bound-state
calculations: Due to the explicit appearance of the additional mixed
energy components the formulation as an eigenvalue problem in
Hamiltonian form such as in the case of a pure instantaneous
three-body kernel is no longer possible.  Moreover further
approximations of the reduced Green's function $\langle\overline{\cal
G}_M^{(2)}\rangle$ are indispensable, which gives rise to the question
of how to approximate $\langle\overline{\cal G}_M^{(2)}\rangle$
systematically.  One would expect that a perturbative
approximation of $\langle\overline{\cal G}_M^{(2)}\rangle$ simply by
cutting the Neumann series of $\overline{\cal G}_M^{(2)}$ at finite
order ({\it e.g.} in the so far considered Born approximation), would
not be sufficient to describe accurately the effects of the
two-particle interaction within a three-body bound state, as {\it
e.g.} two particle correlations. In order to take non-perturbatively
at least an infinite subset of diagrams of $\langle\overline{\cal
G}_M^{(2)}\rangle$ into account, one could follow an idea of Phillips
and Wallace \cite{PhWa96,PhWa98,LaAf97}, who investigated the
three-dimensional reduction of the {\it two-fermion} Bethe-Salpeter
equation with general four-dimensional ({\it i.e.} non-instantaneous)
interaction kernels. Their method provides a generalization of a
former formalism of Klein \cite{Kl53,Kl54} using the quasi-potential
approach of Logunov and Tavkhelidze \cite{LoTa63} and has a close
connection to standard time-ordered perturbation theory. Applying
their idea to our three-body case, their method essentially consists
in a systematical prescription to determine order-by-order (in the
coupling of $\overline K^{(2)}_M$) an instantaneous three-particle
irreducible kernel $\overline K^{(2)}_{M\;{\rm inst}}$, where
irreducibility is now defined with respect to the Salpeter propagator
$\langle{G_0}_M\rangle$, such that
\begin{equation}
\label{Kinst_first_attempt}
\langle\overline{\cal G}_M^{(2)}\rangle
=\langle{G_0}_M\rangle
-{\rm i}\;\langle{G_0}_M\rangle\;
\overline K^{(2)}_{M\;{\rm inst}}
 \langle\overline{\cal G}_M^{(2)}\rangle.
\end{equation}
This would allow for a formulation of the Salpeter equation
(\ref{SE_V3_V2}) in a form that is the same as in the previously
discussed case of vanishing two body interactions, {\it i.e.}
\begin{equation}
\label{Kinst_first_attempt_SE}
\Phi_M = -{\rm i}\;\langle{G_0}_M\rangle\left[ 
\overline K^{(2)}_{M\;{\rm inst}}
+
V^{(3)}\right]\;\Phi_M.
\end{equation}
However, the method of \cite{PhWa96,PhWa98} has an inconsistency pointed out
by the authors themselves:
Obviously, eqs. (\ref{Kinst_first_attempt}) and
(\ref{Kinst_first_attempt_SE}) are in clear contradiction to the occurrence 
of mixed energy components discussed above: due to the
projector property of $\langle{G_0}_M\rangle$, eq.
(\ref{Kinst_first_attempt}) would restrict
$\langle\overline{\cal G}_M^{(2)}\rangle$ and consequently $\Phi_M$
to purely positive and negative components only. In the next subsection we will
therefore improve our reduction procedure such that this method nevertheless
becomes applicable without revealing such inconsistencies.

\subsubsection{Reduction to a Salpeter equation in Hamiltonian form}
In this subsection we present a systematical reduction procedure, which avoids
the difficulties and inconsistencies of the foregoing first attempt and allows
for a formulation of the Salpeter equation with a structure quite
similar to that of sect. \ref{sec_reduc_V3}, where the dynamics was given by
a connected instantaneous three-body interaction alone. Furthermore, this
procedure will provide a systematic approximation of the exact reduced
equation that is still manageable in practice and appropriate for explicit
calculations.  Our aim is to get a reduction of the Bethe-Salpeter
equation which even in the presence of unconnected two-quark kernels exhibits
the same form and properties as the Salpeter equation
(\ref{salpeter_V3}) in the case of vanishing two-body terms. Consequently
it then can likewise be formulated as an eigenvalue problem (or at least as a
generalized eigenvalue problem) in Hamiltonian form as discussed in
subsect. \ref{subsec:hamilton_form_V3}. In other words:
\begin{itemize}  
\item The free three-quark propagation shall be given by the
  Salpeter propagator $\langle{G_0}_{M}\rangle$. Accordingly, we 
  search for an instantaneous three-particle irreducible kernel $V^{\rm
    eff}_{M}$ (a quasi potential) with irreducibility defined with respect to
  the propagator $\langle{G_0}_{M}\rangle$, which covers all the complexity
  arising from the unconnected two-particle interactions and adds to the
  genuine instantaneous three-quark kernel $V^{(3)}$.
\item Due to the projector property of the Salpeter propagator
\begin{equation}
\label{project_property_G0_cov}
\Lambda\langle{G_0}_{M}\rangle  =
\langle{G_0}_{M}\rangle \overline\Lambda =
\langle{G_0}_{M}\rangle
\end{equation}
where $\Lambda$ and $\overline\Lambda$ are the Salpeter projectors
defined in eq. (\ref{salpeter_proj_CMS}) and
(\ref{bar_salpeter_proj_CMS}), the reduced equation then merely involves the
purely positive and purely negative energy components.  Consequently the
reduced amplitudes emerging from the Salpeter equation and its adjoint
must be eigenstates of the Salpeter projector $\Lambda$ and
$\overline\Lambda$, respectively.
\item
However, as demonstrated in the previous discussion of subsect.
\ref{subsec:reduction_V2_first}, the Salpeter amplitude itself is no
longer an eigenstate of the Salpeter projector when two-particle
interactions are taken into account. We found that in this case also the mixed energy components
occur. Consequently, the reduced amplitude resulting from our desired reduced
equation can not be the full Salpeter amplitude $\Phi_{M}$ but only its
projected part $\Phi_{M}^\Lambda := \Lambda\Phi_{M}$. To summarize, we are looking for
a reduction of the Bethe-Salpeter equation of the form
\begin{equation}
\label{searched_reduced_SE}
\Phi_{M}^\Lambda =  
-\textrm{i}\;\langle{G_0}_{M}\rangle\;\left[V^{(3)} + V^{\rm eff}_{M}\right]\;\Phi^\Lambda_{M}
\quad \textrm{with}\quad\Phi_{M}^\Lambda :=
\Lambda\Phi_{M}.
\end{equation}
Equivalence to the Salpeter equation (\ref{SE_V3_V2}) then requires that
all interactions via the mixed components must be effectively taken into
account in the quasi-potential $V^{\rm eff}_{M}$ and moreover there must be an
interconnection which allows to regain the full Salpeter amplitude
$\Phi_{M}$ and finally the full Bethe-Salpeter amplitude
$\chi_{M}$ from the projected amplitude
$\Phi_{M}^\Lambda$.
\end{itemize}
Now let us become specific and show how such a kind of reduction can
indeed be achieved.  To this end we split the instantaneous three-body
kernel $V^{(3)}$ according to
\begin{equation}
V^{(3)} = V^{(3)}_{\Lambda} + V^{(3)}_{\rm R}, 
\end{equation}
with $V^{(3)}_{\Lambda}$ that part of $V^{(3)}$ which couples exclusively to purely positive and purely
negative energy states, {\it i.e.}
\begin{eqnarray}
V^{(3)}_{\Lambda}({\bf p_\xi},{\bf p_\eta};\;{\bf p_\xi'},{\bf p_\eta'})
&:=& \overline\Lambda ({\bf p_\xi},{\bf p_\eta})\; 
V^{(3)}({\bf p_\xi},{\bf p_\eta};\;{\bf p_\xi'},{\bf p_\eta'})\;
\Lambda (\;{\bf p_\xi'},{\bf p_\eta'})
\end{eqnarray}
and the  residual part $V^{(3)}_{\rm R}:= V^{(3)} - V^{(3)}_{\Lambda}$, which
couples also to the mixed energy components
and  has the property
\begin{equation}
\label{prop_V3R}
\overline\Lambda\;V^{(3)}_{\rm R}\;\Lambda \equiv 0.
\end{equation}
Then we have for the Bethe-Salpeter equation and its adjoint:
\begin{eqnarray}
\chi_{M} &=&  
-\textrm{i}\;{G_0}_{M}\; 
\left[V^{(3)}_{\Lambda}  + V^{(3)}_{\rm R} + \overline
  K^{(2)}_{M}\right]\;\chi_{M},\\[2mm]
\overline\chi_{M} &=&  
-\textrm{i}\;\overline\chi_{M}\;  \left[V^{(3)}_{\Lambda} + V^{(3)}_{\rm R} + \overline K^{(2)}_{M}\right]\;{G_0}_{M}.
\end{eqnarray}
Recall that in the case of vanishing two-particle kernels only the first
part $V^{(3)}_{\Lambda}$ contributes to the Salpeter equation, while the
residual part $V^{(3)}_{\rm R}$ disappears according to property
(\ref{prop_V3R}), as discussed in subsect.
\ref{subsec:proj_struct_SE_V3}.  But now, in connection with retardation
effects of the unconnected two-particle terms, also the residual part
$V^{(3)}_{\rm R}$ gives contributions to the reduction of the Bethe-Salpeter
equation as has been shown in the previous subsect.
\ref{subsec:reduction_V2_first}.  Keeping this in mind we now want to proceed
in a way similar to our first attempt in subsect.
\ref{subsec:reduction_V2_first}, where we transfered the effects of the
retarded two-particle terms $\overline K^{(2)}_{M}$ into the Green's function
$\overline{\cal G}_M^{(2)}$.  However, our discussion indicates that it is
even more convenient to absorb together with $\overline K^{(2)}_{M}$ also the
instantaneous kernel $V^{(3)}_{\rm R}$ into a Green's function
${\cal G}_{M}$, since the contributions of $V^{(3)}_{\rm R}$ occur
exclusively in connection with $\overline K^{(2)}_{M}$.  In this way we
achieve that really \textit{all} complications resulting from the unconnected
two body-terms are gathered in the resolvent ${\cal G}_{M}$, which now is
defined by
\begin{equation}
\label{resolvent_cal_GP}
{\cal G}_{M}\;
\left[{G^{-1}_0}_{M}\; + \textrm{i}\;V^{(3)}_{\rm R} + \textrm{i}\; \overline K^{(2)}_{M}\right] 
= 
\left[{G^{-1}_0}_{M}\; + \textrm{i}\;V^{(3)}_{\rm R} + \textrm{i}\; \overline K^{(2)}_{M}\right]
\;{\cal G}_{M}
= \Id
\end{equation}
and thus is the solution of the inhomogeneous integral equations
\begin{equation}
\label{inhom_IE_GP}
{\cal G}_{M}
=
{G_0}_{M} -\textrm{i}\;{G_0}_{M}\; \left[V^{(3)}_{\rm R} + \overline K^{(2)}_{M}\right]\;{\cal G}_{M}
=
{G_0}_{M} -\textrm{i}\;{\cal G}_{M}\; \left[V^{(3)}_{\rm R} +\overline K^{(2)}_{M}\right]\;{G_0}_{M}.
\end{equation}
With ${\cal G}_{M}$ the Bethe-Salpeter equation and its adjoint
can be rewritten as before in the equivalent form
\begin{eqnarray}
\chi_{M} &=&  
-\textrm{i}\;{\cal G}_{M}\; V^{(3)}_{\Lambda}  \;\chi_{M},\\[2mm]
\overline\chi_{M} &=&  
-\textrm{i}\;\overline\chi_{M}\; V^{(3)}_{\Lambda}\;{\cal G}_{M},
\end{eqnarray}
which is suited for the six-dimensional reduction, because the new
three-body kernel $V^{(3)}_{\Lambda}$ is instantaneous. The reduction
is performed as before. Similar to eq.  (\ref{Vchi_eq_Vphi}) we obtain
first
\begin{eqnarray}
\left[V^{(3)}_\Lambda\;\chi_{M}\right]( p_\xi, p_\eta)
&=&\left[V^{(3)}_\Lambda\;\Phi_{M}\right]({\bf p_\xi},{\bf p_\eta})\\[2mm]
\left[\overline\chi_{M}\;V^{(3)}_\Lambda\right]( p_\xi, p_\eta)
&=&
\left[\overline\Phi_{M}\;V^{(3)}_\Lambda\right]({\bf p_\xi},{\bf p_\eta}),
\end{eqnarray}
where the Salpeter amplitudes $\Phi_{M}$ and $\overline\Phi_{M}$
are the reductions of the corresponding Bethe-Salpeter
amplitudes as defined in eqs. (\ref{def:salpeter_ampl_CMS}) and
(\ref{def:salpeter_ampl_ad_CMS}).  Inserting this back into the Bethe-Salpeter 
equations we get the prescription how to
reconstruct the full eight-dimensional Bethe-Salpeter
amplitudes from the Salpeter amplitudes:
\begin{eqnarray}
\chi_{M} 
&=&  
-\textrm{i}\;{\cal G}_{M}\; V^{(3)}_{\Lambda}  \;\Phi_{M},\\[2mm]
\overline\chi_{M} 
&=&  
-\textrm{i}\;\overline\Phi_{M}\; V^{(3)}_{\Lambda}\;{\cal G}_{M}.
\end{eqnarray}
However, with the definition $V^{(3)}_{\Lambda}=\overline \Lambda V^{(3)}\Lambda$, we 
in fact have
\begin{eqnarray}
\label{reconstruct_SA_Lam}
\chi_{M} 
&=&   
-\textrm{i}\;{\cal G}_{M}\; \overline\Lambda V^{(3)}\;\Phi_{M}^{\Lambda}
\\[2mm]
\label{reconstruct_adSA_Lam}
\overline\chi_{M} 
&=&  
-\textrm{i}\;\overline\Phi_{M}^{\Lambda}\;V^{(3)} \Lambda\;{\cal G}_{M}
\end{eqnarray}
showing that for a reconstruction of the full eight-dimensional
Bethe-Salpeter amplitudes it indeed suffices to know only the
projected components
\begin{eqnarray}
\Phi_{M}^{\Lambda}\left({\bf p_\xi},{\bf p_\eta}\right) 
&:=& 
\Lambda \left({\bf p_\xi},{\bf p_\eta}\right)\; \Phi_{M}\left({\bf p_\xi},{\bf p_\eta}\right),\\[1mm]
\overline\Phi_{M}^{\Lambda}\left({\bf p_\xi},{\bf p_\eta}\right) 
&:=&   
\overline\Phi_{M}\left({\bf p_\xi},{\bf p_\eta}\right)\;\overline\Lambda \left({\bf p_\xi},{\bf p_\eta}\right)
\end{eqnarray}
of the Salpeter amplitudes.  Performing now the integration over
${p_\xi}^0$ and ${p_\eta}^0$ on both sides of
eqs. (\ref{reconstruct_SA_Lam}) and (\ref{reconstruct_adSA_Lam}), the
Bethe-Salpeter amplitudes $\chi_{M}$ and $\overline \chi_{M}$ on the
left hand side reduce to the Salpeter amplitudes $\Phi_{M}$ and
$\overline\Phi_{M}$ and on the right hand side we obtain the reduction
$\langle{\cal G}_{M}\rangle$ of the resolvent ${\cal G}_{M}$ leading
to the interconnection between the full Salpeter amplitudes $\Phi_{M}$
and $\overline\Phi_{M}$ and their corresponding projected parts
$\Phi_{M}^{\Lambda}$ and $\overline\Phi_{M}^{\Lambda}$, respectively,
{\it i.e.}
\begin{eqnarray}
\label{SAreconstruct_SA_Lam}
\Phi_{M} &=& -\textrm{i}\;\langle{\cal G}_{M}\rangle\;\overline\Lambda V^{(3)}\;\Phi^\Lambda_{M},\\[4mm]
\label{adSAreconstruct_adSA_Lam}
\overline\Phi_{M} &=&
-\textrm{i}\;\overline\Phi_{M}^\Lambda \;V^{(3)}
\Lambda\;\langle{\cal G}_{M}\rangle.
\end{eqnarray}
Here the mixed energy components of the full amplitudes $\Phi_{M}$ and
$\overline\Phi_{M}$ reenter via the mixed energy components of
$\langle{\cal G}_{M}\rangle$.  To get the
equations for the components $\Phi_{M}^{\Lambda}$ and
$\overline\Phi_{M}^{\Lambda}$, we finally have to perform the
projection on purely positive and purely negative energy components via
the Salpeter projectors $\Lambda$ and $\overline \Lambda$ on both
sides of the eqs. (\ref{SAreconstruct_SA_Lam}) and
(\ref{adSAreconstruct_adSA_Lam}), respectively.
We then find
\begin{eqnarray}
\label{GP_SE_V2V3_COV}
\Phi_{M}^\Lambda &=&  
-\textrm{i}\;
\langle{\cal G}_{M}\rangle_\Lambda\; V^{(3)}\;\Phi^\Lambda_{M},\\[4mm]
\label{adGP_SE_V2V3_COV}
\overline\Phi_{M}^\Lambda &=&  
-\textrm{i}\;\overline\Phi_{M}^\Lambda \;V^{(3)} \;
\langle{\cal G}_{M}\rangle_\Lambda,
\end{eqnarray}
where we introduced the symbol $\langle{\cal G}_{M}\rangle_\Lambda$
to denote the corresponding projection on $\langle{\cal G}_{M}\rangle$,
\begin{eqnarray}
\label{proj_red_calG}
\langle{\cal G}_{M}\rangle_\Lambda
&:=&
\Lambda\langle{\cal G}_{M}\rangle\overline\Lambda,
\end{eqnarray}
which cuts off the mixed energy components on both sides of $\langle{\cal G}_{M}\rangle$.
Thus, due to the Neumann series of ${\cal G}_M$, the reduced propagator
$\langle{\cal G}_{M}\rangle_\Lambda$ may be represented as power series
which starts in lowest order with the free Salpeter propagator
$\langle {G_0}_{M}\rangle$ and consists of an infinite number of reduced
Feynman diagrams, which all are restricted to purely positive and negative
energy components, as the Salpeter propagator $\langle {G_0}_{M}\rangle$
itself:
\begin{eqnarray}
\label{series_reduced_GP}
\langle{\cal G}_{M}\rangle_\Lambda
&=&
\langle{G_0}_{M}\rangle 
+ 
\;\Lambda\left\langle {G_0}_{M}\; (-\textrm{i})\left[V^{(3)}_{\rm R} 
+ \overline K^{(2)}_{M}\right]\;{G_0}_{M}\right\rangle\overline\Lambda\\[2mm]
&&\;+\; 
\Lambda\left\langle {G_0}_{M}\; (-\textrm{i})\left[V^{(3)}_{\rm R} 
+ \overline K^{(2)}_{M}\right]\;{G_0}_{M}\; (-\textrm{i})\left[V^{(3)}_{\rm R} 
+ \overline K^{(2)}_{M}\right]\;{G_0}_{M}\right\rangle\overline\Lambda 
\;+\; 
\ldots\; .\nonumber
\end{eqnarray}
The idea is now to classify the reducible and irreducible diagrams in
this infinite reduced series in the same way as done in sect.
\ref{sec:bsequation}, where the quantum field theoretical six-point
Green's function $G$ was non-perturbatively constructed as a solution
of an inhomogeneous eight-dimensional integral equation. But now this
classification is done on the reduced level where irreducibility is
understood with respect to the 'free' Salpeter propagator
$\langle{G_0}_{M}\rangle$.  This means that we are looking for an
irreducible kernel $V^{\rm eff}_{M}$ such that $\langle{\cal
G}_{M}\rangle_\Lambda$ is the solution of the following
six-dimensional integral equation:
\begin{equation}
\label{reduced_IE_Veff}
\langle{\cal G}_{M}\rangle_\Lambda
\;\stackrel{!}{=}\;
\langle{G_0}_{M}\rangle 
-\textrm{i}\;\langle{G_0}_{M}\rangle\;V^{\rm eff}_{M}\;
\langle{\cal G}_{M}\rangle_\Lambda
\;=\;
\langle{G_0}_{M}\rangle 
-\textrm{i}\;
\langle{\cal G}_{M}\rangle_\Lambda
\;V^{\rm eff}_{M}\;\langle{G_0}_{M}\rangle.
\end{equation}
Note that in contrast to our first attempt, where this ansatz due to
the restrictive action of the Salpeter propagator
$\langle{G_0}_{M}\rangle$ on purely positive and negative energy
components led to inconsistencies, here the ansatz becomes possible
now, because $\langle{\cal G}_{M}\rangle_\Lambda$ itself and thus all
terms of the series (\ref{series_reduced_GP}) have by construction the
same restriction as $\langle{G_0}_{M}\rangle$ to these components
only.  Formally the determination of $V^{\rm eff}_{M}$ corresponds to
the inversion of $\langle{\cal G}_{M}\rangle_\Lambda$, which due to
the projector properties is restricted to the subspace of positive and
negative energy components. In particular, this requires the inversion
of the Salpeter propagator $\langle{G_0}_{M}\rangle$ in this
subspace. For this purpose we introduce the operator ${h_0}_{M}$ by
\begin{eqnarray}
\label{pseudo_breit_hamilton}
{h_0}_{M}({\bf p_\xi}, {\bf p_\eta};\; {\bf p_\xi'}, {\bf p_\eta'})
&:=&
-\textrm{i}\;
\gamma^0
\tens
\gamma^0
\tens
\gamma^0\;
\left[M\;\Id - {\cal H}_0({\bf p_\xi}, {\bf p_\eta})\right]\nn
&&
\phantom{\textrm{i}}\times\;
(2 \pi)^3\;\delta^{(3)}({\bf p_\xi} -{\bf p_\xi'})\;\; 
(2 \pi)^3\;\delta^{(3)}({\bf p_\eta}-{\bf p_\eta'}),
\end{eqnarray}
with ${\cal H}_0$ the free three-fermion Hamiltonian defined in
eq. (\ref{free_breit_hamilt_CMS}) such that the 'inverse' of
$\langle{G_0}_{M}\rangle$ in this subspace is given by\footnote{Note that according to our
  concise operator notation $\Lambda$ and $\overline\Lambda$ here have the meaning of
an integral operator {\it i.e.}:
$
\Lambda ({\bf p_\xi},{\bf p_\eta};\;{\bf p_\xi'},{\bf p_\eta'}) 
:=
\Lambda ({\bf p_\xi},{\bf p_\eta})\;\; 
(2 \pi)^3\;\delta^{(3)}({\bf p_\xi}-{\bf p_\xi'})\;\; 
(2 \pi)^3\;\delta^{(3)}({\bf p_\eta}-{\bf p_\eta'})$} 
\begin{equation}
\label{inversion_G0}
\langle{G_0}_{M}\rangle\;{h_0}_{M}\; =\; \Lambda,
\quad
{h_0}_{M}\;\langle{G_0}_{M}\rangle\; =\; \overline\Lambda,
\end{equation}
and the 'inversion' of $\langle{\cal G}_{M}\rangle_\Lambda$ can now be
expressed by 
\begin{equation}
\label{calG_inv}
\left\langle{\cal G}_{M}\right\rangle_\Lambda\;
\left[\;{h_0}_M  + {\rm i}\;V^{\rm eff}_{M}\; \right]
\;=\;
\Lambda,\quad
\left[\;{h_0}_M  + {\rm i}\;V^{\rm eff}_{M}\; \right]\;
\left\langle{\cal G}_{M}\right\rangle_\Lambda\;
\;\;=
\overline\Lambda.
\end{equation} 
A unique definition of the effective, irreducible kernel $V^{\rm eff}_{M}$ 
then requires its restriction to positive and negative components according to 
\begin{equation}
\label{restriction_Veff}
\overline\Lambda V^{\rm eff}_{M}\;\;=\;\;
V^{\rm eff}_{M}\Lambda \;\;=\;\; 
V^{\rm eff}_{M}.
\end{equation}

Finally, having found this quasi potential $V^{\rm eff}_{M}$, we can use
eqs. (\ref{calG_inv}) in order to transform the Salpeter equation
(\ref{GP_SE_V2V3_COV}) and its adjoint (\ref{adGP_SE_V2V3_COV}) into the
desired form as indicated in the beginning of this subsection in eq.
(\ref{searched_reduced_SE}):
\begin{eqnarray}
\label{SE_V2_V3_cov}
\Phi_{M}^\Lambda &=&  
-\textrm{i}\;\langle{G_0}_{M}\rangle\;\left[V^{(3)} + V^{\rm eff}_{M}\right]\;\Phi^\Lambda_{M},\\[4mm]
\overline\Phi_{M}^\Lambda &=&  
-\textrm{i}\;\overline\Phi_{M}^\Lambda \;\left[V^{(3)} + V^{\rm eff}_{M}\right]\;\langle{G_0}_{M}\rangle.
\end{eqnarray}
The form and therefore the properties of this reduced bound state equation
are indeed exactly the same as in the case of vanishing two-particle kernels
discussed in the sect. \ref{sec_reduc_V3}.  The only extension is the
effective quasi potential $V^{\rm eff}_{M}$ which occurs in addition to
the genuine instantaneous three-body kernel $V^{(3)}$ and absorbs all the
complexities entering via the retardation effects from the unconnected
two-quark interactions. Note, however,
that $V^{\rm eff}_{M}$, in contrast to $V^{(3)}$, in general is
energy-, {\it i.e.} $M$-dependent as indicated by the subscript $M$.\\

To obtain a Hamiltonian formulation of the Salpeter equation,
we multiply eq. (\ref{SE_V2_V3_cov}) by
$[\;\textrm{i}\;\gamma^0\tens\gamma^0\tens\gamma^0\;{h_0}_{M}\;]$ and thus end up with a the generalized\footnote{'generalized'
means that now the Salpeter Hamiltonian ${\cal H}_{M}$
itself depends on the eigenvalue $M$ due to the
$M$-dependence of $V^{\rm eff}_{M}$} eigenvalue problem which
determines the bound state mass $M$ and the corresponding
amplitude $\Phi_{M}^\Lambda$:
\begin{equation}
\label{cov_hamilton_form_SE_V2V3}
{\cal H}_{M}\;\Phi_{M}^\Lambda = M \;\Phi_{M}^\Lambda,
\end{equation}
where the Salpeter Hamiltonian ${\cal H}_{M}$ now explicitly
reads
\begin{eqnarray}
\lefteqn{\left[{\cal H}_{M}\;\Phi_{M}^\Lambda\right]({\bf p_\xi}, {\bf p_\eta})\;\;:=\;\;
{\cal H}_0({\bf p_\xi}, {\bf p_\eta})\;\Phi_{M}^\Lambda({\bf p_\xi}, {\bf p_\eta})}&&\nn[3mm]
&+& \Lambda({\bf p_\xi}, {\bf p_\eta})\;\gamma^0\tens\gamma^0\tens\gamma^0\;
\int 
\frac{\textrm{d}^3 {\bf p_\xi'}}{(2\pi)^3}\;
\frac{\textrm{d}^3 {\bf p_\eta'}}{(2\pi)^3}\;
V^{(3)}({\bf p_\xi},{\bf p_\eta};\;
{\bf p_\xi'},{\bf p_\eta'})\;
\Phi_{M}^\Lambda({\bf p_\xi'}, {\bf p_\eta'})\nn[3mm]
&+&
\Lambda({\bf p_\xi}, {\bf p_\eta})\;\gamma^0\tens\gamma^0\tens\gamma^0\;
\int 
\frac{\textrm{d}^3 {\bf p_\xi'}}{(2\pi)^3}\;
\frac{\textrm{d}^3 {\bf p_\eta'}}{(2\pi)^3}\;
V^{\rm eff}_{M}({\bf p_\xi},{\bf p_\eta};\;
{\bf p_\xi'},{\bf p_\eta'})\;
\Phi_{M}^\Lambda({\bf p_\xi'}, {\bf p_\eta'}).
\end{eqnarray}

The next step is to determine $V^{\rm eff}_{M}$. According to eq.
 (\ref{proj_red_calG}) we have 
\begin{equation}
\label{proj_red_calG_2}
\langle{\cal G}_{M}\rangle_\Lambda
\stackrel{!}{=}
\Lambda\langle{\cal G}_{M}\rangle\overline\Lambda,
\end{equation}
where on the left $\langle{\cal G}_{M}\rangle_\Lambda$ is given by the
integral equation (\ref{reduced_IE_Veff}) which defines $V^{\rm eff}_{M}$ and
on the right we insert ${\cal G}_{M}$ as given by the integral equation
(\ref{inhom_IE_GP}) with kernel $V^{(3)}_{\rm R} + \overline K^{(2)}_{M}$.
This equation then has to be solved for $V^{\rm eff}_{M}$. As shown in
detail in appendix \ref{sec:det_eff_K}, with the restriction (\ref{restriction_Veff}) the
effective instantaneous kernel $V^{\rm eff}_{M}$ can be uniquely determined
order-by-order as an infinite power series expansion
\begin{equation}
\label{Veff_series}
V^{\rm eff}_{M} 
\;\;=\;\; 
\sum_{k=1}^\infty\;\;{V^{\rm eff}_{M}}^{(k)}
\end{equation}
of irreducible interaction terms ${V^{\rm eff}_{M}}^{(k)}$
in powers $k$ of the kernel $V^{(3)}_{\rm R} + \overline K^{(2)}_{M}$.
The explicit expressions in arbitrary order $k$ are
then constructed according to the following prescription
\begin{eqnarray}
\label{Veff_born}
{V^{\rm eff}_{M}}^{(1)}
&=&
{h_0}_M\;\Lambda\left\langle{G_0}_{M}\; \overline K^{(2)}_{M}\;{G_0}_{M}\right\rangle\overline\Lambda\;{h_0}_M,\\[4mm]
\label{Veff_order_k}
{V^{\rm eff}_{M}}^{(k)}
&=&
\textrm{i}\;{h_0}_M\;\Lambda\left\langle{G_0}_{M} 
\underbrace{
(-\textrm{i})\left[V^{(3)}_{\rm R} + \overline K^{(2)}_{M}\right] {G_0}_{M}
\;\ldots\; 
(-\textrm{i})\left[V^{(3)}_{\rm R} + \overline K^{(2)}_{M}\right]{G_0}_{M}}_
{k\;\; {\rm times}}\right\rangle\overline\Lambda\;{h_0}_M\nn[3mm]
\hspace*{10mm}&&
-\;\textrm{i}\;\sum_{r=2}^k\!\!\!\!\!
\sum_{\footnotesize
\begin{array}{c}
k_1, k_2, \ldots, k_r < k\\ 
k_1 + k_2 + \ldots + k_r = k
\end{array}}\!\!\!\!\!\!\!
\left[-\textrm{i}\;{V^{\rm eff}_{M}}^{(k_1)}\right]\langle{G_0}_{M}\rangle
\left[-\textrm{i}\;{V^{\rm eff}_{M}}^{(k_2)}\right]\langle{G_0}_{M}\rangle
\ldots 
\left[-\textrm{i}\;{V^{\rm eff}_{M}}^{(k_r)}\right].
\end{eqnarray}
Notice the emerging structure of these equations: The reduced Feynman
diagram of $k$th order ({\it i.e.} the first term on the right hand side of
eq. (\ref{Veff_order_k})) consists on the one hand of the irreducible
part ${V^{\rm eff}_{M}}^{(k)}$ of order $k$, which we are in fact interested
in, and on the other hand it contains an order-$k$ reducible part, built from
all possible iterations of reducible diagrams ${V^{\rm eff}_{M}}^{(k_i)}$ of
lower order $k_i < k$ with $\sum_i k_i = k$, as given by the second term in
(\ref{Veff_order_k}), which obviously has to be subtracted to get the desired
${V^{\rm eff}_{M}}^{(k)}$. 

\subsubsection{The normalization condition for the reduced amplitudes}
The solutions $\chi_{\bar P}$ of the Bethe-Salpeter equation
(\ref{BSE_V3_V2}) have to satisfy the normalization condition, which may be
formulated in the explicitly covariant form
(\ref{norm_condition_compact_cov}). The covariant framework ensures that the
normalization in the rest-frame implies the correct normalization of the
Bethe-Salpeter 
amplitudes $\chi_{\bar P}$ in any frame.  In this
section we will determine the corresponding normalization condition for the
projected Salpeter amplitudes $\Phi^{\Lambda}_{M}$, {\it i.e.} the
solutions of the Salpeter equation (\ref{SE_V2_V3_cov}) in the
rest-frame. For this purpose we start with the Bethe-Salpeter norm
which in the center-of-mass frame reads
\begin{equation}
\label{norm_cond_V2V3}
-{\rm i}\;\overline\chi_{M}\;
\left[\frac{\partial}{\partial M}\;H_M\right]
\chi_{M}
= 2 M,
\end{equation}
where the pseudo-Hamiltonian $H_M$ is defined by
\begin{eqnarray}
H_M 
&=& 
{G_0}_M^{-1} 
+ \textrm{i}\;\overline K^{(2)}_{M} 
+ \textrm{i}\;V^{(3)}.
\end{eqnarray}
To get the analogous condition for the reduced amplitudes, we have to express
the eight-dimensional Bethe-Salpeter amplitude $\chi_{M}$ and
its adjoint $\overline \chi_{M}$ by the corresponding reduced
six-dimensional amplitudes $\Phi^{\Lambda}_{M}$ and
$\overline\Phi^{\Lambda}_{M}$, respectively. We do this by using the 
relations (\ref{reconstruct_SA_Lam}) and
(\ref{reconstruct_adSA_Lam}), {\it i.e.}  the prescription how to reconstruct
the Bethe-Salpeter amplitudes from the Salpeter amplitudes:
\begin{eqnarray}
\chi_{M} 
&=&  
-\textrm{i}\;{\cal G}_{M}\; \overline\Lambda V^{(3)}\;\Phi_{M}^{\Lambda},
\\[4mm]
\overline\chi_{M} 
&=&  
-\textrm{i}\;\overline\Phi_{M}^{\Lambda}\;V^{(3)} \Lambda\;{\cal G}_{M}.
\end{eqnarray}
Recall that in eq. (\ref{resolvent_cal_GP}) the Green's
function ${\cal G}_{M}$ was defined as the resolvent of the pseudo-Hamiltonian
\begin{equation}
\label{resolvent_equation_calGP}
H_M^{\rm R} \;:=\; {G_0}_M^{-1} + \textrm{i}\;\overline K^{(2)}_{M} 
+ \textrm{i}\;V^{(3)}_{\rm R},
\quad\textit{i.e.}\quad 
H_M^{\rm R}\;{\cal G}_{M}
\;= \;
{\cal G}_{M}\;H_M^{\rm R}
\;= \;
\Id.
\end{equation}
Accordingly, with the decomposition $V^{(3)} = V^{(3)}_\Lambda +  V^{(3)}_{\rm R}$, we write
\begin{equation}
H_M
=
H_M^{\rm R} + \textrm{i}\;V^{(3)}_\Lambda,
\end{equation}
where the projected part $V^{(3)}_\Lambda$ of the instantaneous kernel $V^{(3)}$ has no explicit
$M$-dependence as $V^{(3)}$ itself.
Therefore, it gives no contribution to the normalization condition (\ref{norm_cond_V2V3})
which thus becomes
\begin{equation}
\label{norm_cond_V2V3_b}
{\rm i}\;
\overline\Phi_{M}^{\Lambda}\;V^{(3)} \Lambda\;\;
{\cal G}_{M}\;
\bigg[\frac{\partial}{\partial M}\;H_M^{\rm R}\bigg]{\cal G}_{M}\;\; 
\overline\Lambda\; V^{(3)}\;\Phi_{M}^{\Lambda}
\;\;=\;\; 2 M.
\end{equation}
Using the resolvent equation (\ref{resolvent_equation_calGP}) for the Green's function ${\cal G}_{M}$
the derivative of $H_M^{\rm R}$ can be rewritten as a derivative of
${\cal G}_{M}$:
\begin{equation}
{\cal G}_{M}\;
\bigg[\frac{\partial}{\partial M}\;H_M^{\rm R} \bigg]
\;{\cal G}_{M}
\;\;=\;\;
- \frac{\partial}{\partial M}\;{\cal G}_{M}.
\end{equation}
Substitution into (\ref{norm_cond_V2V3_b}) then yields
\begin{equation}
-{\rm i}\;\;
\overline\Phi_{M}^{\Lambda}\;V^{(3)} \Lambda\;\;
\left[\frac{\partial}{\partial M}\;{\cal G}_{M}\right]
\;\overline\Lambda\; V^{(3)}\;\Phi_{M}^{\Lambda}=2 M,
\end{equation}
which is convenient for the further reduction, since the 'vertex' amplitudes
$\overline\Phi_{M}^{\Lambda}\;V^{(3)} \Lambda$ and $\overline\Lambda\;
V^{(3)}\;\Phi_{M}^{\Lambda}$ are six-dimensional and the only remaining full
eight-dimensional quantity is the derivative of the Green's function
${\cal G}_{M}$ which can be reduced by the internal integrations over
${p_\xi}^0$ and ${p_\eta}^0$ according to eq. (\ref{reduc_sixpoint}):
\begin{equation}
-{\rm i}\;\;
\overline\Phi_{M}^{\Lambda}\;V^{(3)}\;\; 
\Lambda
\left\langle\frac{\partial}{\partial M}\;{\cal G}_{M}\right\rangle
\overline\Lambda
\;\;\; V^{(3)}\;\Phi_{M}^{\Lambda}= 2 M.
\end{equation}
The next step is to get rid of the three-body kernel $V^{(3)}$ by absorbing it
into the Salpeter amplitudes $\Phi_{M}^{\Lambda}$ and
$\overline\Phi_{M}^{\Lambda}$ by means of the Salpeter equations
(\ref{GP_SE_V2V3_COV}) and (\ref{adGP_SE_V2V3_COV}). We start with
\begin{equation}
\Lambda
\left\langle\frac{\partial}{\partial M}\;{\cal G}_{M}\right\rangle
\overline\Lambda
\;\;=\;\;
\frac{\partial}{\partial M}\;
\left[\;\Lambda\left\langle{\cal G}_{M}\right\rangle\overline\Lambda\; 
\right]
\;\;=\;\;
\frac{\partial}{\partial M}\;
\left\langle{\cal G}_{M}\right\rangle_\Lambda,
\end{equation}
where we used that the partial derivative $\partial/\partial M$ can be
commuted with the integrations over ${p_\xi}^0$ and ${p_\eta}^0$ and also with
the Salpeter projector $\Lambda$. With the resolvent equation
(\ref{calG_inv}) for the reduced propagator
$\langle{\cal G}_{M}\rangle_\Lambda$, we can then rewrite the derivative of
$\langle{\cal G}_{M}\rangle_\Lambda$ as
\begin{equation}
\frac{\partial}{\partial M}\;
\left\langle{\cal G}_{M}\right\rangle_\Lambda
\;\;=\;\;
-\;\left\langle{\cal G}_{M}\right\rangle_\Lambda\;
\frac{\partial}{\partial M}\;
\left[\;{h_0}_M  + {\rm i}\;V^{\rm eff}_{M}\; \right]
\left\langle{\cal G}_{M}\right\rangle_\Lambda,
\end{equation}
which finally enables us to apply the Salpeter equation (\ref{GP_SE_V2V3_COV})
and its adjoint (\ref{adGP_SE_V2V3_COV}) in order to absorb the three-body
kernel $V^{(3)}$ into the Salpeter amplitudes:
\begin{eqnarray}
{\rm i}\;\;
\overline\Phi_{M}^{\Lambda}\;V^{(3)}\;
\left\langle{\cal G}_{M}\right\rangle_\Lambda\;\;\;
\frac{\partial}{\partial M}\;
\left[\;{h_0}_M  + {\rm i}\;V^{\rm eff}_{M}\; \right]
\;\;\;
\left\langle{\cal G}_{M}\right\rangle_\Lambda
\; V^{(3)}\;\Phi_{M}^{\Lambda}
&=&
2 M\nn[2mm]
\Leftrightarrow\quad\quad
-{\rm i}\;\;
\overline\Phi_{M}^{\Lambda}\;\;
\frac{\partial}{\partial M}\;
\left[\;{h_0}_M  + {\rm i}\;V^{\rm eff}_{M}\; \right]
\;\Phi_{M}^{\Lambda}
&=&
2 M.
\end{eqnarray}
The explicit expression for the derivative of the operator ${h_0}_{M}$, owing
to its definition (\ref{pseudo_breit_hamilton}), is given by
\begin{equation}
\left[\frac{\partial}{\partial M}{h_0}_{M}\right]({\bf p_\xi}, {\bf p_\eta};\; {\bf p_\xi'}, {\bf p_\eta'})
=
-\textrm{i}\;
\gamma^0
\tens
\gamma^0
\tens
\gamma^0\;\;(2 \pi)^3\;\delta^{(3)}({\bf p_\xi}-{\bf p_\xi'})\;\; 
(2 \pi)^3\;\delta^{(3)}({\bf p_\eta}-{\bf p_\eta'}),
\end{equation}
and one readily shows that the general relation (\ref{interconnectPhi_Phiad_mom}) between the
rest-frame Salpeter amplitude $\Phi_M$ and its corresponding adjoint
$\overline\Phi_M$ holds likewise for the projected amplitudes $\Phi_{M}^{\Lambda}
= \Lambda\Phi_{M}$ and $\overline\Phi_{M}^{\Lambda} =
\overline\Phi_{M}\overline\Lambda$, {\it i.e.}
\begin{equation}
\label{interconnectPhiLam_PhiLam_ad_mom}
\overline \Phi_M^\Lambda({\bf p_\xi}, {\bf p_\eta})
=
- {\Phi_M^\Lambda}^\dagger ({\bf p_\xi}, {\bf p_\eta})\;
\gamma^0
\tens
\gamma^0
\tens
\gamma^0,
\end{equation}
so that we finally end up with the following form of the normalization
condition for the reduced amplitudes $\Phi_{M}^{\Lambda}$ 
\begin{equation}
\label{norm_cond_V3V2_CMS}
\SP{\Phi_M^\Lambda} {\Phi_M^\Lambda}
 \;\;-\;\; \SP{\Phi_M^\Lambda} {\;\gamma^0\tens
 \gamma^0\tens \gamma^0\; \Big(\frac{\partial}{\partial M}\;V^{\rm
 eff}_{M}\Big) \;\Phi_{M}^{\Lambda}} \;\;=\;\; 2 M.
\end{equation}
Here $\SP{\cdot} {\cdot}$ denotes the positive definite scalar product
(\ref{SP_V3}), which is induced by the ${\cal L}^2$-normalization
condition of the Salpeter amplitude in the case of vanishing
two-particle kernels. In comparison to the case where the dynamics is
determined by an instantaneous three body kernel alone we thus find
that (owing to its explicit energy dependence) the effective kernel
$V^{\rm eff}_{M}$ in general leads to an additional contribution to
the norm.

\subsubsection{Lowest order contributions to the effective kernel}
\label{lowest_order_contrib_Veff}
In equations (\ref{Veff_series} -- \ref{Veff_order_k}) we displayed the
general order-by-order prescription to construct $V^{\rm eff}_{M}$ . 
In practice, we have to approximate the effective kernel $V^{\rm eff}_{M}$,
which itself consists of an infinite number of terms. A systematical
approximation is now given by truncating the series (\ref{Veff_series}) at
some finite order $k < \infty$, {\it i.e.}
\begin{equation}
\label{Veff_series_approx}
V^{\rm eff}_{M} 
\simeq 
{V^{\rm eff}_{M}}^{(1)}
+
{V^{\rm eff}_{M}}^{(2)}
+
\ldots
+
{V^{\rm eff}_{M}}^{(k)},
\end{equation}
thus yielding an approximation of the  Salpeter
amplitude $\Phi^\Lambda_M \simeq{\Phi^\Lambda_M}^{\!\!(k)}$ by the solution of
\begin{equation}
{\Phi^\Lambda_M}^{\!\!(k)}
=-{\rm i}\langle {G_0}_M\rangle\left(V^{(3)}+\sum_{i=1}^k {V^{\rm eff}_{M}}^{(i)} \right){\Phi^\Lambda_M}^{\!\!(k)}.
\end{equation}
Note that such a finite order approximation of $V^{\rm eff}_{M}$ means
for the original reduced propagator $\langle{\cal G}_M\rangle_\Lambda$
(and thus also for the Salpeter equation) an approximation beyond
perturbation theory, due to the infinite iteration of $V^{\rm
eff}_{M}$ in $\langle{\cal G}_M\rangle_\Lambda$.  It is worthwhile to
mention that this subsequent approximation of the Salpeter equation
(within the CMS frame) still preserves formal
covariance. Our reduction procedure and thus the construction of
$V^{\rm eff}_{M}$ can be covariantly formulated in any arbitrary
reference frame according to the covariant replacements
$p^0\rightarrow p_\|$ and ${\bf p} \rightarrow p_\bot$ as mentioned
previously. Consequently, also the truncation
(\ref{Veff_series_approx}) of the effective kernel $V^{\rm eff}_{M}$
can in fact be performed frame-independently. Another aspect
concerning the approximation (\ref{Veff_series_approx}) requires
attention.  For the calculation of transition matrix elements we need
the full Bethe-Salpeter amplitude $\chi_{M}$ which (if $V^{\rm
eff}_{M}$ and $\Phi^\Lambda_M$ are known exactly) can be reconstructed
by the prescription (\ref{reconstruct_SA_Lam}) via the Green's
function ${\cal G}_M$. To be consistent we need an approximation
${\cal G}_M^{(k)}$ that corresponds to the approximation
(\ref{Veff_series_approx}) of the effective kernel. In other words, we
require the corresponding order $k$ approximation $\chi_{M}^{(k)}$ of
the Bethe-Salpeter amplitude $\chi_{M}$ such that its reduction yields
the order $k$ approximation ${\Phi^\Lambda_M}^{\!\!(k)}$ of the
Salpeter amplitude. As shown in ref. \cite{Kr01} a consistent prescription for an
approximated reconstruction of the Bethe-Salpeter amplitude can indeed be
found.\\

With regard to explicit calculations let us now become specific and compute
the explicit expressions for the contributions to $V^{\rm eff}_{M}$ up to second order.

\subsubsection*{The Born term ${V^{\rm eff}_{M}}^{(1)}$}
\begin{figure}[!h]
  \begin{center}
    \epsfig{file={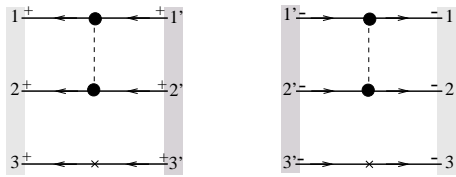},width=60mm}
    \end{center}
\caption{Time ordered graphs for the Born term
  $\left\langle{G_0}_{M}\right\rangle \; {V^{\rm eff}_{M}}^{(1)}\;
  \left\langle{G_0}_{M}\right\rangle$ due to eq.
  (\ref{Veff_Born_explicit}). The instantaneous two-body kernel (shown for the quark pair (12) only) 
is represented by the vertical dashed
  line.}
\label{fig:VeffBorn}
\end{figure}

Concerning the Born term ${V^{\rm eff}_{M}}^{(1)}$ we can refer to a
former result of subsect. \ref{subsec:reduction_V2_first} given in eq.
(\ref{reduction_Born}). According to eq. (\ref{Veff_born}) the projectors $\Lambda$ and
$\overline\Lambda$ select the purely positive and negative energy components,
{\it i.e.}  the terms (a) and (b) of eq. (\ref{reduction_Born}), and cut off
the mixed contributions, which correspond to the terms (c) to (f). We then find
the following, rather simple result, which is a sum of three unconnected
two-fermion potentials for each quark pair (see also fig. \ref{fig:VeffBorn}
for the diagrammatic representation):
\begin{eqnarray}
\lefteqn{
{V^{\rm eff}_{M}}^{(1)} ({\bf p_\xi}, {\bf p_\eta};\;{\bf p_\xi'}, {\bf p_\eta'})\;\;=\;\; 
\left[{h_0}_M\;\Lambda\left\langle {G_0}_M\;\overline K^{(2)}_M\;{G_0}_M\right\rangle\overline\Lambda\;{h_0}_M\right]
({\bf p_\xi}, {\bf p_\eta};\;{\bf p_\xi'}, {\bf p_\eta'})}&&\nn[2mm]
&=&\gamma^0\tens\gamma^0\tens\gamma^0\;\;\times\nn[3mm]
&&
\Bigg\{\Lambda^+_1\tens\Lambda^+_2\tens\Lambda^+_3\;
\left[\gamma^0\tens\gamma^0\;V^{(2)}({\bf p_\xi}, {\bf p_\xi'})\right]\tens\Id
\;(2 \pi)^3\;\delta^{(3)}({\bf p_\eta}\!-\!{\bf p_\eta'})\;
{\Lambda^+_1}'\!\tens{\Lambda^+_2}'\!\tens{\Lambda^+_3}'\nn[1mm]
&&-
\Lambda^-_1\tens\Lambda^-_2\tens\Lambda^-_3\;
\left[\gamma^0\tens\gamma^0\;V^{(2)}({\bf p_\xi}, {\bf p_\xi'})\right]\tens\Id
\;(2 \pi)^3\;\delta^{(3)}({\bf p_\eta}\!-\!{\bf p_\eta'})\;
{\Lambda^-_1}'\!\tens{\Lambda^-_2}'\!\tens{\Lambda^-_3}'\Bigg\}\nn[1mm]
\label{Veff_Born_explicit}
&&
\;+\;
\begin{array}{l}
\textrm{cyclic. perm. of $(12)\;3$ corresponding to the}\\
\textrm{interacting quark pairs (23) and (31)} 
\end{array}
\end{eqnarray}
where $\Lambda^\pm_i \equiv \Lambda^\pm_i ({\bf p_i})$ and ${\Lambda^\pm_i}' \equiv \Lambda^\pm_i ({\bf p_i'})$.
Note that this Born term ${V^{\rm eff}_{M}}^{(1)}$ in fact is  $M$-independent.
\subsubsection*{The second order term ${V^{\rm eff}_{M}}^{(2)}$}
Already in second order the expressions become much more complex:
Using the recipe given by eqs. (\ref{Veff_series} -- \ref{Veff_order_k}) we obtain for the second order term ${V^{\rm
eff}_{M}}^{(2)}$:
\begin{eqnarray}
\lefteqn{{V^{\rm eff}_{M}}^{(2)} \;=}&&\nn
&&
-{\rm i}\;
\Bigg\{
{h_0}_M\;\Lambda
\left\langle {G_0}_M\;\overline K^{(2)}_M\;{G_0}_M\right\rangle
(\Id-\overline\Lambda)\;
V^{(3)}\Lambda
+
\overline\Lambda\; 
V^{(3)}
(\Id-\Lambda)
\left\langle {G_0}_M\;\overline K^{(2)}_M\;{G_0}_M\right\rangle
\overline\Lambda\;{h_0}_M
\nn
&&
\phantom{-{\rm i}\;\Bigg\{\Bigg.}
+
{h_0}_M\;\bigg[\;\phantom{-}
\Lambda\left\langle {G_0}_M\;\overline K^{(2)}_M\;{G_0}_M\;\overline K^{(2)}_M\;{G_0}_M\right\rangle\overline\Lambda\nn[-2mm]
&&
\phantom{-{\rm i}\;\Bigg\{\Bigg.+{h_0}_M\bigg[\bigg.}
-
\Lambda\left\langle {G_0}_M\;\overline K^{(2)}_M\;{G_0}_M\right\rangle\overline\Lambda 
\;{h_0}_M\;
\Lambda\left\langle {G_0}_M\;\overline K^{(2)}_M\;{G_0}_M\right\rangle\overline\Lambda\;
\bigg]\;{h_0}_M
\Bigg\}.
\label{Veff_order_2}
\end{eqnarray}
Analyzing eq. (\ref{Veff_order_2}) in more detail, ${V^{\rm eff}_{M}}^{(2)}$ essentially
consists of three structurally different contributions, 
\begin{equation}
\label{Veff_order_2_contrib}
{V^{\rm eff}_{M}}^{(2)} \;=\; W_M^{(2)} + U_M^{(2)} + C_M^{(2)}.
\end{equation}

\begin{figure}[!h]
  \begin{center}
    \epsfig{file={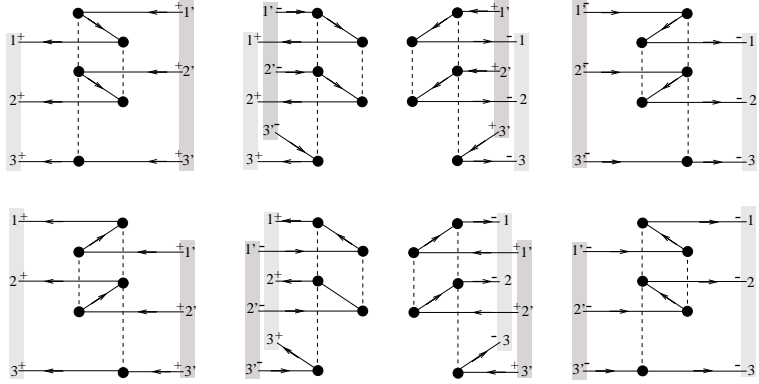},width=100mm}
    \end{center}
\caption{Time ordered graphs of the contribution $W_M^{(2)}$ to the second
  order term ${V^{\rm eff}_{M}}^{(2)}$ given in eq. (\ref{Veff_order2_W}). The
  instantaneous three-body kernel is represented by the vertical dashed line
  that connects three quark lines (indicated by the dots). The instantaneous
  two-particle kernel is shown for the pair (12) only.}
\label{fig:Veff2_V2V3}
\end{figure}

The term $W_M^{(2)}$
is given by the first term on the right hand side of eq.
(\ref{Veff_order_2}):
\begin{eqnarray}
\label{Veff_order2_W}
W_M^{(2)} 
&:=& 
-\;{\rm i}\;
{h_0}_M \Lambda\;\left\langle {G_0}_M\;\overline K^{(2)}_M\;{G_0}_M\right\rangle
(\Id-\overline\Lambda)\;V^{(3)}\;\Lambda
\nn
&&
-\;{\rm i}\;
\overline\Lambda\;V^{(3)}\;(\Id-\Lambda)
\left\langle {G_0}_M\;\overline K^{(2)}_M\;{G_0}_M\right\rangle
\overline\Lambda\;{h_0}_M.
\end{eqnarray}
It is of first order in the instantaneous two-particle kernel (as the
Born term ${V^{\rm eff}_{M}}^{(1)}$) and again we can go back to
subsect. \ref{subsec:reduction_V2_first} and use the result
(\ref{reduction_Born}) to compute the explicit expression of
$W_M^{(2)}$.  However, in contrast to the Born term, now the terms (c)
to (f) of the expression (\ref{reduction_Born}) that couple also to
the mixed energy components, enter only.  They are attached in a
symmetrical way from the right and left hand side to the residual part
$V^{(3)}_{\rm R}$ of $V^{(3)}$, such that the mixed energy components
appear internally in $W_M^{(2)}$.  Notice that the terms of
$W_M^{(2)}$ are suppressed with respect to the corresponding reducible
terms $V^{(3)}_\Lambda\langle {G_0}_M\rangle{V^{\rm eff}_{M}}^{(1)} +
{V^{\rm eff}_{M}}^{(1)}\langle {G_0}_M\rangle V^{(3)}_\Lambda$, built
by iteration of $V^{(3)}_\Lambda$ and the first order Born term
${V^{\rm eff}_{M}}^{(1)}$ by means of the Salpeter equation.  This is
apparent from the different non-singular internal energy denominators
in $W_M^{(2)}$ in comparison to the singular expression $\langle
{G_0}_M\rangle$ that emerge in the reducible terms (see eq.
(\ref{reduction_Born}) and the subsequent discussion of the
corresponding terms (a), (b) $\leftrightarrow$ (c), (d), (e), (f) in
subsect. \ref{subsec:reduction_V2_first}). A diagrammatic
representation of the corresponding time-ordered Feynman graphs is
shown in fig. \ref{fig:Veff2_V2V3}.

\begin{figure}[!h]
  \begin{center}
    \epsfig{file={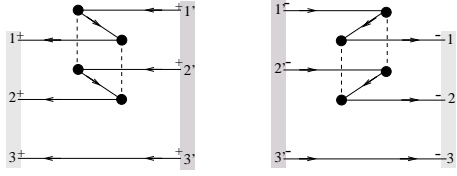},width=60mm}
    \end{center}
\caption{Time ordered graphs of the unconnected irreducible
two-particle kernel $U_M^{(2)}$ as defined in eq.
(\ref{Veff_order2_U}). Here only the term (12)3 is shown.}
\label{fig:Veff2_V12V12}
\end{figure}

Furthermore, the second irreducible part on the right hand side of eq.
(\ref{Veff_order_2}), which is of second order in the two-body kernel, may be
split into two terms of different structure. According to the
decomposition $\overline K^{(2)}_M = K^{12}_M + K^{23}_M + K^{31}_M$, with
\begin{eqnarray}
K^{12}_M(p_\xi,p_\eta;\;p_\xi',p_\eta')
=       
V^{(2)}({\bf p_{\xi}}, {\bf p_{\xi}'}) \tens
  {S_F^3}^{-1}\!\left(\mbox{$\frac{1}{3}M-p_{\eta}$}\right)\;
  (2\pi)^4 \; \delta^{(4)}(p_{\eta}-p'_{\eta_3})
\end{eqnarray}
and $K^{23}_M$, $K^{31}_M$ the corresponding cyclic permutations of
$K^{12}_M$, firstly we find an unconnected part $U_M^{(2)}$ that consists
of a sum of irreducible two-body loops in each quark pair, 
{\it i.e.}
\begin{eqnarray}
\label{Veff_order2_U}
U_M^{(2)} &:=& -{\rm i}\;{h_0}_M\;\bigg[
\Lambda\left\langle {G_0}_M\; K^{12}_M\;{G_0}_M\;K^{12}_M\;{G_0}_M\right\rangle\overline\Lambda\\
&&
\phantom{-{\rm i}\;{h_0}_M\;\bigg[\bigg.}
-
\Lambda\left\langle {G_0}_M\;K^{12}_M\;{G_0}_M\right\rangle\overline\Lambda 
\;{h_0}_M\;
\Lambda\left\langle {G_0}_M\;K^{12}_M\;{G_0}_M\right\rangle\overline\Lambda\;
\bigg]\;{h_0}_M\nn
&& +\quad\textrm{corresponding terms with interacting quark pairs (23) and (31).}
\nonumber
\end{eqnarray}
Remember that the second term in eq. (\ref{Veff_order2_U}) just
subtracts the reducible part of $\Lambda\left\langle {G_0}_M\;
  K^{12}_M\;{G_0}_M\;K^{12}_M\;{G_0}_M\right\rangle\overline\Lambda$ that is
built up by a two-fold iteration of the corresponding Born graphs
(\ref{Veff_Born_explicit}), so that we are left with an irreducible double
Z-loop graph of the corresponding time-ordered Feynman diagram as shown in
fig. \ref{fig:Veff2_V12V12}. 

\begin{figure}[!h]
  \begin{center}
    \epsfig{file={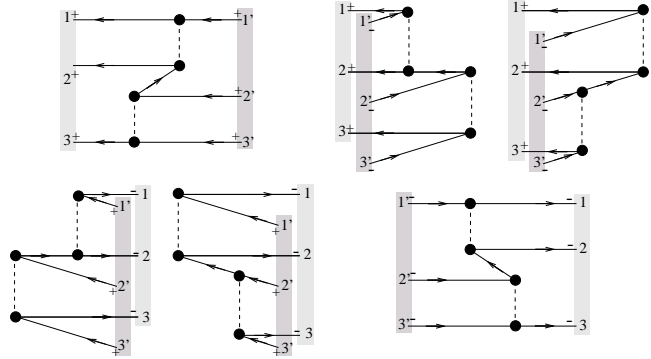},width=85mm}
    \end{center}
\caption{Time ordered graphs of the connected irreducible
quark-exchange interaction $C_M^{(2)}$ as defined in eq.
(\ref{Veff_order2_C}). As an example only term (12, 23) is shown.}
\label{fig:Veff2_V12_V23}
\end{figure}

Secondly, we find a connected part $C_M^{(2)}$ that contains the sum
of all possible irreducible quark-exchange diagrams. This term is given by 
\begin{eqnarray}
\label{Veff_order2_C}
C_M^{(2)} &:=& -{\rm i}\;{h_0}_M\;\bigg[
\Lambda\left\langle {G_0}_M\; K^{12}_M\;{G_0}_M\;K^{23}_M\;{G_0}_M\right\rangle\overline\Lambda\\
&&
\phantom{-{\rm i}\;{h_0}_M\;\bigg[\bigg.}
-
\Lambda\left\langle {G_0}_M\;K^{12}_M\;{G_0}_M\right\rangle\overline\Lambda 
\;{h_0}_M\;
\Lambda\left\langle {G_0}_M\;K^{23}_M\;{G_0}_M\right\rangle\overline\Lambda\;
\bigg]\;{h_0}_M\nn
&& +\quad
\begin{array}{l}
\textrm{corresponding terms with other quark pairings in the incoming and}\\
\textrm{outgoing channels: (23, 12), (12, 31), (31, 12), (31, 23) and (23, 31)}
\end{array}
\nonumber
\end{eqnarray}
In fig. \ref{fig:Veff2_V12_V23} the different time-ordered graphs contributing 
to the irreducible second order quark-exchange interaction are shown diagrammatically.\\

The explicit calculation of the second order terms $\Lambda\langle
{G_0}_M\;\overline K^{(2)}_M\;{G_0}_M\;\overline
K^{(2)}_M\;{G_0}_M\rangle\overline\Lambda$, needed for the determination of
$U_M^{(2)}$ and $C_M^{(2)}$, is lengthy but straightforward and can be
performed by making again elaborate use of the residue theorem. Owing to the
increasing number of quark-lines and the increasing number of contributing
poles in the relative energy variables $p_\xi^0$ and $p_\eta^0$, the structure
and coordinate dependence of these explicit expressions is rather complicated
in comparison to the rather simple structure of the Born term
(\ref{Veff_Born_explicit}).  Moreover $U_M^{(2)}$ and $C_M^{(2)}$ exhibit an
explicit $M$-dependence. We restrict our explicit calculations to the leading
Born term (Born approximation).

\section{Bound-states in Born approximation of the quasi-potential}
\label{bound_states_born}
The discussion of the lowest order contributions to the effective
quasi-potential ${V^{\rm eff}_{M}}$ in the last section clearly showed 
that with increasing order of
the contributions to ${V^{\rm eff}_{M}}$ the explicit expressions rapidly
become more complicated. While the dominant leading order Born 
term ${{V^{\rm eff}_{M}}}^{(1)}$ is still rather simple in 
structure, already the second order contribution ${{V^{\rm eff}_{M}}}^{(2)}$
contains a lot of different irreducible terms whose structure is quite complex
and thus impedes an efficient numerical treatment.
Therefore, expecting these contributions to be small in comparison to the
leading Born term, we consider the Born
approximation 
\begin{equation}
{V^{\rm eff}_{M}} \simeq {V^{\rm eff}_{M}}^{(1)} 
\end{equation}
only. For the sake
of completeness let us summarize the corresponding expressions for the Salpeter
equation and the normalization condition in this approximation. These
equations shall constitute the basis of our quark model for baryons. 

\subsection{Salpeter equation and normalization condition}
\label{SE_and_norm_born}
Approximating the series (\ref{Veff_series}) by the
leading Born term (\ref{Veff_Born_explicit}), the approximated Salpeter 
equation (\ref{cov_hamilton_form_SE_V2V3}) can still be formulated
as an ordinary eigenvalue problem,
\begin{equation}
\label{sequation_hamilt_born}
{\cal H}\Phi_M^\Lambda = M\;\Phi_M^\Lambda,
\end{equation}
since, due to eq. (\ref{Veff_Born_explicit}), the Born term
${V^{\rm eff}_{M}}^{(1)}$ in fact is $M$-independent.
The $M$-independent Salpeter Hamiltonian ${\cal H}$ then reads explicitly:
\begin{eqnarray}
\label{Salp_Hamilt_V3_V2_born}
\left[{\cal H}\Phi_M^\Lambda\right]({\bf p_\xi}, {\bf p_\eta})
&=& 
{\cal H}_0({\bf p_\xi}, {\bf p_\eta})\;\Phi_M^\Lambda({\bf p_\xi}, {\bf p_\eta})\\[3mm]
&+&
\left[ 
\Lambda_1^{+}({\bf p_1})
\tens
\Lambda_2^{+}({\bf p_2})
\tens
\Lambda_3^{+}({\bf p_3})
+
\Lambda_1^{-}({\bf p_1})
\tens
\Lambda_2^{-}({\bf p_2})
\tens
\Lambda_3^{-}({\bf p_3})
\right]\nn
&& \times\;\;
\gamma^0 \tens\gamma^0 \tens\gamma^0\;
\int 
\frac{\textrm{d}^3 p_\xi'}{(2\pi)^3}\;
\frac{\textrm{d}^3 p_\eta'}{(2\pi)^3}\;
V^{(3)}({\bf p_\xi},{\bf p_\eta};\;{\bf p_\xi'},{\bf
  p_\eta'})\;
\Phi_M^\Lambda({\bf p_\xi'}, {\bf p_\eta'})\nn[3mm]
&+&
\left[ 
\Lambda_1^{+}({\bf p_1})
\tens
\Lambda_2^{+}({\bf p_2})
\tens
\Lambda_3^{+}({\bf p_3})
-
\Lambda_1^{-}({\bf p_1})
\tens
\Lambda_2^{-}({\bf p_2})
\tens
\Lambda_3^{-}({\bf p_3})
\right]\nn
&& \times\;\;
\gamma^0 \tens\gamma^0 \tens\Id\;
\int 
\frac{\textrm{d}^3 p_\xi'}{(2\pi)^3}\;
V^{(2)}({\bf p_\xi},{\bf p_\xi'})\tens\Id\;
\Phi_M^\Lambda({\bf p_\xi'}, {\bf p_\eta})\nn
&+& \textrm{corresponding terms with interacting quark pairs (23) and (31)}.\nonumber
\end{eqnarray}
Note the striking structural difference between the connected
three-body part and the unconnected two-body part: The two body term
shows a relative sign between the positive and negative energy
projectors and occurrence of the identity (instead of $\gamma^0$) in the Dirac
space of the spectator quark.\\

To be consistent, the same approximation of ${V^{\rm eff}_{M}}$ must also be
used in the normalization condition (\ref{norm_cond_V3V2_CMS}).  In Born
approximation, the second term in the normalization condition
(\ref{norm_cond_V3V2_CMS}) vanishes, owing to the explicit $M$-independence of
the Born term ${V^{\rm eff}_{M}}^{(1)}$ and we arrive at
\begin{equation}
\frac{\partial}{\partial M}\;{V^{\rm eff}_{M}}^{(1)} = 0
\quad
\Rightarrow
\quad 
\SP{\Phi_M^\Lambda} {\Phi_M^\Lambda} \;\;=\;\; 2 M.
\end{equation}
Consequently, the solutions $\Phi^\Lambda_M$ of the Salpeter equation (\ref{Salp_Hamilt_V3_V2_born}) in Born 
approximation of ${V^{\rm eff}_{M}}$ have to fulfill the same
${\cal L}^2$-normalization condition (\ref{SAnorm_V3_CMS}) as in the case where the
dynamics was determined by the instantaneous three-body kernel alone (see
  subsect. \ref{subsec:norm_SA_V3}), {\it i.e.}
\begin{equation}
\label{norm_cond_V2V3_born}
\SP{\Phi_M^\Lambda} {\Phi_M^\Lambda}
=
\int 
\frac{\textrm{d}^3 p_\xi}{(2\pi)^3}\;
\frac{\textrm{d}^3 p_\eta}{(2\pi)^3}\;
\sum_{a_1, a_2, a_3}
{\Phi_M^\Lambda}_{a_1 a_2 a_3}^* ({\bf p_\xi}, {\bf p_\eta})\;
{\Phi_M^\Lambda}_{a_1 a_2 a_3} ({\bf p_\xi}, {\bf p_\eta})
= 2 M.
\end{equation}
The Salpeter equation for $\Phi_M^\Lambda$ and the corresponding adjoint
equation for $\overline \Phi_M^\Lambda$ must be consistent with the
relation (\ref{interconnectPhiLam_PhiLam_ad_mom}) between
$\Phi_M^\Lambda$ and $\overline \Phi_M^\Lambda$. This leads to the following
condition for the instantaneous interaction kernels $V^{(3)}$ and $V^{(2)}$,
\begin{eqnarray}
\gamma^0
\tens
\gamma^0
\tens
\gamma^0\;
\left[{V^{(3)}}
({\bf p_\xi'},{\bf p_\eta'};\;
{\bf p_\xi},{\bf p_\eta})\right]^\dagger\;
\gamma^0
\tens
\gamma^0
\tens
\gamma^0
&\stackrel{!}{=}&
V^{(3)}
({\bf p_\xi},{\bf p_\eta};\;
{\bf p_\xi'},{\bf p_\eta'})
\nn[2mm]
\label{hermit_cond_V3V2}
\gamma^0
\tens
\gamma^0\;
\left[{V^{(2)}}
({\bf p_\xi'},{\bf p_\xi})\right]^\dagger\;
\gamma^0
\tens
\gamma^0
&\stackrel{!}{=}&
V^{(2)}
({\bf p_\xi}, {\bf p_\xi'}),
\end{eqnarray}
which implies that the Salpeter Hamiltonian
(\ref{Salp_Hamilt_V3_V2_born}) in Born approximation of the effective
kernel is hermitean with respect to the scalar product $\SP{\cdot} {\cdot}$,
{\it i.e.}
\begin{equation}
\SP {\Phi_1} {{\cal H}\;\Phi_2} =  \SP {{\cal H}\;\Phi_1} {\Phi_2}
\quad \forall\quad \Phi_1,\Phi_2 \quad\textrm{with}\quad  \Lambda\Phi_{1,2} = \Phi_{1,2} 
\end{equation}
As in the case of vanishing two-quark kernels this again guarantees that
\begin{itemize}
\item the eigenvalues (bound-state masses) $M$ of ${\cal H}$ are real, 
{\it i.e.} $M^* = M$;
\item the Salpeter amplitudes $\Phi^\Lambda_{M_1}$ and
  $\Phi^\Lambda_{M_2}$ corresponding to different eigenvalues $M_1 \not = M_2$
  are mutually orthogonal: $\SP {\Phi^\Lambda_{M_1}} {\Phi^\Lambda_{M_2}} = 0$.
\end{itemize}
\subsection{Symmetries of the Salpeter equation}
\label{sec:sym_SE}
So far we discussed the constraints (\ref{hermit_cond_V3V2}) of the
instantaneous two- and three-quark interaction kernels $V^{(3)}$ and
$V^{(2)}$ that followed from the interconnection of the amplitude
$\Phi^\Lambda_M$ and its adjoint $\overline\Phi^\Lambda_M$ and
guarantee the hermiticity of the Salpeter Hamiltonian ${\cal H}$ with
respect to the positive definite scalar product $\SP{\cdot} {\cdot}$.
We are led to further conditions on the kernels if we regard the
symmetries which the strong interaction of the quarks has to
respect. Since the underlying theory, quantum chromodynamics (QCD), is
invariant under parity transformations (${\cal P}$), time-reversal
(${\cal T}$) and charge conjugation (${\cal C}$), these symmetry
properties must be incorporated in the Salpeter equation. This means
specifically: If $\Phi^\Lambda_M$ is a solution of the Salpeter
equation, the same must also hold for ${\cal D}\;\Phi^\Lambda_M$ with
${\cal D}\in\{{\cal P},{\cal T},{\cal C}\}$ the representation of the
corresponding transformation on the (projected) Salpeter amplitudes
$\Phi^\Lambda_M = \Lambda\Phi_M$. Below we shall investigate the
corresponding constraints on the interaction kernels $V^{(3)}$ and
$V^{(2)}$ that follow from these invariance conditions. Instead of
${\cal P}$, ${\cal T}$ and ${\cal C}$ we alternatively consider ${\cal
P}$, ${\cal T}$ and ${\cal CPT}$.
\subsubsection{Parity invariance}
The representation of the parity transformation ${\cal P}(x^0,{\bf x}):=(x^0,{\bf -x})$ 
on the full momentum space Salpeter amplitudes $\Phi_M$ is given by
\begin{equation}
\label{repres_P_Phi}
\left[{\cal P}\Phi_M\right]({\bf p_\xi},{\bf p_\eta})
=
\gamma^0\tens\gamma^0\tens\gamma^0\; \Phi_M({\bf -p_\xi},{\bf -p_\eta}).
\end{equation}
Owing to the intertwining relation $\Lambda^\pm_i({\bf p_i})\gamma^0 =
\gamma^0\Lambda^\pm_i({-\bf p_i})$ the different energy components of $\Phi_M$
represent invariant subspaces under the parity transformation ${\cal P}$,
such that ${\cal P}$ decomposes into irreducible representations on these
different subspaces. In particular, the Salpeter projector $\Lambda$
commutes\footnote{The brackets $[\cdot,\cdot]$ denote the commutator $[A,B]:=
  A B - B A$} with ${\cal P}$, {\it i.e} $[{\cal P},\;\Lambda]=0$
such that 
\begin{equation}
\label{represent_P_Phi_Lam}
\left[{\cal P}\Phi_M^\Lambda\right]({\bf p_\xi},{\bf p_\eta})
=
\gamma^0\tens\gamma^0\tens\gamma^0\; \Phi_M^\Lambda({\bf -p_\xi},{\bf -p_\eta}) 
\quad \textrm{with}\quad \Lambda\; {\cal P}\Phi_M^\Lambda = {\cal P}\Phi_M^\Lambda
\end{equation}
is the representation of ${\cal P}$ on the projected Salpeter
amplitudes $\Phi_M^\Lambda = \Lambda\Phi_M$, which actually appear in
the Salpeter equation (\ref{sequation_hamilt_born}).  Parity
invariance implies that with $\Phi_M^\Lambda$ also ${\cal
P}\Phi_M^\Lambda$ is a solution of the Salpeter equation, {\it i.e.}
the Salpeter Hamiltonian has to commute with the representation ${\cal
P}$ of the parity transformation, {\it i.e.}  $\left[{\cal P},\;{\cal
H}\right] = 0$.  With $[{\cal P},\;\Lambda]=0$ and the invariance of
the free Hamiltonian ${\cal H}_0$ under ${\cal P}$, {\it i.e} $[{\cal P},\;{\cal
H}_0]=0$, one readily deduces the following conditions for the three-
and two-quark interaction kernels:
\begin{eqnarray}
\gamma^0
\tens
\gamma^0
\tens
\gamma^0\;
{V^{(3)}}
(-{\bf p_\xi},-{\bf p_\eta};\;
-{\bf p_\xi'},-{\bf p_\eta'})\;
\gamma^0
\tens
\gamma^0
\tens
\gamma^0
&\stackrel{!}{=}&
V^{(3)}
({\bf p_\xi},{\bf p_\eta};\;
{\bf p_\xi'},{\bf p_\eta'}),
\nn[2mm]
\label{Pinv_cond_V3V2}
\gamma^0
\tens
\gamma^0\;
V^{(2)}
(-{\bf p_\xi},-{\bf p_\xi'})\;
\gamma^0
\tens
\gamma^0
&\stackrel{!}{=}&
V^{(2)}
({\bf p_\xi}, {\bf p_\xi'}).
\end{eqnarray}
As usual, $\left[{\cal P},\;{\cal H}\right] = 0$
also implies that the solutions $\Phi^\Lambda_M$ of the Salpeter
equation simultaneously are eigenstates of ${\cal P}$, {\it i.e.} 
\begin{equation}
{\cal P} \Phi^\Lambda_{M,\;\pi} = \pi\; \Phi^\Lambda_{M,\;\pi},
\end{equation}
with definite parity $\pi = \pm 1$.
\subsubsection{Time-reversal invariance}
The representation of the time-reversal transformation ${\cal T}(x^0,{\bf
  x}) := (-x^0,{\bf x})$ on the full momentum space Salpeter amplitudes reads
\begin{equation}
\left[{\cal T}\Phi_M\right]({\bf p_\xi},{\bf p_\eta})
=
-\gamma^1\gamma^3\tens\gamma^1\gamma^3\tens\gamma^1\gamma^3\; \Phi_M^*({\bf -p_\xi},{\bf -p_\eta}). 
\end{equation}
Again the different energy components of $\Phi_M$ define invariant subspaces
under the time-reversal transformation according to the intertwining 
relation $\Lambda^\pm_i({\bf p_i})\;\gamma^1\gamma^3 =
\gamma^1\gamma^3\;{\Lambda^\pm_i}^*({-\bf p_i})$.
In particular, we find that the Salpeter projector is time-reversal
invariant, {\it i.e.} $[{\cal T},\;\Lambda]=0$, 
such that we have a representation of the time-reversal transformation on the
subspace of purely positive and negative components, {\it i.e.} for the
projected amplitudes $\Phi_M^\Lambda = \Lambda \Phi_M$ holds:
\begin{equation}
\left[{\cal T}\Phi_M^\Lambda\right]({\bf p_\xi},{\bf p_\eta})
=
-\gamma^1\gamma^3\tens\gamma^1\gamma^3\tens\gamma^1\gamma^3\; {\Phi_M^\Lambda}^*({\bf -p_\xi},{\bf -p_\eta}) 
\quad \textrm{with}\quad \Lambda\; {\cal T}\Phi_M^\Lambda = {\cal T}\Phi_M^\Lambda.
\end{equation}
To respect time-reversal invariance of the strong interaction, we must impose that
$\left[{\cal T},\;{\cal H}\right] = 0$.
Using the invariance property $[{\cal T},\;\Lambda]=0$ of $\Lambda$ and the
time-reversal invariance of the free Hamiltonian
${\cal H}_0$, {\it i.e.} $[{\cal T},\;{\cal H}_0]=0$
we end up with the conditions
\begin{eqnarray}
\label{Tinv_cond_V3V2}
-\gamma^1\gamma^3\!\tens\!\gamma^1\gamma^3\!\tens\!\gamma^1\gamma^3\;
{V^{(3)}}^*\!\!
(-{\bf p_\xi},-{\bf p_\eta};
 -{\bf p_\xi'},-{\bf p_\eta'})\;
\gamma^1\gamma^3\!\tens\!\gamma^1\gamma^3\!\tens\!\gamma^1\gamma^3\;
&\stackrel{!}{=}&
V^{(3)}
({\bf p_\xi},{\bf p_\eta};
{\bf p_\xi'},{\bf p_\eta'}),
\nn[2mm]
\gamma^1\gamma^3\tens\gamma^1\gamma^3
\;
{V^{(2)}}^*
(-{\bf p_\xi},-{\bf p_\xi'})\;
\gamma^1\gamma^3\tens\gamma^1\gamma^3
&\stackrel{!}{=}&
V^{(2)}
({\bf p_\xi}, {\bf p_\xi'}).
\end{eqnarray}

\subsubsection{${\cal CPT}$-symmetry -- Interpretation of negative bound-state masses}
\label{subsubsec:CPT}
\begin{figure}[!h]
  \begin{center} 
        \epsfig{file={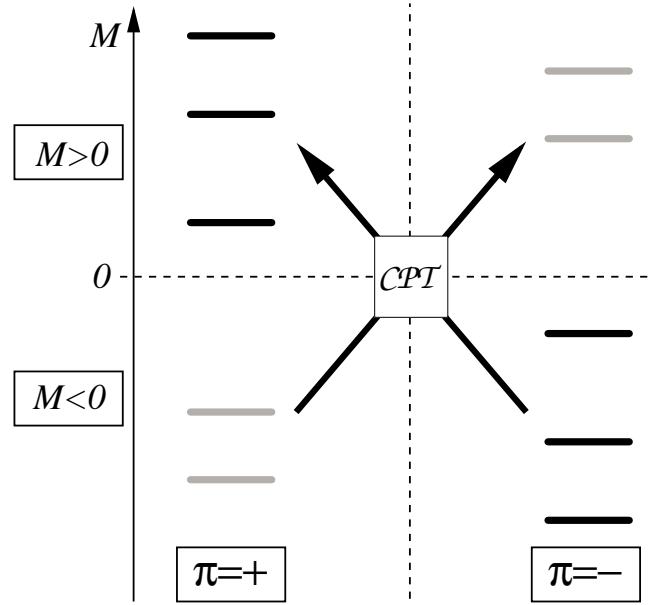},width=85mm}
    \end{center}
\caption{Interpretation of the negative energy solutions due to the
  ${\cal CPT}$-symmetry of the Salpeter equation represented by
  eq. (\ref{CPT_sym_SA}). See text for explanation.}
\label{fig:CPTsym}
\end{figure}

The Salpeter Hamiltonian ${\cal H}$ being hermitean with
respect to the positive definite scalar product (\ref{SP_V3}) guarantees
that the eigenvalues $M$, {\it i.e.} the bound-state masses, are real, as
one imposes for physically acceptable solutions.  However, ${\cal H}$ is
not positive definite, since even the free Hamiltonian
${\cal H}_0$ is not positive.  Accordingly, ${\cal H}$ in general possesses
both positive and negative eigenvalues and the spectrum might be even unbound from below.
These negative eigenvalues, at first face seem physically
unacceptable and the corresponding amplitudes also contradict the
normalization (\ref{norm_cond_V2V3_born}) via the positive definite
${\cal L}^2$-norm. Nevertheless, these negative energy solutions with $M<0$ can
 be interpreted physically meaningful. In fact, since our
covariant Salpeter approach is based on relativistic quantum field
theory, it should reveal a particle-antiparticle symmetry as a
characteristic feature due to ${\cal CPT}$-invariance.  Accordingly, we
demand that the instantaneous two- and three-quark interaction kernels
commute with the Dirac-space operator $\btens_{i=1}^3 \gamma^0\gamma^5$, {\it i.e.}
\begin{eqnarray}
\label{CPT_condV3}
\left[
\gamma^0\gamma^5\tens\gamma^0\gamma^5\tens\gamma^0\gamma^5,\;\; 
V^{(3)}({\bf p_\xi},{\bf p_\eta};\;{\bf p_\xi'},{\bf p_\eta'}) 
\right] 
&=& 
0\\
\label{CPT_condV2}
\left[\gamma^0\gamma^5\tens\gamma^0\gamma^5,\;\;
V^{(2)}({\bf p_\xi}, {\bf p_\xi'})
\right] 
&=& 
0
\end{eqnarray}
in order to ensure that the three-quark Salpeter equation in
fact respects the ${\cal CPT}$-symmetry of the strong interaction.
In this manner the negative bound-state masses get a well defined
physical interpretation as we will see in the following discussion.
The conditions (\ref{CPT_condV3}) and (\ref{CPT_condV2}) on $V^{(3)}$
and $V^{(2)}$ imply that the Salpeter Hamiltonian
${\cal H}$ given in eq. (\ref{Salp_Hamilt_V3_V2_born})
anticommutes with $\btens_{i=1}^3 \gamma^0\gamma^5$, {\it i.e.}
\begin{equation}
\label{anticom_CPT_H}
\left\{\gamma^0\gamma^5\tens\gamma^0\gamma^5\tens\gamma^0\gamma^5,\;
{\cal H}\right\} = 0
\end{equation}
as can easily be shown with the
anticommutator\footnote{The brackets $\{\cdot,\cdot\}$ denote the
anticommutator $\{A,B\}:= A B + B A$} and intertwining relations
\begin{equation}
\label{relation_g0g5}
\left\{\gamma^0\gamma^5,\; \gamma^0\right\} = 0,\quad
\left\{\gamma^0\gamma^5,\; H_i({\bf p_i})\right\} = 0\quad\textrm{and}\quad
\gamma^0\gamma^5\;\Lambda_i^\pm({\bf p_i})
= 
\Lambda_i^\mp({\bf p_i})\;\gamma^0\gamma^5.
\end{equation}
Moreover, it follows from (\ref{relation_g0g5}) that $\gamma^0\tens\gamma^0\tens\gamma^0$ and
hence also the representation ${\cal P}$ of the parity transformation
(\ref{repres_P_Phi}) anticommutes with $\btens_{i=1}^3 \gamma^0\gamma^5$:
\begin{equation}
\label{anticom_CPT_P}
\left\{\gamma^0\gamma^5\tens\gamma^0\gamma^5\tens\gamma^0\gamma^5,\;
{\cal P}\right\} = 0.
\end{equation}
Now let $\Phi_{-M,\;\pi}^\Lambda$ be a solution of the Salpeter equation
with negative mass $-M < 0$ and parity $\pi$ which obeys:
\begin{equation}
{\cal H}\;\Phi_{-M,\;\pi}^\Lambda = -M\;\Phi_{-M,\;\pi}^\Lambda,\quad {\cal P}\;\Phi_{-M,\;\pi}^\Lambda = \pi\;\Phi_{-M,\;\pi}^\Lambda.
\end{equation}
We consider the transformation $\Phi_{-M,\;\pi}^\Lambda
\mapsto \widetilde\Phi_{-M,\;\pi}^\Lambda$ of the amplitude $\Phi_{-M,\;\pi}^\Lambda$, given by
\begin{equation}
\widetilde\Phi_{-M,\;\pi}^\Lambda({\bf p_\xi}, {\bf p_\eta}) :=
\gamma^0\gamma^5\tens\gamma^0\gamma^5\tens\gamma^0\gamma^5\; \Phi_{-M,\;\pi}^\Lambda({\bf p_\xi}, {\bf p_\eta}).
\end{equation}
Then, due to eq. (\ref{anticom_CPT_H}), also this ${\cal CPT}$-transformed
amplitude $\widetilde\Phi_{-M,\;\pi}^\Lambda$ is a solution of the Salpeter
equation, but now with the positive bound-state mass $+M>0$. At the
same time eq. (\ref{anticom_CPT_P}) implies that $\widetilde\Phi_{-M,\;\pi}^\Lambda$ has parity
$-\pi$ opposite to $\Phi_{-M,\;\pi}^\Lambda$. Thus, we have
\begin{equation}
{\cal H}\;\widetilde\Phi_{-M,\;\pi}^\Lambda =
+M\;\widetilde\Phi_{-M,\;\pi}^\Lambda
\quad 
{\cal P}\;\widetilde\Phi_{-M,\;\pi}^\Lambda = -\pi\;\widetilde\Phi_{-M,\;\pi}^\Lambda
\end{equation}
Consequently, the eigenvalues come in pairs with opposite sign, but with 
eigenfunctions (Salpeter amplitudes) having opposite parity.  This
symmetry indeed allows the interpretation of the negative energy solutions for
a given set of quantum numbers as antibaryon states, which after the transformation
$\Phi_{-M,\;\pi}^\Lambda \mapsto \widetilde\Phi_{-M,\;\pi}^\Lambda$ yield positive
energy solutions of opposite parity but otherwise with the same quantum numbers:
\begin{equation}
\label{CPT_sym_SA}
\widetilde\Phi_{-M,\;\pi}^\Lambda
=
\btens_{i=1}^3 \gamma^0\gamma^5\; \Phi_{-M,\;\pi}^\Lambda
\;\;\equiv\;\; \Phi_{M,\;-\pi}^\Lambda.
\end{equation}
This is a new interesting feature of our Salpeter equation-based baryon
model in contrast to nonrelativistic (or relativized) quark potential models,
which are usually based on the ordinary Schr\"odinger equation: Solving
the Salpeter equation for fixed spin $J$ yields at the same time both the positive and the
negative parity bound state spectrum of the baryons, see fig.
\ref{fig:CPTsym} for a diagrammatical illustration of this feature.
Furthermore the positive and negative parity states are coupled in this way
and are not independent as in the ordinary nonrelativistic potential models.

Notice that, owing to the intertwining relation
$\gamma^0\gamma^5\;\Lambda_i^\pm({\bf p_i}) = \Lambda_i^\mp({\bf
p_i})\;\gamma^0\gamma^5$, the roles of the positive and negative energy
components are interchanged by the ${\cal CPT}$-transformation:
\begin{equation}
\Phi_{M,\;-\pi}^{+++} \;=\;  \btens_{i=1}^3 \gamma^0\gamma^5\;\;\Phi_{-M,\;\pi}^{---}
\quad\textrm{and}\quad
\Phi_{M,\;-\pi}^{---} \;=\;  \btens_{i=1}^3 \gamma^0\gamma^5\;\;\Phi_{-M,\;\pi}^{+++}.
\end{equation}
Consequently, only both subspaces of purely positive {\it and}
negative energy components together (but {\it not} separately) define
an invariant subspace under the ${\cal CPT}$-transformation. Thus
really both subspaces are necessary to get an (irreducible)
representation of ${\cal CPT}$. In particular the so-called reduced
Salpeter equation, in which the negative components are {\it a priori}
neglected (Tamm-Dancoff approximation), violates in general the
${\cal CPT}$-symmetry.

\subsection{Eigenstates of the Salpeter projector}
\label{subsec:struct_soluSE}
Let us come back to the specific projector structure of the Salpeter equation
and discuss in some more detail the corresponding induced structure of the solutions,
{\it i.e.} of the (projected) Salpeter amplitudes $\Phi_M^\Lambda =
\Lambda\Phi_M$, which obviously are eigenstates of the Salpeter projector.
These projection properties of the solutions reduce the number of independent
functions necessary to describe the baryon state.  Due to the form of the
Salpeter projector $\Lambda$ the solutions split into the two orthogonal
purely positive and purely negative energy components and thus the (in Dirac
space) 64-component function in fact reduces to an effectively 16-component
function only. To perform this reduction of the Salpeter amplitude to a
16-component function we have to determine the general form of the eigenstates
of the Salpeter projector $\Lambda$.  To this end we will consider first the
positive and negative energy solutions of the free Dirac equation for a
spin-$1/2$ particle. These four-component Dirac spinors, which are the
eigenstates of the energy projectors $\Lambda^\pm$, can be constructed in the
usual way by the embedding map of two-component Pauli spinors. This scheme can
then be generalized to the (projected) three-fermion Salpeter amplitudes
$\Phi_M^\Lambda$ which are eigenstates of the Salpeter projector $\Lambda$ and
accordingly are formed by a three-fermion embedding map.\\

For the following considerations it is convenient to adjust our notation to
the symmetry properties of the Salpeter amplitude under permutations of the
quarks, especially in the case of different quark masses.  So far we used a
simplified notation suppressing the flavor dependencies of the single-quark
operators $H_i$ and $\Lambda^\pm_i$ and we assigned to each quark $i$ an
individual quark mass $m_i$. With the replacements of $H_i$ and
$\Lambda^\pm_i$ given by
\begin{eqnarray}
H({\bf p_i}) 
&:=& \sum_f\; H_{m_f}({\bf p_i})\tens{\cal P}^{\cal F}_f,\quad\textrm{with}\quad
H_{m_f}({\bf p}) := \gamma^0\;({\bfgrk \gamma \cdot p } + m_f)\\
\label{redef_Lam}
\Lambda^\pm({\bf p_i}) 
&:=& 
\sum_f\; \Lambda^\pm_{m_f}({\bf p_i})\tens{\cal P}^{\cal F}_f,\quad\textrm{with}\quad
\Lambda_{m_f}^\pm({\bf p}) := \frac{\omega_{m_f}({\bf p}) \pm H_{m_f}({\bf p})}{2 \omega_{m_f}({\bf p})}\\
&&
\phantom{
\sum_f\; \Lambda^\pm_{m_f}({\bf p_i})\tens{\cal P}^{\cal F}_f,}\quad\;\textrm{and}\;\;\quad
\omega_{m_f}({\bf p}):= \sqrt{|{\bf p}|^2 + m_f^2},
\end{eqnarray}
the correct assignment of quark masses according to their flavor $f=u,d,s$ is
realized by the flavor projectors ${\cal P}^{\cal F}_f := \ket{f}\bra{f}$ such
that the free Hamiltonian ${\cal H}_0$ and the Salpeter projector
\begin{eqnarray}
\Lambda({\bf p_\xi}, {\bf p_\eta}) &=& \Lambda^{+++}({\bf p_\xi}, {\bf p_\eta}) + \Lambda^{---}({\bf p_\xi}, {\bf p_\eta}),\\[2mm]
\Lambda^{\pm\pm\pm}({\bf p_\xi}, {\bf p_\eta}) 
&:=&
\Lambda^\pm({\bf p_1})\tens 
\Lambda^\pm({\bf p_2})\tens 
\Lambda^\pm({\bf p_3})
\end{eqnarray}
become  permutationally invariant operators.

\subsubsection{Dirac spinors as embedded Pauli spinors}
Let us first consider the Dirac spinors
$\psi_m^\pm:\textrm{R}_m^\pm\rightarrow \C^4$ for a single spin-$1/2$ particle
with mass $m$, {\it i.e.} the positive and negative energy solutions of the
free Dirac equation on the positive/negative mass shell
$\textrm{R}_m^\pm:= \{p\in \R^4:\langle p,p \rangle=m^2,\;p =
(\pm\omega_m({\bf p}),{\bf p})\}$.  These are eigenstates of the
positive and negative energy projectors $\Lambda_m^\pm$:
\begin{eqnarray}
\Lambda_m^+({\bf p})\; \psi_m^+(p) &=& \psi_m^+(p)
\quad\textrm{with}\quad p=(+\omega_m({\bf p}), {\bf p})\in \textrm{R}_m^+\\[3mm]
\Lambda_m^-({\bf p})\; \psi_m^-(\tilde p) &=& \psi_m^-(\tilde p)
\quad\textrm{with}\quad \tilde p=(-\omega_m({\bf p}), {\bf p})\in \textrm{R}_m^-.
\end{eqnarray} 
As usual, the positive and negative energy solutions
$\psi_m^\pm:\textrm{R}_m^\pm\rightarrow \C^4$ of the Dirac equation may be
written in the Weyl representation as
\begin{eqnarray}
\label{embed_Tp}
\psi_m^+(p) \;=\; T^+_m({\bf p})\;\varphi_m^+(p)
\quad\textrm{with}\quad
T^+_m({\bf p}) &:=& \frac{1}{\sqrt{2\;\omega_m{({\bf p})}}}
\left(
\begin{array}{c}
\sqrt{\sigma({\cal P} p)}\\
\sqrt{\sigma(p)}
\end{array}
\right)
\\[2mm]
\label{embed_Tm}
\psi_m^-(\tilde p) \;=\; T^-_m({\bf p})\;\varphi_m^-(\tilde p)
\quad\textrm{with}\quad
T^-_m({\bf p}) &:=& \frac{1}{\sqrt{2\;\omega_m{({\bf p})}}}
\left(
\begin{array}{c}
-\sqrt{\sigma(p)}\\
\sqrt{\sigma({\cal P} p)}
\end{array}
\right)
\end{eqnarray}
where $\varphi_m^\pm: \textrm{R}_m^\pm\rightarrow \C^2$ are 
two-component Pauli spinors, ${\cal P}$ is the parity
transformation, {\it i.e.} ${\cal P}\;p= (\omega_m({\bf p}),-{\bf p})$ for
$p=(p^0,{\bf p})=(\omega_m({\bf p}),{\bf p})$, and
\begin{equation}
\sigma(p) := \sigma_\mu\;p^\mu\quad
\Rightarrow\quad
\sqrt{\sigma(p)} 
= 
\frac{\sigma(p)+m}{\sqrt{2(\omega_m{({\bf p})+m)}}},
\end{equation} 
with $\sigma_i$ the Pauli matrices and $\sigma_0 =
\sigma^0=\Id_{\C^2}$.  We wrote these relations already in a form that
defines the so-called \textbf{embedding operations} $T^\pm_m({\bf p}):\C^2\mapsto
\C^4$. They map arbitrary two-component Pauli spinors $\varphi_m^\pm$ into
four-component orthogonal eigenstates $\psi_m^\pm$ of the energy projectors
$\Lambda_m^\pm$. This is also apparent from the properties
$\Lambda_m^\pm({\bf p})\;T^\pm_m({\bf p}) = T^\pm_m({\bf p})$ and
$\Lambda_m^\mp({\bf p})\;T^\pm_m({\bf p}) = 0$ of these embedding maps.
On the other hand the mappings $\psi_m^\pm\leftrightarrow\varphi_m^\pm$ are
also unique, since they satisfy 
\begin{equation}
\label{isometry_prop_embed}
\left[T^\pm_m({\bf p})\right]^\dagger\;T^\pm_m({\bf p}) = \Id_{\C^2}
\quad\textrm{and thus}\quad
\varphi_m^\pm = {T^\pm_m}^\dagger\;\psi_m^\pm
\end{equation}
and, in particular, they are isometric operations: ${\psi_m^\pm}^\dagger
\psi_m^\pm = {\varphi_m^\pm}^\dagger \varphi_m^\pm$. Finally, we define single-quark
embedding operations
\begin{equation}
\label{single_quark_embed}
T^\pm({\bf p}) := \sum_f\; T^\pm_{m_f}({\bf p})\tens {\cal P}^{\cal F}_f
\end{equation}
which account for the correct mass assignment for each flavor $f$ 
and accordingly map to eigenstates of the
energy projectors $\Lambda^\pm({\bf p})$ defined in eq.
(\ref{redef_Lam}):
\begin{equation}
\Lambda^\pm({\bf p})\;T^\pm({\bf p}) = T^\pm({\bf p})
\quad\textrm{and}\quad
\Lambda^\mp({\bf p})\;T^\pm({\bf p}) = 0
\end{equation}

\subsubsection{Embedding map for Salpeter amplitudes $\Phi_M^\Lambda$}
Now we use this result for the construction of 
the solutions 
\begin{equation}
\Phi_M^\Lambda({\bf p_\xi}, {\bf p_\eta}) 
= 
\Phi_M^{+++}({\bf p_\xi}, {\bf p_\eta})
+
\Phi_M^{---}({\bf p_\xi}, {\bf p_\eta})
\end{equation}
of the Salpeter equation, whose components
\begin{equation}
\Phi_M^{\pm\pm\pm}({\bf p_\xi}, {\bf p_\eta}) 
\;=\;
\Lambda^{\pm\pm\pm}({\bf p_\xi}, {\bf p_\eta})\; 
\Phi_M({\bf p_\xi}, {\bf p_\eta})
\end{equation}
are eigenstates of the positive and  negative
energy projectors $\Lambda^{\pm\pm\pm}$. According to the above discussion we
now define the three-quark embedding maps by the tensor products
of single quark embedding operators (\ref{single_quark_embed})
\begin{equation}
\label{3q_embed_op}
T^{\pm\pm\pm}({\bf p_\xi}, {\bf p_\eta}) 
:=
T^\pm({\bf p_1})\tens 
T^\pm({\bf p_2})\tens 
T^\pm({\bf p_3}).
\end{equation}
Then the  positive and negative energy contributions
$\Phi_M^{\pm\pm\pm}$ to $\Phi_M^\Lambda$ can be uniquely written as
\begin{equation}
\Phi_M^{\pm\pm\pm}({\bf p_\xi}, {\bf p_\eta}) 
\;=\;
T^{\pm\pm\pm}({\bf p_\xi}, {\bf p_\eta})\;
\varphi_M^{\pm}({\bf p_\xi}, {\bf p_\eta})
\end{equation}
in terms of the embedded three-particle amplitudes $\varphi_M^{\pm}$ 
which involve triple tensor products of Pauli spinors only. 
Finally, we have for the Salpeter amplitude $\Phi_M^\Lambda$
the unique orthogonal decomposition
\begin{equation}
\Phi_M^\Lambda({\bf p_\xi}, {\bf p_\eta}) 
=  
T^{+++}({\bf p_\xi}, {\bf p_\eta})\;
\varphi_M^{+}({\bf p_\xi}, {\bf p_\eta})
\;+\;
T^{---}({\bf p_\xi}, {\bf p_\eta})\;
\varphi_M^{-}({\bf p_\xi}, {\bf p_\eta})
\end{equation}
and thus the determination of the (in Dirac space originally
$4\tens4\tens4=$) 64-component solution $\Phi_M^\Lambda$ of the Salpeter equation reduces via
the embedding map to finding the two (only $2\tens2\tens2=$) 8-component amplitudes
$\varphi_M^{\pm}$. Due to the isometry of the embedding maps $T^{\pm\pm\pm}$ which follows from
eq. (\ref{isometry_prop_embed}), the normalization condition
for the Salpeter amplitudes $\Phi_M^\Lambda$ can be expressed in terms of
these Pauli-spinors $\varphi_M^{\pm}$ according to 
\begin{equation}
\SP{\Phi_M^\Lambda} {\Phi_M^\Lambda} 
\;\;=\;\; 
\SP{\varphi_M^{+}} {\varphi_M^{+}} + \SP{\varphi_M^{-}} {\varphi_M^{-}}
\;\;=\;\; 2 M,
\end{equation}
where $\SP{\varphi_M^{\pm}} {\varphi_M^{\pm}}$ denotes the usual
nonrelativistic (positive definite) $\mathcal{L}^2$ norm.  The next step is to
investigate the structure of these three-quark Pauli amplitudes for a given
set of quantum numbers specifying a baryon, in order to find a proper basis.

\subsection{General decomposition of the Salpeter amplitudes}
The three-quark Salpeter amplitude $\Phi^\Lambda_M$ of a baryon is
characterized by a set of quantum numbers that are conserved under the strong
interaction. We consider in this work light baryons, which are built up by
quarks with flavors up (u), down (d) and strange (s). The flavor-$SU(3)$ symmetry
is explicitly broken to $SU(2)\tens U(1)$ by the different constituent
quark masses $m_u = m_d < m_s$.  With $m_u = m_d \equiv m_n$ only the $SU(2)$
isospin symmetry shall be assumed to be exact. Due to parity invariance,
rotational invariance and this (broken) flavor invariance a baryon is then
characterized by the parity $\pi$, the total spin $J$ with 3-component $M_J$,
isospin $T$ with 3-component $M_T$ and strangeness $S^*$. Moreover, according
to Pauli's principle, the (projected) Salpeter amplitude, together with its
positive and negative energy components, must be totally antisymmetric under
permutations $\sigma\in S_3$.\\

Consider the three-quark Salpeter amplitude
$\Phi^\Lambda_{M}\equiv {\Phi^\Lambda_{M\;\;}}_{J^\pi M_J T M_T S^*}$ describing a baryon with
the quantum numbers listed above. In order to determine the structure
of its embedded Pauli spinors, we have to investigate, how the
corresponding transformation properties of the Salpeter amplitude
$\Phi^\Lambda_{M}$ transfer to the Pauli
spinors $\varphi_M^{\pm}$ via the embedding maps $T^{+++}$ and
$T^{---}$:
\begin{itemize}
\item 
The representation ${\cal P}$ of the parity transformation 
is given for the Salpeter amplitudes $\Phi_M^\Lambda$ by
\begin{equation}
\left[{\cal P}\Phi_M^\Lambda\right]({\bf p_\xi},{\bf p_\eta})
=
\gamma^0\tens\gamma^0\tens\gamma^0\; \Phi_M^\Lambda({\bf -p_\xi},{\bf -p_\eta}). 
\end{equation}
With our special choice (\ref{embed_Tp}) and (\ref{embed_Tm}) for the
embedding operations $T_m^\pm$, we find the following simple intertwining
relations
\begin{equation}
{\cal P}\;T^{\pm\pm\pm} =  T^{\pm\pm\pm}\;\left[\pm{\cal P}'\right].
\end{equation}
On the right hand side, $\pm{\cal P}'$ is the corresponding induced
representation for the Pauli amplitudes $\varphi^{\pm}_M$, where
the symbol ${\cal P}'$ is used to denote the usual nonrelativistic
representation of the parity transformation, {\it i.e.}
\begin{equation}
\left[{\cal P}'\;\varphi^{\pm}_M\right]({\bf p_\xi}, {\bf p_\eta}) 
:= 
\varphi^{\pm}_M(-{\bf p_\xi}, -{\bf p_\eta}).
\end{equation}
Hence $T^{+++}$ preserves parity, whereas $T^{---}$ reverses parity, and,
consequently, for a Salpeter amplitude with parity $\pi$, {\it i.e.}
${\cal P} \Phi_M^\Lambda = \pi\;\Phi_M^\Lambda$, the positive energy Pauli 
amplitude has the same parity $\pi$, whereas the negative energy amplitude
has the opposite parity $-\pi$:
\begin{equation}
{\cal P}' \varphi^{\pm}_M =  \pm\pi\;\varphi^{\pm}_M.
\end{equation}
\item
The Salpeter amplitudes $\Phi_M^\Lambda$  transform under rotations $R_{\omega}\in
SO(3)$, with rotation vector $\bfgrk \omega \in \R^3$, as 
\begin{equation}
\left[{\cal D}_{R_\omega}\;\Phi_M^\Lambda\right]({\bf p_\xi},{\bf p_\eta})
=
S_u\tens S_u\tens S_u\;
\Phi_M^\Lambda(R_\omega^{-1}{\bf p_\xi},R_\omega^{-1}{\bf p_\eta}).
\end{equation}
In the Weyl representation we have
\begin{equation}
S_u
=
\left(
\begin{array}{cc}
u & 0\\
0 & u
\end{array}
\right)
\quad\textrm{where}\quad
u \;=\; \exp\;(-{\rm i}\; {\bfgrk\sigma}\cdot {\bfgrk \omega})\in SU(2),
\end{equation}
with $u\;\sigma(p)\;u^\dagger = \sigma(R_\omega\;p)$ and we find
the intertwining relations
\begin{equation}
{\cal D}_{R_\omega}\;T^{\pm\pm\pm} = T^{\pm\pm\pm}\;{\cal D}'_{R_\omega}.
\end{equation}
Here the induced representation ${\cal D}'_{R_\omega}$ of $R_\omega$, which acts
on the Pauli amplitudes $\varphi^{\pm}_M$, is exactly the usual
nonrelativistic representation of the rotation $R_\omega$ for a system of three
spin-$\frac{1}{2}$ fermions:
\begin{equation}
\left[{\cal D}'_{R_\omega}\;\varphi^{\pm}_M\right]({\bf p_\xi},{\bf p_\eta})
=
u\tens u\tens u\;
\varphi^{\pm}_M (R_\omega^{-1}{\bf p_\xi},R_\omega^{-1}{\bf p_\eta}).
\end{equation}
Thus, to get the irreducible subspaces $\{J,\; M_J = -J,\ldots,J\}$ of
${\cal D}_{R_\omega}$, {\it i.e.} the Salpeter amplitudes
$\Phi_M^\Lambda$ with definite total spin $J$ and 3-component $M_J$,
the Pauli amplitudes $\varphi^{\pm}_M$ have simply to be
the usual eigenstates of the total angular momentum operator ${\bf \hat J} =
{\bf \hat L} + {\bf \hat S}$ as in a nonrelativistic system of three
spin-$\frac{1}{2}$ fermions.
\item As mentioned already, the three-quark embedding operators
  $T^{\pm\pm\pm}$ explicitly break the flavor-$SU(3)$ symmetry by the
  different quark masses $m_n < m_s$ and only a $SU(2)\tens U(1)$ invariance
  remains. The operators of isospin and strangeness hence commute with the
  embedding maps $T^{\pm\pm\pm}$, and accordingly the Pauli amplitudes
  $\varphi^{\pm}_M$ are their eigenstates with quantum numbers $T$,
  $M_T$ and $S^*$.
\item   
  The three-quark embedding operations $T^{\pm\pm\pm}$ apparently are
  completely symmetric under arbitrary permutations $\sigma \in S_3$ of quarks
  by their construction (\ref{3q_embed_op}), {\it i.e.}
\begin{equation}
{\cal D}_\sigma\;T^{\pm\pm\pm}
= 
T^{\pm\pm\pm}\;{\cal D}'_\sigma,
\end{equation}
where $\mathcal{D}_\sigma$ and $\mathcal{D}'_\sigma$ are the representations
of the permutation $\sigma \in S_3$ on the Salpeter and Pauli amplitudes,
respectively. 
This is a crucial point that permits to reduce the symmetry considerations
of the Salpeter amplitudes to the embedded Pauli spinors. As the 
baryon Salpeter amplitude $\Phi^\Lambda_M$ must be totally antisymmetric, {\it i.e.}
\begin{equation}
{\cal D}_\sigma\; \Phi_M^\Lambda 
= 
\textrm{sign}(\sigma)\; \Phi_M^\Lambda\quad \forall \sigma\in S_3,
\end{equation}
the Pauli spinors themselves
must have this symmetry:
\begin{equation}
{\cal D}'_\sigma\; \varphi_M^{\pm} 
= 
\textrm{sign}(\sigma)\; \varphi_M^{\pm}\quad \forall \sigma\in S_3.
\end{equation}
\end{itemize}
In summary, the relativistic baryon Salpeter amplitude
${\Phi^\Lambda_{M\;\;}}_{J^\pi M_J T M_T S^*}$ with specific quantum
numbers $J$, $\pi$, $T$, $M_T$ and $S^*$ can be formed by embedding
ordinary totally antisymmetric nonrelativistic baryon wave functions
${\varphi_{M\;}}_{J^\pi M_J T M_T S^*}$:
\begin{eqnarray}
\label{embed_non_rel_bar_wf}
{\Phi^\Lambda_{M\;\;}}_{J^\pi M_J T M_T S^*}({\bf p_\xi}, {\bf p_\eta})
&=&
\phantom{+\;\;}
T^{+++}({\bf p_\xi}, {\bf p_\eta})\;
{\varphi^{+}_{M\;}}_{J^\pi M_J T M_T S^*}({\bf p_\xi}, {\bf p_\eta})\nn[2mm]
&&+\;\;
T^{---}({\bf p_\xi}, {\bf p_\eta})\;
{\varphi^{-}_{M\;}}_{J^{-\pi} M_J T M_T S^*}({\bf p_\xi}, {\bf p_\eta})
\end{eqnarray}
To define a basis for the totally antisymmetric three-quark Salpeter
amplitude we can thus proceed in the same manner  as in the nonrelativistic quark model, where
the baryon wave functions ${\varphi_{M\;}}_{J^{\pm\pi} M_J T M_T S^*}$
have the generic form
\begin{equation}
\label{NR_baryon_wf}
{\varphi_{M\;}}_{J^\pi M_J T M_T S^*}({\bf p_\xi},
  {\bf p_\eta})\;\; =
\sum_{
{\cal R}_{L},
{\cal R}_{S},
{\cal R}_{F}}
\Bigg\{
\left\{
\left[ 
\left[\psi^\pi_{L}({\bf p_\xi}, {\bf p_\eta})\right]_{{\cal R}_{L}} 
\tens
\left[\chi_S\right]_{{\cal R}_{S}}
\right]^J_{M_J}
\tens
\left[\phi^{T\;S^*}_{M_T}\right]_{{\cal R}_{F}} 
\right\}_{\cal S}
\tens 
{\cal C}_{\cal A}
\Bigg\}_{\cal A}\nonumber
\end{equation}
with
\begin{itemize}
\item
$\left[\psi^\pi_{L}({\bf p_\xi}, {\bf p_\eta})\right]_{{\cal R}_{L}}$
the momentum space wave function with total orbital angular momentum $L$, parity
$\pi$ and permutational symmetry ${\cal R}_{L}\in\{{\cal S},{\cal M}_{\cal S},{\cal M}_{\cal A}, {\cal A}\}$;
\item
$\left[\chi_S\right]_{{\cal R}_{S}}$
the spin function of three Pauli spinors coupled to total spin $S$
with permutational symmetry ${\cal R}_{S}\in\{{\cal S},{\cal M}_{\cal S},{\cal M}_{\cal A}\}$;
\item 
$\left[\phi^{T\;S^*}_{M_T}\right]_{{\cal R}_{F}}$ the flavor
function with total isospin $T$, $T_3$-component $M_T$ and 
strangeness $S^*$ which is of permutational symmetry
${\cal R}_{F}\in\{{\cal S},{\cal M}_{\cal S},{\cal M}_{\cal A}, {\cal A}\}$;
\item
${\cal C}_{\cal A}$
the totally antisymmetric color-singlet state given by the 
Levi-Civit\`{a} tensor:  
${\cal C}_{\cal A} =
\frac{1}{\sqrt{6}}\;\epsilon_{c_1 c_2 c_3}\;\ket{c_1}\tens\ket{c_2}\tens\ket{c_3}$.
\end{itemize}
The momentum space wave function $\psi^\pi_{L}$ and the spin function $\chi_S$ are coupled as usually to states of
total angular momentum $J,\;M_J$ according to
\begin{equation}
\left[
\left[\psi^\pi_{L}({\bf p_\xi}, {\bf p_\eta})\right]_{{\cal R}_{L}}
\tens
\left[\chi_S\right]_{{\cal R}_{S}}
\right]^J_{M_J}
=
\sum_{M_L, M_S}
\langle L M_L, S M_S | J M_J \rangle\;
\psi^\pi_{L M_L}({\bf p_\xi}, {\bf p_\eta})\;
\chi_{S M_S}
\end{equation} 
with Clebsch-Gordan coefficients $\langle L M_L, S M_S | J M_J
\rangle$; the sum over the symmetries ${\cal R}_{L}$, ${\cal R}_{S}$
and ${\cal R}_{F}$ in (\ref{NR_baryon_wf}) is such that the combined
momentum-, spin-, flavor wave function is totally symmetric,
${\cal R}_{L}\tens{\cal R}_{S}\tens{\cal R}_{F}= {\cal S}$, and, finally, the
baryon amplitude becomes totally antisymmetric with the totally antisymmetric
color-singlet state ${\cal C}_{\cal A}$.\\

Finally, let us discuss the implications of the present covariant approach
with respect to the baryonic spectrum and the nonrelativistic quark model.
According to the preceding discussion, the structure of the Salpeter
amplitudes seems to be very similar to that usually considered in the
nonrelativistic quark model: The Pauli amplitude embedded by the
positive energy embedding map $T^{+++}$ is  of exactly the same structure as
the usual nonrelativistic wave function for a given set of quantum numbers.
But note the additional negative energy contribution to the Salpeter
amplitude: For a specific parity $\pi$ of the baryon the embedding operator
$T^{---}$ brings also the nonrelativistic wave functions with the opposite
('wrong') parity $-\pi$ into play (due to the different behavior of $T^{+++}$
and $T^{---}$ under parity transformations). Thus, at a first glance, our
approach seems to posses a larger number of states than the nonrelativistic
approach.  But recall that the negative mass $(-M<0)$ solutions
of the Salpeter equation can be
interpreted as the antibaryon states to baryons of just the opposite parity
$-\pi$, due to the ${\cal CPT}$-symmetry of the Salpeter equation.
Exactly these negative mass states, which after ${\cal CPT}$ transformation
become the baryon states $(M>0)$ with opposite parity, correspond to the
additional states with the 'wrong' parity $-\pi$. This feature becomes even
more apparent, if we analyze the effect of the ${\cal CPT}$-transformation
on the embedded Pauli wave functions.  For the embedding maps we find
the relations
\begin{equation}
\label{prop_embed_CPT}
\gamma^0\gamma^5\tens \gamma^0\gamma^5\tens \gamma^0\gamma^5\;
T^{\pm\pm\pm}({\bf p_\xi}, {\bf p_\eta}) = \mp\; T^{\mp\mp\mp}({\bf
p_\xi}, {\bf p_\eta})
\end{equation}
such that the ${\cal CPT}$ transformation essentially switches the
embedding maps of positive and negative energy and hence also the
parity. Considering the decompositions of the negative mass solution
${\Phi^\Lambda_{-M\;\;}}_{J^\pi M_J T M_T S^*}$
\begin{equation}
{\Phi^\Lambda_{-M\;\;}}_{J^\pi M_J T M_T S^*}
=
T^{+++}\;
{\varphi^{+}_{-M\;}}_{J^\pi M_J T M_T S^*}
+
T^{---}\;
{\varphi^{-}_{-M\;}}_{J^{-\pi} M_J T M_T S^*}
\end{equation}
and of its related, ${\cal CPT}$-transformed positive mass solution of
opposite parity ${\Phi^\Lambda_{M\;\;}}_{J^{-\pi} M_J T M_T S^*}$
\begin{eqnarray}
{\Phi^\Lambda_{M\;\;}}_{J^{-\pi} M_J T M_T S^*}
&=&
T^{+++}\;
{\varphi^{+}_{M\;}}_{J^{-\pi} M_J T M_T S^*}
+
T^{---}\;
{\varphi^{-}_{M\;}}_{J^{\pi} M_J T M_T S^*}\nn
&=&
\gamma^0\gamma^5\tens
\gamma^0\gamma^5\tens
\gamma^0\gamma^5\;
{\Phi^\Lambda_{-M\;\;}}_{J^\pi M_J T M_T S^*}
\end{eqnarray}
the ${\cal CPT}$-transformation together with the property
(\ref{prop_embed_CPT}) of the embedding maps then yields the following
relations for the corresponding Pauli amplitudes,
\begin{eqnarray}
{\varphi^{+}_{M\;}}_{J^{-\pi} M_J T M_T S^*} 
&=& 
-\; {\varphi^{-}_{-M\;}}_{J^{-\pi} M_J T M_T S^*},\nn
{\varphi^{-}_{M\;}}_{J^{\pi} M_J T M_T S^*}
&=&
+\;{\varphi^{+}_{-M\;}}_{J^\pi M_J T M_T S^*},
\end{eqnarray}
which means that the ${\cal CPT}$-transformation just interchanges the
roles of both Pauli amplitudes. Thus we find in our present covariant
approach exactly the {\it same} number of states as in the nonrelativistic
quark model, a feature that in general can not be taken for granted in a
relativistic approach: Consider {\it e.g.}  the naive flavor-$SU(3)$
quark model. To explain the lowest lying multiplet it is assumed that the
ground state orbital wave function is a totally symmetric $S$-wave, the color
state is completely antisymmetric and hence the spin-flavor state has to be
totally symmetric, which in the nonrelativistic approximation restricts the
possible multiplets to a flavor octet with spin-$1/2$ and a flavor
decuplet with spin-$3/2$.  In a relativistic quark model, however,
the number of spin-degrees of freedom is doubled for each quark, due to the
presence of the lower components, which means that in the relativistic
flavor-$SU(3)$ model the number of possible symmetric spin-flavor
multiplets is much higher than in the nonrelativistic approach, see {\it
  e.g.}  \cite{BoMe79,Mey75,Ca93,HKM75}, in contrast to the experimental
findings, which can be explained qualitatively by the naive nonrelativistic
model very well.  Our approach does not reveal this problem, owing to the Salpeter
projector $\Lambda$ and the ${\cal CPT}$-symmetry of the Salpeter
equation, which circumvents such a proliferation of the number of states.
Note that this feature of our model is a direct consequence of the
instantaneous approximation or, more precisely, of the instantaneous ansatz
for the genuine three-body (confinement) kernel. In view of the success of
nonrelativistic quark models to account for the correct number of baryon
excitations we in fact consider this to be one of the main empirical
arguments to use this instantaneous ansatz.
  
\section{Summary and conclusion}
\label{sec:concl}
In this paper we presented how a relativistically covariant constituent quark
model for baryons can be constructed within the general framework of
quantum field theory. We started with the basic field theoretical quantities
describing bound states of three fermions -- the Bethe-Salpeter amplitudes and
their adjoints -- which form the residua at the bound-state poles of the
six-point Green's function. The Bethe-Salpeter amplitudes, which
might be considered as the covariant analogues of 'wave functions' in the
ordinary nonrelativistic approach, obey a homogeneous eight-dimensional
integral equation in momentum space -- the so-called Bethe-Salpeter equation.
In principle, this is the basic equation for the covariant description of
bound states of three quarks in the framework of QCD, {\it i.e.} solving this
equation for given single quark propagators and irreducible interaction
kernels the discrete spectrum of baryons is then determined by the
normalization condition.

However, neither the full quark propagators nor the interaction
vertices are reliably known functions in case of QCD such that
reasonable phenomenological approximations for these basic ingredients
of the Bethe-Salpeter equation were necessary. In order to remain as
close as possible in contact with the features of the non-relativistic
quark model we adopted the concept of constituent quark masses using
free quark propagators with effective quark masses and the concept to
describe the quark interactions by instantaneous, unretarded
potentials. Although both replacements are chosen purely phenomenologically
they are justified reasonably well by the apparent success of
nonrelativistic potential models. As in the corresponding framework
for mesons both assumptions then allowed a reduction of the full
(eight-dimensional) Bethe-Salpeter equation to a reduced
six-dimensional equation (Salpeter equation) in the case of
instantaneous three-quark forces. In this case we obtained an equation
for the reduced amplitudes (Salpeter amplitudes) with a structure
quite similar to the ordinary Schr\"odinger equation and the
normalization condition for the Bethe-Salpeter amplitudes reduced to
the ordinary ${\cal L}^2$ normalization condition inducing a positive
definite scalar product.  Complications arose, when two-particle
interactions appeared, since these unconnected forces within the
three-body system prevented a straightforward reduction as in the case
of a pure three-body interaction alone.  However, a reasonable
treatment of these forces within the Salpeter framework is important
since in quark models the three-body confinement forces are naturally
supplemented by two-body residual interactions like the
one-gluon-exchange or instanton-induced forces. We presented a method
how in connection with the genuine instantaneous three-body kernel a
reduction to a Salpeter equation of the same structure can
nevertheless be achieved by deriving an effective instantaneous
three-body kernel which parameterizes all effects of the two-body
interactions.

As a crucial property of the instantaneous approximation we found that it
leads to a one-to-one correspondence with the states of the non-relativistic
quark model, a fact which generally can not be taken for granted in
relativistic approaches according to the doubling of the spin-degrees of
freedom. In this respect the special projector structure of the
Salpeter equation reduces the number of functions necessary to describe the
bound state and thus circumvents a proliferation of the number of states: The
Salpeter amplitudes, which still contain the full Dirac structure with positive
and negative energy components, can be formed by an isometric embedding map of
ordinary non-relativistic three-quark Pauli wave functions.

We found the appearance of the negative energy components to be related to
the particle-antiparticle symmetry due to the ${\cal CPT}$ invariance: The
spectrum of the Salpeter equation contains antiparticle solutions
corresponding to particles with charge conjugated quantum numbers. This is a new
feature of our Salpeter model for baryons and quite in contrast to ordinary
nonrelativistic or relativized quark models. Solving the Salpeter equation for
fixed spin $J$ yields at the same time both the positive- {\it and}
negative-parity bound-state spectrum and in particular positive and negative
parity states are coupled in this way and are not independent as in
nonrelativistic approaches.

The fully relativistic kinematics and the formal covariance of our
approach overcomes the old difficulties of nonrelativistic approaches
which in fact should be completely inadequate for small constituent
quark masses.  We expect the three-quark Salpeter equation to provide
a more reasonable framework for quark models of baryons that should be
superior to other treatments such as the nonrelativistic potential model
\cite{IsKa78,IsKa79} or its simple so-called ''relativized'' extension
\cite{CaIs86}. In particular it offers the possibility to investigate
the effects of the full Dirac structure of residual forces like the
one-gluon-exchange or instanton-induced interaction and moreover it
allows for the first time a reliable test of possible assumptions
concerning the Dirac structure of three-body confining forces.  In two
subsequent papers \cite{Loe01b,Loe01c} we will therefore investigate explicit quark models
based on the purely theoretical results of this paper and present
concrete calculations of the complete non-strange and strange baryon
spectrum up to 3 GeV.

\begin{acknowledgement}
  \textbf{Acknowledgments:} We have profited very much from scientific
  discussions with V.~V.~Anisovich, G.~E.~ Brown, E.~Klempt, A.~V.~Sarantsev
  and E.~V.~Shuryak to whom we want to express our gratitude. We also thank
  the Deutsche Forschungsgemeinschaft (DFG) for financial support.
\end{acknowledgement}

\appendix
\section{Appendix: Determination of the effective kernel $V^{\rm eff}_{M}$}
\label{sec:det_eff_K}
In this appendix we derive a prescription to construct the effective quasi
potential $V^{\rm eff}_{M}$ which has been introduced in sect. \ref{sec:reduc_V3V2}. According to
eq. (\ref{proj_red_calG}), $V^{\rm eff}_{M}$ is defined by
\begin{equation}
\label{proj_red_calG_app}
\langle{\cal G}_{M}\rangle_\Lambda
\stackrel{!}{=}
\Lambda\langle{\cal G}_{M}\rangle\overline\Lambda,
\end{equation}
where on the left $\langle{\cal G}_{M}\rangle_\Lambda$ is given by
eq. (\ref{reduced_IE_Veff}) which defines $V^{\rm eff}_{M}$, {\it i.e.}
\begin{equation}
\label{reduced_IE_Veff_app}
\langle{\cal G}_{M}\rangle_\Lambda
\;\stackrel{!}{=}\;
\langle{G_0}_{M}\rangle 
-\textrm{i}\;\langle{G_0}_{M}\rangle\;V^{\rm eff}_{M}\;
\langle{\cal G}_{M}\rangle_\Lambda
\end{equation}
and on the right ${\cal G}_{M}$ is the solution of the integral equation
(\ref{inhom_IE_GP}) with the integral kernel $K^{\rm R}_{M} := V^{(3)}_{\rm R}
+ \overline K^{(2)}_{M}$, {\it i.e.}
\begin{equation}
\label{inhom_IE_GP_app}
{\cal G}_{M}
=
{G_0}_{M} -\textrm{i}\;{G_0}_{M}\;K^{\rm R}_{M} \;{\cal G}_{M}.
\end{equation}
Now the goal is to solve eq. (\ref{proj_red_calG_app}) for $V^{\rm
eff}_{M}$.  For power counting purposes purpose we multiply the kernel
$K^{\rm R}_{M}$ by a parameter $\lambda\in [0,1]$,
\begin{equation}
\label{replace_KR_lamKR}
K^{\rm R}_{M} \longrightarrow \lambda\; K^{\rm R}_{M},
\end{equation}
such that the Neumann series of  ${\cal G}_M$
becomes a power series in $\lambda$, and thus
\begin{equation}
\label{neumann_series_calG}
\Lambda\langle{\cal G}_{M}\rangle\overline\Lambda
=
\langle{G_0}_{M}\rangle 
+
\sum_{k=1}^\infty \lambda^k\; 
\Lambda\langle{G_0}_{M} 
\underbrace{\left[-\textrm{i}\;K^{\rm R}_{M}\right] {G_0}_{M}\;\ldots\; \left[-\textrm{i}\;K^{\rm R}_{M}\right]{G_0}_{M}}_
{k\;\; {\rm times}}\rangle\overline\Lambda.
\end{equation}
The effective kernel $V^{\rm eff}_{M}$ becomes a function of
$\lambda$ which is expanded into a Taylor series according to
\begin{equation}
\label{series_ansatz_Veff}
V^{\rm eff}_{M} 
:= 
\sum_{k=1}^\infty \lambda^k\;  {V^{\rm eff}_{M}}^{(k)}.
\end{equation}
Inserting this into eq. (\ref{reduced_IE_Veff_app}) the Neumann
series of $\langle{\cal G}_{M}\rangle_\Lambda$ yields a multiple
power series in $\lambda$:
\begin{eqnarray}
\label{series_reduced_calG_lam}
\left\langle{\cal G}_{M}\right\rangle_\Lambda&=&\langle{G_0}_{M}\rangle 
-{\rm i}\;\sum_{k=1}^\infty \lambda^k\; \langle{G_0}_{M}\rangle\;{V^{\rm eff}_{M}}^{(k)}\;\langle{G_0}_{M}\rangle\\[4mm]
&+&
\sum_{r=2}^\infty\;\;\sum_{k_1 = 1}^\infty\;\ldots\;\sum_{k_r = 1}^\infty
\lambda^{k_1+k_2+\ldots +k_r}\;
\langle{G_0}_{M}\rangle\left[-\textrm{i}\;{V^{\rm eff}_{M}}^{(k_1)}\right]\langle{G_0}_{M}\rangle
\ldots 
\left[-\textrm{i}\;{V^{\rm eff}_{M}}^{(k_r)}\right]\langle{G_0}_{M}\rangle.\nonumber
\end{eqnarray}
Collecting all terms of equal power in the third term (which is of the order
$\lambda^{\geq 2}$), the multiple power series can be transformed into an ordinary series
\begin{eqnarray}
\label{series_reduced_calG_lam2}
\left\langle{\cal G}_{M}\right\rangle_\Lambda&=&\langle{G_0}_{M}\rangle 
-{\rm i}\;\sum_{k=1}^\infty \lambda^k\; \langle{G_0}_{M}\rangle\;{V^{\rm eff}_{M}}^{(k)}\;\langle{G_0}_{M}\rangle\\[4mm]
&+&
\sum_{k=2}^\infty\;\; \lambda^k\;\;\; \sum_{r=2}^k\!\!\!
\sum_{\footnotesize
\begin{array}{c}
k_1, k_2, \ldots, k_r < k\\ 
k_1 + k_2 + \ldots + k_r = k
\end{array}}
\!\!\!\!\!
\langle{G_0}_{M}\rangle\left[-\textrm{i}\;{V^{\rm eff}_{M}}^{(k_1)}\right]\langle{G_0}_{M}\rangle
\ldots 
\left[-\textrm{i}\;{V^{\rm eff}_{M}}^{(k_r)}\right]\langle{G_0}_{M}\rangle.\nonumber
\end{eqnarray}
Finally, we insert the resulting series (\ref{neumann_series_calG}) and
(\ref{series_reduced_calG_lam2}) into eq. (\ref{proj_red_calG_app}) and we arrive at
\begin{eqnarray}
\label{series_defining_cond}
\lefteqn{
\sum_{k=1}^\infty \lambda^k\; \langle{G_0}_{M}\rangle\;{V^{\rm eff}_{M}}^{(k)}\;\langle{G_0}_{M}\rangle 
\;\;=\;\;
\textrm{i}\;\sum_{k=1}^\infty \lambda^k\; 
\Lambda\left\langle{G_0}_{M} 
\underbrace{\left[-\textrm{i}\;K^{\rm R}_{M}\right] {G_0}_{M}\;\ldots\; \left[-\textrm{i}\;K^{\rm R}_{M}\right]{G_0}_{M}}_
{k\;\; {\rm times}}\right\rangle\overline\Lambda
}&&\nn[4mm]
&-\;\textrm{i}&
\sum_{k=2}^\infty\;\; \lambda^k\;\;\; \sum_{r=2}^k\!\!\!
\sum_{\footnotesize
\begin{array}{c}
k_1, k_2, \ldots, k_r < k\\ 
k_1 + k_2 + \ldots + k_r = k
\end{array}}
\!\!\!\!\!
\langle{G_0}_{M}\rangle\left[-\textrm{i}\;{V^{\rm eff}_{M}}^{(k_1)}\right]\langle{G_0}_{M}\rangle
\ldots 
\left[-\textrm{i}\;{V^{\rm eff}_{M}}^{(k_r)}\right]\langle{G_0}_{M}\rangle
\end{eqnarray}
which now enables us to solve for $V^{\rm eff}_{M}(\lambda)$ order-by-order 
by comparing the expansion coefficients of each power $k$ of $\lambda$. 
Thus, we find the following reduced terms which are irreducible with
respect to $\langle{G_0}_{M}\rangle$:
\begin{itemize}
\item
In lowest order, {\it i.e.} ${\bf k=1}$, we obtain the Born term ${V^{\rm eff}_{M}}^{(1)}$:
\begin{eqnarray}
\langle{G_0}_{M}\rangle\;{V^{\rm eff}_{M}}^{(1)}\;\langle{G_0}_{M}\rangle
=
\Lambda\left\langle{G_0}_{M}\; K^{\rm R}_{M}\;{G_0}_{M}\right\rangle\overline\Lambda
\end{eqnarray}
\item
and in $k$th order, ${\bf k\geq 2}$, we get ${V^{\rm eff}_{M}}^{(k)}$,
determined by:
\begin{eqnarray}
\lefteqn{
\langle{G_0}_{M}\rangle\;{V^{\rm eff}_{M}}^{(k)}\;\langle{G_0}_{M}\rangle
\;=\;
\textrm{i}\;\Lambda\left\langle{G_0}_{M} 
\underbrace{\left[-\textrm{i}\;K^{\rm R}_{M}\right] {G_0}_{M}\;\ldots\; \left[-\textrm{i}\;K^{\rm R}_{M}\right]{G_0}_{M}}_
{k\;\; {\rm times}}\right\rangle\overline\Lambda}&&\\[2mm]
&&
-\;\textrm{i}\;\sum_{r=2}^k\!\!\!
\sum_{\footnotesize
\begin{array}{c}
k_1, k_2, \ldots, k_r < k\\ 
k_1 + k_2 + \ldots + k_r = k
\end{array}}\!\!\!\!\!
\langle{G_0}_{M}\rangle\left[-\textrm{i}\;{V^{\rm eff}_{M}}^{(k_1)}\right]\langle{G_0}_{M}\rangle
\ldots 
\left[-\textrm{i}\;{V^{\rm eff}_{M}}^{(k_r)}\right]\langle{G_0}_{M}\rangle\nonumber
\end{eqnarray}
\end{itemize}
Finally, we amputate the free Salpeter propagators
$\langle{G_0}_{M}\rangle$ using the Hamiltonian ${h_0}_M$ with eq.
(\ref{inversion_G0}) and thus, with the restriction (\ref{restriction_Veff})
for $V^{\rm eff}_{M}$, we then can solve uniquely for ${V^{\rm
    eff}_{M}}^{(k)}$. Consequently, we get the effective kernel $V^{\rm
  eff}_{M}$ as the following infinite sum of irreducible interaction terms
${V^{\rm eff}_{M}}^{(k)}$:
\begin{equation}
\label{Veff_series_app}
V^{\rm eff}_{M} 
\;\;=\;\; 
\sum_{k=1}^\infty\;\;{V^{\rm eff}_{M}}^{(k)},
\end{equation}
where
\begin{eqnarray}
\label{Veff_born_app}
{V^{\rm eff}_{M}}^{(1)}
&=&
{h_0}_M\;\Lambda\left\langle{G_0}_{M}\; \overline K^{(2)}_{M}\;{G_0}_{M}\right\rangle\overline\Lambda\;{h_0}_M,\\[2mm]
\label{Veff_order_k_app}
{V^{\rm eff}_{M}}^{(k)}
&=&
\textrm{i}\;{h_0}_M\;\Lambda\left\langle{G_0}_{M} 
\underbrace{
(-\textrm{i})\left[V^{(3)}_{\rm R} + \overline K^{(2)}_{M}\right] {G_0}_{M}
\;\ldots\; 
(-\textrm{i})\left[V^{(3)}_{\rm R} + \overline K^{(2)}_{M}\right]{G_0}_{M}}_
{k\;\; {\rm times}}\right\rangle\overline\Lambda\;{h_0}_M\nn
\hspace*{10mm}&&
-\;\textrm{i}\;\sum_{r=2}^k\!\!\!\!\!
\sum_{\footnotesize
\begin{array}{c}
k_1, k_2, \ldots, k_r < k\\ 
k_1 + k_2 + \ldots + k_r = k
\end{array}}\!\!\!\!\!\!\!
\left[-\textrm{i}\;{V^{\rm eff}_{M}}^{(k_1)}\right]\langle{G_0}_{M}\rangle
\left[-\textrm{i}\;{V^{\rm eff}_{M}}^{(k_2)}\right]\langle{G_0}_{M}\rangle
\ldots 
\left[-\textrm{i}\;{V^{\rm eff}_{M}}^{(k_r)}\right].\!\!\!\nn
\end{eqnarray}
Finally, let us discuss how the instantaneous term $V^{(3)}_{\rm R}$, {\it
i.e.} that part of $V^{(3)}$ that couples to the mixed energy components,
enters in ${V^{\rm eff}_{M}}^{(k)}$.  As we mentioned already, this part of
$V^{(3)}$ appears solely in connection with $K^{(2)}_{M}$. More specifically:
\begin{enumerate}
\item In the Born term ${V^{\rm eff}_{M}}^{(1)}$ the isolated
  contribution of $V^{(3)}_{\rm R}$ vanishes due to its instantaneity and its
  projector property $\overline \Lambda V^{(3)}_{\rm R} \Lambda = 0$:
\begin{equation}
\left\langle{G_0}_{M}\; V^{(3)}_{\rm R} \;{G_0}_{M}\right\rangle 
=
\left\langle{G_0}_{M}\right\rangle \; V^{(3)}_{\rm R} \;\left\langle{G_0}_{M}\right\rangle 
=
\left\langle{G_0}_{M}\right\rangle \; \overline \Lambda V^{(3)}_{\rm R} \Lambda \;\left\langle{G_0}_{M}\right\rangle 
= 0,
\end{equation}
where we used $\langle{G_0}_{M}\rangle=\langle{G_0}_{M}\rangle\overline \Lambda = \Lambda\langle{G_0}_{M}\rangle$; 
\item
For the same reason, reduced Feynman diagrams with more than two direct iterations
of $V^{(3)}_{\rm R}$ disappear:
\begin{eqnarray}
\left\langle{G_0}_{M}
\ldots {G_0}_{M}
\; V^{(3)}_{\rm R} \;{G_0}_{M}
\; V^{(3)}_{\rm R} \;{G_0}_{M}
\; V^{(3)}_{\rm R} \;{G_0}_{M}
\ldots{G_0}_{M}\right\rangle&=&\nn 
\left\langle{G_0}_{M}
\ldots {G_0}_{M}\right\rangle 
\; V^{(3)}_{\rm R}
\;
\underbrace{\left\langle{G_0}_{M}\right\rangle 
\; V^{(3)}_{\rm R} 
\;\left\langle{G_0}_{M}\right\rangle}_{= 0} 
\; V^{(3)}_{\rm R} \;\left\langle{G_0}_{M}
\ldots{G_0}_{M}\right\rangle 
&=& 0.
\end{eqnarray}
\item
Also, the reduced irreducible kernel $V^{\rm eff}_{M}$ does not contain two  direct iterations
of $V^{(3)}_{\rm R}$ either, since such terms are reducible with respect to 
$\langle{G_0}_{M}\rangle$, because of
\begin{eqnarray}
\left\langle{G_0}_{M}\ldots {G_0}_{M}
\; V^{(3)}_{\rm R} \;{G_0}_{M}\; V^{(3)}_{\rm R} 
\;{G_0}_{M}\ldots{G_0}_{M}\right\rangle\nn 
=\left\langle{G_0}_{M} \ldots {G_0}_{M}\right\rangle 
\; V^{(3)}_{\rm R}\;
\left\langle{G_0}_{M}\right\rangle 
\; V^{(3)}_{\rm R}\;
\left\langle{G_0}_{M} \ldots {G_0}_{M}\right\rangle 
\end{eqnarray}
and thus are built by iterating two reduced Feynman diagrams of lower order.
\end{enumerate}
Therefore, we conclude that $V^{(3)}_{\rm R}$ emerges in $V^{\rm eff}_{M}$
only such that $K^{(2)}_{M}$ is always directly attached to $V^{(3)}_{\rm R}$
from the left and/or the right hand side. This means that in the Green's
function $\langle{\cal G}_{M}\rangle_\Lambda$ (and even in
${\cal G}_{M}$) at most two direct iterations of $V^{(3)}_{\rm R}$ can
occur.  We want to remark here that this limitation of the number of direct
iterations of $V^{(3)}_{\rm R}$ offers an alternative counting scheme for the
determination of $V^{\rm eff}_{M}$ via power series expansion, namely in
powers of $\overline K^{(2)}_{M}$ instead of powers of
$K^{\rm R}_{M} = V^{(3)}_{\rm R} + \overline K^{(2)}_{M}$.

%
%
%
%
%

\end{document}